\documentclass[a4paper,11pt]{article}
\usepackage{jheppub}

\usepackage[T1]{fontenc} 

\def\ket#1{\left|\,#1\,\right\rangle}

\def\ket#1{\left|\,#1\,\right\rangle}

\newcommand{\s}{{\rm\scriptscriptstyle S}}
\newcommand{\h}{{\rm\scriptscriptstyle IS}}

\newcommand{\jL}{j_{\scriptscriptstyle {\rm L}}}
\newcommand{\jR}{j_{\scriptscriptstyle {\rm R}}}
\newcommand{\nL}{n_{\scriptscriptstyle {\rm L}}}
\newcommand{\nR}{n_{\scriptscriptstyle {\rm R}}}
\def\bh{\hat b}

\def\bea{\begin{eqnarray*}}
\def\eea{\end{eqnarray*}}
\newcommand{\sa}{^{\scriptscriptstyle {\rm A}}}
\newcommand{\sB}{^{\scriptscriptstyle {\rm B}}}
\newcommand{\sL}{^{\scriptscriptstyle {\rm L}}}
\newcommand{\sIL}{^{\scriptscriptstyle {\rm IL}}}
\newcommand{\+}{ {\scriptscriptstyle\dot{+}}}
\newcommand{\m}{ {\scriptscriptstyle\dot{-}}}

\title{\boldmath Non-rational $\widehat{\rm su}(2)$ cosets and Liouville field theory}

\author[a]{Zbigniew Jask\'{o}lski }
\author[a,b]{Paulina Suchanek}

\affiliation[a]{ Institute of Theoretical Physics,
	 University of Wroc{\l}aw,\\
	 pl. M. Borna 1, 95-204~Wroc{\l}aw, Poland}
\affiliation[b]{Theory Group, SLAC National Accelerator Laboratory,\\
Menlo Park, CA 94025, USA
}

\emailAdd{jask@ift.uni.wroc.pl}
\emailAdd{suchanek@slac.stanford.edu}

\abstract{
We propose an $\widehat{\rm su}(2)$ WZNW model with a non-rational level and a continuous spectrum based on the non-unitary hermitian
representations of the chiral algebra $\widehat{\rm su}(2)_\kappa$. It is conjectured that for this model
the continuous spectra counterpart of the Goddard-Kent-Olive (GKO)  coset construction yields the Liouville
and the imaginary Liouville field theories. We support the conjecture by a number of nontrivial tests based on analytic calculations.
}
\notoc

\begin{document}
\maketitle

\hspace{50pt}

\section{Introduction}

It was conjectured several years ago  that
partition functions of ${\cal N} = 2$ superconformal ${\rm SU}(N)$ gauge theories in four dimensions are directly related to
correlation functions of the two-dimensional Liouville/Toda field theories \cite{Alday:2009aq,Wyllard:2009hg}.
One of generalizations of the AGT correspondence was the  relation between ${\cal N}=2$ ${\rm SU}(N)$ gauge theories on $\mathbb{R}^4/\mathbb{Z}_p$ and para-Liouville/Toda theories \cite{Belavin:2011pp,Nishioka:2011jk}.
It was in particular observed  \cite{Bonelli:2011jx,Bonelli:2011kv,Wyllard:2011mn,Belavin:2011sw} that in the case $N=p=2$ the blow-up formula for the Nekrasov partition function suggests the SL-LL relation of
the schematic form
\begin{equation}
\label{mrel}
\hbox{free fermion}\,\otimes\,{\cal N} =1\ \hbox{super-Liouville}
\hskip 8pt \sim \hskip 8pt
\hbox{Liouville} \ \otimes_P\, \hbox{Liouville},
\end{equation}
where the symbol $\otimes_P$ denotes a projected tensor product in which only selected pairs of conformal families are present.
\newpage

An explanation of relations
  of this kind was given in \cite{Wyllard:2011mn}.
It was motivated by old results \cite{Crnkovic:1989gy,Crnkovic:1989ug,Lashkevich:1992sb,Lashkevich:1993fb} relating various rational models
realized as quotients
\begin{equation}
\label{rationalcoset}
V(p,m) \sim \frac{\widehat{su}(2)_m  \times  \widehat{su}(2)_p}{\widehat{su}(2)_{m+p}},
\end{equation}
where  $\widehat{su}(2)_p$ denotes the $su(2)$ Kac-Moody algebra of level $p$.
Relation (\ref{mrel}) can be seen as a non-rational counterpart of the relation between the Virasoro minimal models $V(m)=V(1,m)$ and the ${\cal N}=1$  superconformal minimal models $SV(m)=V(2,m)$  \cite{Crnkovic:1989gy,Crnkovic:1989ug,Lashkevich:1992sb,Lashkevich:1993fb}:
$$
V(1)\otimes SV(m) \sim V(m)\otimes_P V(m+1), \;\;\;\;m=1,2,\dots.
$$
Essential elements of  the SL-LL relation in the Neveu-Schwarz sector were discussed in \cite{Belavin:2011sw}.
The proof has been recently completed in \cite{SL-LL} and the extension to the Ramond sector
was analyzed in \cite{Schomerus:2012se}.

The exact  SL-LL correspondence raises  the questions about other relations of this type.
The main  aim of the present paper is to  formulate and justify  the relations of  the following schematic form
\begin{equation}
\label{rel}
\begin{array}{rcl}
\widehat{\rm su}(2)_{\kappa} \ \otimes\ \widehat{\rm su}(2)_{1}
\hskip 8pt &\sim &\hskip 8pt
\hbox{Liouville}\ \otimes_P\ \widehat{\rm su}(2)_{\kappa+1},
\\
\widehat{\rm su}(2)_{\kappa} \ \otimes\ \widehat{\rm su}(2)_{1}
\hskip 8pt &\sim & \hskip 8pt
\hbox{imaginary Liouville}\ \otimes_P\ \widehat{su}(\rm 2)_{\kappa+1},
\end{array}
\end{equation}
with continuous level $\kappa\in \mathbb{R}$ related to the central charge of the Liouville theory
and with continuous spectra of $\widehat{su}(2)_{\kappa}$ and $\widehat{su}(2)_{\kappa+1}$ WZNW models.
On the one hand side these relations can be seen as non-rational counterparts of the famous Goddard-Kent-Olive coset construction of minimal models
\cite{GKO} with the  branching functions encoded in the projected tensor product.
On the other  hand they go beyond the standard coset construction.
Being  motivated by the SL-LL equivalence
we propose to regard
 the relations above as exact equivalences of CFT$_2$ models.
This in particular implies exact relations between structure constants and correlation functions of all the  models involved.

Before entering the discussion of the  $\widehat{su}(2)_{\kappa}$ WZNW theory
it is instructive to place relation (\ref{rel}) in a slightly wider context.
The Virasoro minimal models \cite{Belavin:1984vu},  the Dotsenko-Fateev (DF) models\footnote{
By  the DF model we mean a (not unitary) CFT  with nonrational  central charge and the infinite
discrete diagonal spectrum consisting of all degenerate weights \cite{Felder:1989wv}. This model was recently discussed in
 \cite{Ribault:2014hia}
under the name {\it generalized minimal model}. As this term is frequently used for the imaginary Liouville theory
we shall use the  name proposed in \cite{Felder:1989wv}.
}
\cite{Dotsenko:1984nm,Dotsenko:1984ad}, the Liouville theory
\cite{Dorn:1994xn,Zamolodchikov:1995aa,Teschner:1995yf,Teschner:2001rv}
 and the imaginary Liouville theory (the generalized minimal model) \cite{Zamolodchikov:2005fy,Zamolodchikov:2005sj,Kostov:2005kk,Ribault:2015sxa}
form a system of interrelated objects connected by analytic
continuations of different structures.
For instance the structure constants of the minimal models and the DF model are  analytic continuations of
the imaginary Liouville  structure constants \cite{Ribault:2014hia}.
The (real) Liouville structure constants cannot be obtained in this way
which can be seen as yet another manifestation of the  $c=1$ barrier \cite{Zamolodchikov:2005fy,Zamolodchikov:2005sj,Kostov:2005kk}.
They are however unique solutions to
the equations obtained by the analytic continuation of the corresponding equations in the Liouville theory and
the DF structure constants can be identified as their residues at the degenerate weights.
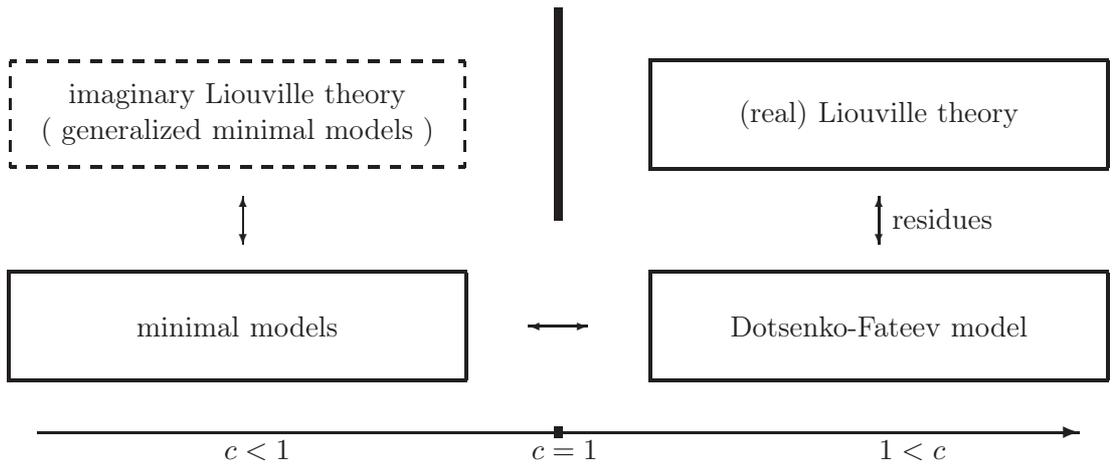
\begin{figure}
\begin{picture}(400,160)(-30,0)
\setlength\fboxsep{0pt}
\linethickness{1pt}

\put(240,110){
{\framebox(170,40)[c]{\shortstack[c]{
            (real) Liouville theory
          }}}}
\put(0,110){\dashbox{4}(170,40)[c]{\shortstack[c]{
            imaginary Liouville
            theory
            \\
           ( generalized minimal models )
          }}}

{\linethickness{3pt}
\put(205,90){\line(0,1){80}}
\put(205,8){\line(0,1){4}}}


\put(0,30){\framebox(170,40)[c]{\shortstack[c]{
            minimal models
          }}}
\put(240,30){\framebox(170,40)[c]{\shortstack[c]{
          Dotsenko-Fateev model
          }}}
\linethickness{.5pt}

\put(87,90){\vector(0,1){9}}
\put(87,90){\vector(0,-1){9}}
\put(325,90){\vector(0,1){9}}
\put(325,90){\vector(0,-1){9}}
\put(205,50){\vector(-1,0){11}}
\put(205,50){\vector(1,0){11}}


 \thicklines \put(10,10){\vector(1,0){390}}
\put(195,0){\shortstack[c]{
          $c=1$
          }}
\put(325,0){\shortstack[c]{
          $1<c$
          }}
\put(80,0){\shortstack[c]{
          $c<1$
          }}
\put(330,87){\shortstack[c]{
        residues
          }}
\end{picture}
\caption{ Virasoro models}
\end{figure}

The underlying fundamental structure of the whole system is the fusion matrix for the Virasoro conformal blocks.
Indeed the finite fusing matrices of the minimal models and of the DF model  can be obtained by analytic continuation
 of the fusion integral kernel of the Liouville theory.
It was shown by Runkel
\cite{Runkel:1998he,Runkel:1999dz}
in the case of minimal models and by Ponsot and Teschner
\cite{Ponsot:2001ng}
in the case of Liouville theory, that
under certain assumptions all structure constants of these models can be reconstructed from their fusing matrices.
In a more general context of non-rational theories the relation between structure constants and fusing matrices was discussed in \cite{Sarkissian:2011tr}.
It has been also recently clarified by Gaiotto that the fusing matrix
is fundamental for the explicit construction of the Verlinde line operators in the Liouville theory \cite{Gaiotto:2014lma}.


In the case of the Kac-Moody algebra $\widehat{\rm su}(2)$ a counterpart of the picture above is less understood.
There are however many elements already known. The relations between them were recently discussed in \cite{Ribault:2014hia}.
They are schematically shown on Fig.2.
The box {\it $\widehat{\it su}\it (2)$ minimal models} represents the unitary rational series of $\widehat{\rm su}(2)$ models with positive integer levels
\cite{Zamolodchikov:1986bd,Dotsenko:1989ui,Dotsenko:1990}
and the non-unitary $\widehat{\rm su}(2)$ models with rational levels
\cite{Furlan:1991mm,Petersen:1996np}
and spectra corresponding to the admissible representations \cite{Kac:1988qc}.

The $\widehat{\rm su}(2)$ models
with non-rational level
were first analyzed in \cite{Christe:1986cy}. In the case
of the diagonal spectrum of the degenerate representations
the structure constants were
found by Andreev  \cite{Andreev:1995bj}. The derivation was based on the Fateev-Zomolodchikov (FZ) relation \cite{Zamolodchikov:1986bd}
between the ${\rm su}(2)$ Knizhnik-Zamolodchikov (KZ) equation \cite{Knizhnik:1984nr} and the differential equations for a Virasoro degenerate field
originally observed in the context of minimal models.
The model has been recently discussed in   \cite{Ribault:2014hia} under the name {\it generalized  $\widehat{\it su}\it (2)$ WZW model}
\footnote{
We are using the name {\it Andreev model}  instead
as the term {\it the generalized  $\widehat{\it su}\it (2)$ WZW model} was already used for the  $\widehat{\rm su} (2)$
counterpart of the imaginary Liouville theory \cite{Dabholkar:2007ey}.
}.

\begin{figure}
\begin{picture}(400,160)(-30,0)
\setlength\fboxsep{0pt}
\linethickness{1pt}

\put(240,110){\framebox(190,40)[c]{\shortstack[c]{
            (real) $\widehat{\rm su}(2)$ WZNW model \\
           ( $H^+_3$ at level $-\kappa$ )
          }}}
\put(-10,110){\dashbox{4}(190,40)[c]{\shortstack[c]{
            imaginary $\widehat{\rm su}(2)$ WZNW model
            \\ ( generalized $\widehat{\rm su}(2)$ minimal models )
          }}}

{\linethickness{3pt}
\put(205,90){\line(0,1){80}}
\put(205,8){\line(0,1){4}}}


\put(-10,30){\framebox(180,40)[c]{\shortstack[c]{
           $\widehat{\rm su}(2)$ minimal models
          }}}
\put(240,30){\framebox(180,40)[c]{\shortstack[c]{
         Andreev model
          }}}
\linethickness{.5pt}

\put(87,90){\vector(0,1){9}}
\put(87,90){\vector(0,-1){9}}
\put(325,90){\vector(0,1){9}}
\put(325,90){\vector(0,-1){9}}
\put(205,50){\vector(-1,0){11}}
\put(205,50){\vector(1,0){11}}


 \thicklines \put(400,10){\vector(-1,0){390}}
\put(190,0){\shortstack[c]{
          $\kappa=-2$
          }}
\put(315,0){\shortstack[c]{
          $-2>\kappa$
          }}
\put(78,0){\shortstack[c]{
          $\kappa >-2$
          }}
\put(328,87){\shortstack[c]{
        residues
          }}
\end{picture}
\caption{ $\widehat{\rm su}(2)$ models}
\end{figure}
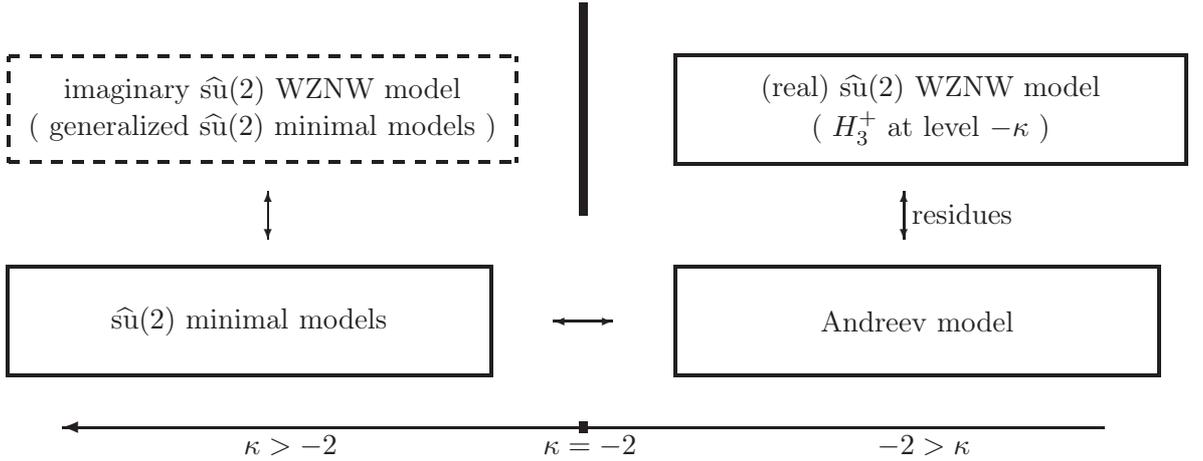
In the continuous spectra part of Fig.2 the counterpart of the (real) Liouville theory is called the (real) $\widehat{\rm su}(2)$ WZNW model. It is defined in
the range of levels $\kappa<-2$ and its spectrum corresponds to principal unitary series of ${\rm sl}(2,\mathbb{C})$ representations with $j\in -{1\over 2}+i\mathbb{R}$.\footnote{This representations can be seen as non-unitary representations of ${\rm su}(2)\oplus {\rm su}(2)$.}
Under the general duality between $G^\mathbb{C}/G $ and $G$ WZNW models \cite{Gawedzki:1988hq,Gawedzki:1988nj,Gawedzki:1989rr,Gawedzki:1991yu} the
 $\widehat{\rm su}(2)$ WZNW model at level $\kappa<-2$ is equivalent to
the $H^+_3= SL(2,\mathbb{C})/ SU(2)$ coset model at level $\kappa'=-\kappa>2$
\footnote{In the present context the equivalence was observed for instance in \cite{Rastelli:2005ph}.}.
The later one has been solved both by the conformal bootstrap method
\cite{Gawedzki:1991yu,Teschner:1997fv,Teschner:1997ft,Teschner:1999ug}
and  by the path integral techniques
\cite{Ishibashi:2000fn,Hosomichi:2000bm,Hosomichi:2001fm,Satoh:2001bi}.
The crossing symmetry of the $H^+_3$ WZNW model was proven in \cite{Teschner:2001gi} with the help
of the suitably extended FZ relation.

The  $\widehat{\rm su}(2)_{\kappa<-2}$ structure constants do not admit an analytic continuation to the range $\kappa> -2$.
This can be seen for instance from the relation between the $H^+_3$ and the Liouville correlators discovered in \cite{Ribault:2005wp} and
later re-derived by the path integral techniques in \cite{Hikida:2007tq}. Under this relation
the $\kappa=-2$ barrier corresponds to the $c=1$ in the Liouville theory.
As in the case of the Virasoro models the  bootstrap difference equations  can be analytically continued
to the range $\kappa >-2$. This was observed in \cite{Dabholkar:2007ey} where also the corresponding structure constants were calculated.
The model was called  there {\it the  generalized  $\widehat{\it su}\it (2)$ WZW model} in parallel to the Virasoro case.
We shall use an alternative name {\it the imaginary  $\widehat{\it su}\it (2)$ WZNW model} to avoid confusion with \cite{Ribault:2014hia}.
The relations between different $\widehat{\rm su}(2)$ models are in perfect analogy to the relations connecting the Virasoro models.
Both the $\widehat{\rm su}(2)$ minimal models and the A model structure constants are analytic continuations of the imaginary
 $\widehat{\rm su}(2)$ WZNW model ones \cite{Ribault:2014hia,Giribet:2007wp}. On the other hand
the residues of the (real)  $\widehat{\rm su}(2)$ WZNW model coincide with
the A model structure constants  \cite{Ribault:2014hia}.

The Virasoro system of fig.1 and its underlying fusing matrix are based on the special class of representations of the Virasoro algebra -
the Virasoro Verma modules. In the case of  $\widehat{\rm su}(2)_{\kappa}$  symmetry with nonrational level there are two possible choices
which we shall briefly describe.

Let us first observe that any complex $ {\rm sl}(2,\mathbb{R})$ representation with the algebra generators $J^3,J^\pm$ is also an
${\rm su}(2)$ representation with the generators $J^3,iJ^\pm$. This simple rescaling changes the hermitian conjugation properties so the
invariant hermitian forms are different.
For instance the principal unitary series representation ${\cal D}_{j,\epsilon}, j\in-{1\over 2}+i\mathbb{R}, \epsilon =0,{1\over 2}$
of ${\rm sl}(2,\mathbb{R})$
\cite{Gelfand,Vilenkin}
is also a series of ${\rm su}(2)$ representations but with indefinite invariant hermitian
forms. We denote by
$\widehat{\cal D}^\kappa_{j,\epsilon}$ the relaxed module of the affine algebra
$\widehat {\rm su}(2)_{\kappa}$ over the representation ${\cal D}_{j,\epsilon}$.
The tensor product
\begin{equation}
\label{representations}
\widehat{\cal D}^\kappa_{j,\epsilon}\otimes \widehat{\cal D}^\kappa_{j,\epsilon},\;\;\;\textstyle j\in-{1\over 2}+i\mathbb{R}, \epsilon =0,{1\over 2},
\end{equation}
provides a representation of the direct sum $\widehat {\rm su}(2)_{\kappa}\oplus\widehat {\rm su}(2)_{\kappa}$
of the left and the right chiral symmetries. This is the class of representation  we are concerned with in the present paper.

The second possibility is to start with the principal continuous series representation ${\cal P}_j, j\in -{1\over 2}+i\mathbb{R}$
of  ${\rm sl}(2,\mathbb{C})$ \cite{Gelfand} which can be seen as a representation of
$ {\rm su}(2)\oplus{\rm su}(2)$. One can then construct
 the relaxed $\widehat {\rm su}(2)_{\kappa}\oplus\widehat {\rm su}(2)_{\kappa}$ module $\widehat{\cal P}_j$ over it \cite{Feigin:1997ha}.
This is the class of representations used in the quantization of the coset model ${\rm H}^+_3={\rm SL}(2,\mathbb{C})/{\rm SU}(2)$
\cite{Teschner:1997fv,Teschner:1997ft,Teschner:1999ug}.
Most of the results mentioned above concerns models based on this class.

It should be emphasized that although
similar in many respects these two classes of representations are essentially different. This implies two
versions of the scheme of Fig.2. Both of them are very similar to the Virsoro system of Fig.1.\footnote{The same pattern emerges for the  N=1 superconformal models as well. }
To what extend the idea of systems of CFT models based on the same class of representations,
conformal blocks and fusing matices
is an
appropriate classifying concept depends on whether
relations between individual models admit extensions to the whole systems.
This is so for instance in the case of the FZ relation \cite{Zamolodchikov:1986bd} originally found in the minimal model corner
of the Virasoro and  $\widehat{\rm su}(2)$ systems and then successfully extended to their continuous spectra parts.
The main idea of the present paper is to analyze whether the GKO construction has this property.

The organization of the paper is as follows. In Section 2 we introduce some basic material concerning hermitian ${\rm su}(2)$
representations ${\cal D}_{j,\epsilon}$. Most of the results are standard and based on
the well developed theory of ${\rm sl}(2,\mathbb{R})$ representations \cite{Gelfand,Vilenkin}.
This concerns the definition of the loop module (Subsection 2.1), the reducibility of the representations (Subsection 2.2),
explicit constructions of the reflection map (Subsection 2.4) and of the bilinear invariants (Subsection 2.5) and the
formulae for the 3-linear invariants (Clebsch-Gordan coefficients) in the spin basis (Subsection 2.8).
A special attention has been payed to the izospin variables introduced in Subsection 2.3 and their role in constructing
the bilinear (Subsection 2.6) and the 3-linear (Subsection 2.7) invariants.
In Subsection 2.9 we discuss the conditions under which a hermitian ${\rm su}(2)$-invariant
pairing between representations ${\cal D}_{j,\epsilon}$ exists.
The constructions of this subsection are not used in the present paper.
They were included to complete the notion of the hermitian non-unitary ${\rm su}(2)$ representations.
We close this section by a short analysis of the tensor product of representations relevant on the l.h.s of
expected relations (\ref{rel}).

In Section 3 we formulate some basic structures of the $\widehat{\rm su}(2)_\kappa$ WZNW models with non-rational levels.
They are based on representations (\ref{representations}) defined in Subsection 3.1. In Subsection 3.2 we introduce
the izospin description of these representations and the associated highest weight modules.
The spectrum of the model and the primary fields in the izospin variables are introduced in Subsection 3.3.
Let us note that  the $J^3_0$ eigenvalue of the highest weight state in the module related to the representation ${\cal D}_{-1-j,\epsilon}$
is $j$. For this reason we denote the chiral primary field related to the representation ${\cal D}_{-1-j,\epsilon}$  by $\Phi_{j,\epsilon}$.
The spectrum is diagonal and consists of representations (\ref{representations}).
Since the representations ${\cal D}_{j,\epsilon}$ and  ${\cal D}_{-1-j,\epsilon}$ are equivalent
we declare an identification of the corresponding primary fields by means of the reflection map.
The transformation properties of the primary fields are described in Subsection 3.4
in terms of the OPE's with the ${\rm su}(2)$ currents and the Ward identities.

In Subsection 3.5 the general form of the 2-point function of primary fields
compatible with the symmetry and the reflection properties is analyzed.
Due to the tensor product of the left and the right representations  (\ref{representations}) the 2-point function factorizes
into chiral parts. The same concerns
the 3-point functions discussed
 in Subsection 3.6.
 The dependence on the field locations ($z$ variables) is determined by the Ward identities in the standard manner.
 Declaring compatibility with the reflection properties of primary fields
 we find an appropriate form of the 3-point invariants which fixes the dependence on the izospin $x$ variables.
 The remaining part depends only on $j$ parameters labeling the representations. We assume it is
 given by the ${\rm H}^+_3$ structure constants
with the normalization chosen in \cite{Dabholkar:2007ey} in the case of $\kappa>-2$ (being in line with the normalization used in the
${\rm su}(2)$ minimal models).
This is motivated by the mentioned above expectation that
the
 $\widehat{\rm su}(2)_\kappa$ WZNW model at level $\kappa<-2$ should be equivalent to
the ${\rm H}^+_3= {\rm SL}(2,\mathbb{C})/ {\rm SU}(2)$ coset model at level $\kappa'=-\kappa>2$
\cite{Gawedzki:1988hq,Gawedzki:1988nj,Gawedzki:1989rr,Gawedzki:1991yu}.
Let us note however that the derivation of the structure constants given in \cite{Dabholkar:2007ey} and based on the Teschner method
\cite{Teschner:1997ft,Teschner:1999ug}
assumes the complex principal ${\rm sl}(2,\mathbb{C})$ representation rather then the tensor product of two
principal ${\rm sl}(2,\mathbb{R})$ representations.
The exact proof that it also applies in the present case
goes beyond the scope of this paper. We hope to come back to this point in subsequent publication.

Both the $z$-dependent part and the $x$-dependent part of the 3-point function naturally split into chiral components.
Motivated by the freedom in the similar splitting of the Liouville structure constants \cite{SL-LL}
we define the chiral component of the $j$ dependent part by an appropriate splitting of the upsilon functions into the Barnes gamma function components.
This choice of the chiral 3-point function is partially confirmed by its properties with respect to the chiral reflection.

As in the case of the SL-LL correspondence one has  off-diagonal spectra
in the projected tensor product on the r.h.s. of  relations (\ref{rel}).
In the Liouville and the imaginary Liouville theory the extension is the same as in the SL-LL equivalence.
In the $\widehat{\rm su}(2)_{\kappa +1}$ WZNW model one has to include the representations
$$
\hat{\cal D}^\kappa_{j,\epsilon},\;\;\;\;j=-{\textstyle{1\over 2}}+n+ is,\;\;\;
s\in \mathbb{R},\;\;\epsilon =0,{\textstyle \frac12},\;\;\;n\in {\textstyle \frac12}\mathbb{Z}.
$$
The spectrum is off-diagonal only with respect to the discrete variable $n$.
The corresponding off-diagonal extension of the structure constants is based on   the
 additional condition
 $$
 n_1+n_2+n_3\;\dot{=}\; 0,
 $$
where $\dot{=}$ denotes equality modulo 1.
As the structure constants are already split into
 chiral components the extension is almost straightforward.  Only the extension of the $x$-dependent part
requires a special care in order to avoid square root ambiguities (Subsection 3.7).

We close Section 3 by a brief description of the structure constants in the Liouville and the imaginary Liouville theories
(Subsection 3.8). We also introduce a nonstandard (but convenient from the point of view of (\ref{rel})) parametrization of the conformal weights:
\bea
c\sL &=& 1+6Q^2,\;\;\;\Delta\sL_j \;=\; - Q^2 \,j(1+j),\;\;\;Q\;=\;b+b^{-1},
\\
c\sIL &=& 1-6\hat Q^2,\;\;\;\Delta\sIL_j \;=\; \hat Q^2 \,j(1+j),\;\;\;\hat Q \;=\; \hat b^{-1}-\hat b.
\eea

Section 4 contains our main results. We start in Subsection 4.1 with the GSO construction of the $\widehat{\rm su}(2)_{\kappa +1}$ and the Virasoro generators in the tensor product $\widehat{\cal D}_{j,\epsilon}^{\kappa}\otimes \widehat{\cal S}_{j=0,{1\over 2}}$ where
$\widehat{\cal S}_{j=0,{1\over 2}}$ are standard unitary representations of $\widehat{\rm su}(2)_{\kappa=1}$.
This yields the  relations between the parameters of the theories involved.
In the case of Liouville theory with the central charge parameterized by $b<1$ the ${\rm su}(2)$ WZNW models are on the opposite sides of the $\kappa=-2$
barrier and one gets  relations (\ref{parameters_b}).
In the case of the imaginary Liouville theory parameterized by $\hat b <1$ the ${\rm su}(2)$ WZNW models are on the same real side of the $\kappa=-2$
barrier and the parameters are related by (\ref{parameters_bIm}).

In Subsection 4.2 we analyze the decomposition of $\widehat{\cal D}_{-j-1,\epsilon}^{\kappa}\otimes \widehat{\cal S}_{j=-1,-{3\over 2}}$ into
irreducible representations of   $\widehat{\rm su}(2)_{\kappa +1}\otimes {\rm Vir}$. Motivated by character decomposition (\ref{decomposition})
and calculations of low level states we conjecture  the following decomposition of representations
\begin{equation}
\label{decorep}
\hat {\cal D}^\kappa_{-1-j,\epsilon} \otimes \hat{\cal S}^1_{-1}
=
\bigoplus\limits_{n\in \mathbb{Z}} \hat {\cal D}^{\kappa+1}_{-1-j-n,\epsilon} \otimes {\cal V}_{\Delta_{n},c}
,
\;\;\;\;
\hat {\cal D}^{\kappa}_{-1-j,\epsilon} \otimes \hat{\cal S}^1_{-{3\over 2}}
\;=\;
\bigoplus\limits_{n\in \mathbb{Z}+{1\over 2}} \hat {\cal D}^{\kappa+1}_{-1-j-n,\bar \epsilon} \otimes {\cal V}_{\Delta_{n},c},
\end{equation}
where $\bar \epsilon =\epsilon \dot{+} {1\over 2}$ and $\dot{+}$ denotes summation modulo 1 and ${\cal V}_{\Delta,c}$ is the Virasoro Verma module with
the highest weight $\Delta$ and the central charge $c$.
In the case of the Liouville theory one has
$$
\Delta_{n} =
\Delta^{\scriptscriptstyle {\rm L}}_{j + \frac{n}{bQ}} =
-Q^2 \left( j + \frac{n}{bQ}\right)  \left( 1 + j + \frac{n}{bQ}\right), \;\;\;c=c\sL.
$$
In the izospin formulation the representations on the right hand sides are generated by the highest weight states $\ket{x}^*_{j+n,\epsilon}$.
The corresponding chiral fields $\Phi^*_{j+n,\epsilon}$ are descendants for the algebra $\widehat{\rm su}(2)_\kappa\otimes \widehat{\rm su}(2)_1$
and are primaries with respect to the algebra $\widehat{\rm su}(2)_{\kappa+1}\otimes {\rm Vir}$. For $n=0,\pm{1\over 2},\pm 1$ they are explicitly
calculated
in Subsection 4.3 up to normalization factors.

Equivalences (\ref{rel}) can be formulated as  equalities of all full correlation functions with an appropriate identification of fields on their opposite sides.
In a slightly stronger version they state equalities of all chiral correlation functions.
In the case of 3-point chiral functions and the Liouville theory we conjecture that one can always adjust the normalization of the fields such that
the following relation holds:
\begin{eqnarray} \label{3-pkt_gen2}
\nonumber
&& \hspace{-10pt}
\langle \Phi^\star_{j_3+n_3, \epsilon_3} (x_3, z_3) \Phi^\star_{j_2+n_2,\epsilon_2 } ( x_2,z_2)
\Phi^\star_{j_1+n_1,\epsilon_1 } ( x_1, z_1) \rangle_\epsilon
 \\[-4pt]
  \\[-4pt]
&& \nonumber =
\langle \Phi_{j_3+\frac{n_3 }{bQ}} (z_3) \Phi_{j_2+\frac{n_2}{bQ} } ( z_2) \Phi_{j_1+\frac{n_1}{bQ}} (z_1)  \rangle\sL
\,
 \langle \Phi\sB_{j_3+n_3,  \epsilon_3} (x_3, z_3) \Phi\sB_{j_2+n_2,  \epsilon_2 } ( x_2,z_2) \Phi\sB_{j_1+n_1, \epsilon_1 } ( x_1, z_1)  \rangle^\h_{\epsilon \+ \frac12 n_{123} }  \, ,
\end{eqnarray}
where
$\Phi_{j+\frac{n }{bQ}}$ are Liouville fields normalized  by setting the reflection amplitude to 1 and
$\Phi\sB_{j+n, \epsilon }$ are primary fields in the $\widehat{\rm su}(2)_{\kappa+1}$ theory normalized by  2-point function (\ref{2point}).

 Due to the simple form of  the structure constants in the $\kappa=1$ model (\ref{structure1})
the condition $ n_1+n_2+n_3\;\dot{=} \;0$ is always satisfied in the 3-point functions.
The r.h.s. of   (\ref{3-pkt_gen2}) can be calculated using shift properties of the Barnes
gamma functions (\ref{shift}). The real difficulty is the calculation of the correlator of descendent fields on the l.h.s.

We have checked relation (\ref{3-pkt_gen2}) and its counterpart for the imaginary Liouville theory
for $n=0,\pm{1\over 2},\pm 1$. In the simplest case of $n=0$ this is done in Subsection 4.3.
The proof is based on new identities for the Barnes gamma functions (\ref{Gamma_LH}), (\ref{Gamma_ImH})
derived in Appendix B. This version of (\ref{3-pkt_gen2}) fixes relative normalization of primary fields on both sides of the correspondence.
It is interesting to observe that the spectrum of the imaginary Liouville theory implied by the quotient construction
coincides with the spectrum recently proposed on the basis of numerical analysis of the bootstrap equations \cite{Ribault:2015sxa}.

In Subsection 4.4 we show that in the general case relation (\ref{3-pkt_gen2}) can be cast in the form of  explicit expression (\ref{cosetfactor})
for the coset factor defined as
$$
{\langle \Phi^\star_{j_3+n_3, \epsilon_3\+n_3} (x_3, z_3) \Phi^\star_{j_2+n_2,\epsilon_2\+ n_2 } ( x_2,z_2) \Phi^\star_{j_1+n_1,\epsilon_1\+n_1 } ( x_1, z_1) \rangle_\epsilon
\over
\langle
   \Phi\sa_{j_3, \epsilon_3} (x_3, z_3)
   \Phi\sa_{j_2, \epsilon_2} (x_2, z_2)
   \Phi\sa_{j_1, \epsilon_1} (x_1, z_1)
\rangle^\s_\epsilon},
$$
where $\Phi\sa_{j, \epsilon }(x,z)$ are primary fields in the $\widehat{\rm su}(2)_{\kappa}$ theory.
An essential advantage of this form of the conjecture is that it is the same for the Liouville and for the imaginary Liouville theory
and reduces calculations to the product of $\widehat{\rm su}(2)_\kappa$ and $\widehat{\rm su}(2)_1$ WZNW models with
$\kappa$ on the real side of the $\kappa=-2$ barrier.

Some explicit calculations of the coset factor involving  $n=\pm {1\over 2}$ and $n=\pm 1$ fields
are given in Subsections 4.5 and 4.6, respectively.  We have also calculated all other coset factors with $n_1,n_2,n_3 \in \{ \pm \frac12, \pm 1\}$.
In all the cases considered we got confirmation of the conjectured form (\ref{cosetfactor}).

These verifications are the main results of the present paper.
They are based on the new nontrivial identities for the Barnes gamma functions with different parameters (\ref{Gamma_LH}), (\ref{Gamma_ImH}),
decomposition (\ref{decorep}) and the properties of the 3-linear invariants of the representations involved.
They provide strong evidence that the  general relations (\ref{rel}) are correct. Some consequences of this result and possible extensions of this paper
are briefly discussed in Section 5.

\section{${\rm su}(2)$  hermitian representations}

\subsection{loop module }

For any $j,\alpha \in \mathbb{C}$ we define  the loop module ${\cal D}_{j,\alpha}$  as an ${\rm sl}(2,\mathbb{C})$
module  with the basis $\left\{\ket{n+\!\alpha}:n\!\in\!\mathbb{Z}\right\}$
and the algebra action
given by \cite{Feigin:1997ha}
\begin{eqnarray*}
J^3\ket{n+\alpha}&=& (n+\alpha) \ket{n+\alpha},
\\
J^+\ket{n+\alpha}&=& (n+\alpha - j)\ket{n+\alpha +1},
\\
J^-\ket{n+\alpha} &=&( -n -\alpha -j)\ket{n+\alpha -1}.
\end{eqnarray*}
The loop module ${\cal D}_{j,\alpha}$ can be seen as a complex representation of the ${\rm su}(2)$ algebra:
\begin{eqnarray*}
\left[ J^3,J^\pm\right] &=& \pm J^\pm,
\\
\left[ J^+,J^-\right] &=& 2 J^3,
\end{eqnarray*}
where
$
J^\pm=J^1\pm iJ^2,\;\left[J^a,J^b\right] \;=\; i\epsilon_{abc}J^c.
$
The eigenvalue of the Casimir operator
$$
C=J^-J^+ + (J^3)^2 +J^3
$$
in this representation is $j(j+1)$. The requirement of integrability
 leads to the condition
 $$
 {\rm e}^{i4\pi J^3} = 1
 $$
hence
$
\alpha \in {1\over 2}\mathbb{Z}.
$
Without loss of generality, one can
assume $\alpha=\epsilon = 0,{1\over 2}$.
In the following we shall restrict ourselves to the two cases $\alpha=\epsilon$ and $\alpha =j$.
In the second case the module ${\cal D}_{j,j}$ contains a highest weight submodule ${\cal E}_j$
generated from  the highest weight state $\ket{j}$
$$
J^+\ket{j}=0,\;\;\;\;\;J^3\ket{j}=j\ket{j}.
$$

\subsection{reducibility of ${\cal D}_{j,\epsilon}$ representations}

The parameters $(j,\epsilon)$ are called integral if $2j$ and $2\epsilon$ are integers of the same parity.
For all non-integral $(j,\epsilon)$ the representation ${\cal D}_{j,\epsilon}$ is irreducible.
For integral $(j,\epsilon)$,  $n_+ +\epsilon=-n_--\epsilon=j$  is integer or half integer and will be denoted by $l$.
One has
\begin{eqnarray*}
J^+ \ket{n_++\epsilon} &=& 0,\;\;\;\;J^3 \ket{n_++\epsilon} \;=\; n_+\ket{n_++\epsilon},
\\
J^-\ket{n_-+\epsilon}&=&0,\;\;\;\;J^3 \ket{n_-+\epsilon} \;=\; n_-\ket{n_-+\epsilon}.
\end{eqnarray*}
There are two invariant subspaces
$$
{\cal D}^+_l= {\rm span}\{\ket{n+\epsilon}:n\leq n_+\},\;\;\;\;\;\;{\cal D}^-_l= {\rm span}\{\ket{n+\epsilon}:n\geq n_-\}.
$$
If $n_->n_+$ ($l<0$) the intersection ${\cal D}^+_l\cup {\cal D}^-_l$
is null and the sum  is an invariant subspace.
Then the quotient ${\cal D}_{l,\epsilon}/{\cal D}^+_l\cup {\cal D}^-_l$
is equivalent to the
finite-dimensional, spin $l$
representations ${\cal S}_l$ of ${\rm su}(2)$.
For $l=-{1\over 2}$,  ${\cal D}_{l,\epsilon}= {\cal D}^+_l\cup {\cal D}^-_l$ hance the quotient representation is trivial
${\cal S}_{-{1\over 2}}=\{0\}$.
The representations ${\cal S}^\pm_l$ induced on subspaces ${\cal D}^\pm_l$ are irreducible.
The simplest nontrivial quotient is ${\cal S}_{-{3\over 2}}=\{\ket{{1\over 2}},\ket{-{1\over 2}}\}$
with the action of generators given by
\footnote{For the sake of future convenience we shall denote the ${\rm su}(2)$ generators
in the finite dimensional representations by symbol $K^a$ rather then $J^a$.}
$$
K^3
=\left(\begin{array}{cc}
{1\over 2} &0 \\
0 & -{1\over 2}
\end{array}\right)
\;,\;\;\;\;\;\;
K^+
=\left(\begin{array}{cc}
0 &1 \\
0 & 0
\end{array}\right)
\;,\;\;\;\;\;\;
K^-
=\left(\begin{array}{cc}
0 &0 \\
1& 0
\end{array}\right).
$$
If $n_-<n_+$ ($l>0$) also the intersection ${\cal D}^+_l\cap {\cal D}^-_l$ is invariant and is equivalent to the standard
finite-dimensional, spin $l$
representations ${\cal S}_l$ of ${\rm su}(2)$. The quotients ${\cal D}^\pm_l/{\cal D}^+_l\cap {\cal D}^-_l$ are
isomorphic to the representations ${\cal S}^\pm_l$, respectively.
In the simplest case
${\cal S}_{{1\over 2}}=\{\ket{{1\over 2}},\ket{-{1\over 2}}\}$
the action of generators takes the form
$$
K^3
=\left(\begin{array}{cc}
{1\over 2} &0 \\
0 & -{1\over 2}
\end{array}\right)
\;,\;\;\;\;\;\;
K^+
=-\left(\begin{array}{cc}
0 &1 \\
0 & 0
\end{array}\right)
\;,\;\;\;\;\;\;
K^-
=-\left(\begin{array}{cc}
0 &0 \\
1& 0
\end{array}\right).
$$

\subsection{associated highest weight module and the  izospin variables}

We shall now describe a construction of the highest weight module ${\cal E}_j$ associated to the
module ${\cal D}_{-1-j,\epsilon}$. To this end let us
define a new set of generators
\begin{eqnarray*}
J^+(x)&=& J^+ -2xJ^3 -x^2J^- ,\\
J^3(x)&=& J^3+ xJ^- ,\\
J^-(x)&=& J^-,
\end{eqnarray*}
where $x$ is a complex parameter called the izospin variable \cite{Zamolodchikov:1986bd}.
For any $x$ they satisfy the commutation relations of the $ {\rm su}(2)$  algebra:
\begin{eqnarray*}
\left[ J^3(x), J^\pm(x)\right]
&=&
\pm J^\pm(x),
\\
\left[ J^+(x), J^-(x)\right]
&=&
2J^3(x).
\end{eqnarray*}
Consider now a formal series \cite{Feigin:1997ha}
\begin{equation}
\label{hws}
\ket{x}_{j, \epsilon} = \sum\limits_{m=-\infty}^\infty x^{j-m-\epsilon} \ket{m+\epsilon}_{-1-j}
\end{equation}
which can be seen as a generating function for $J^3_0$ eigenstates in  ${\cal D}_{-1-j,\epsilon}$.
This is the highest weight state with respect to the algebra $J^a(x)$:
\begin{eqnarray*}
J^+(x)\ket{x}_{j,\epsilon} &=&0,
\;\;\;\;\;
J^3(x)\ket{x}_{j,\epsilon}  \;=\;
j\ket{x}_{j,\epsilon} .
\end{eqnarray*}
One also has
\begin{eqnarray}
\nonumber
J^+\ket{x}_{j,\epsilon}
&=&\Big( -x^2\partial_x +2 j x\Big)\ket{x}_{j,\epsilon} ,
\\
\label{xrep}
J^3\ket{x}_{j,\epsilon} &=&
\Big( -x\partial_x +j\Big)\ket{x}_{j,\epsilon}  ,
\\
J^-\ket{x}_{j,\epsilon} &=&\partial_x\ket{x}_{j,\epsilon} .
\nonumber
\end{eqnarray}
An advantage of this construction is that it treats the whole ${\cal D}_{-1-j,\epsilon}$ representation as a single
highest weight state $\ket{x}_{j,\epsilon} $. This is also an efficient tool in analyzing multi-linear invariants.

\subsection{equivalent representations and the reflection map}

Representations ${\cal D}_{j,\epsilon}, {\cal D}_{j',\epsilon'}$ are equivalent if there exists an operator
$Q:{\cal D}_{j,\epsilon}\to{\cal D}_{j',\epsilon'}$ such that
$$
Q J^a = J^aQ,\;\;\;\;a=3,\pm.
$$
In both representations the operator $J^3$ is diagonal in the basis $\ket{n+\epsilon}$. As all the diagonal matrix elements of $J^3$  are different
the operator $Q$ has to be diagonal as well:
$$
Q \ket{n+\epsilon} = q_{n+\epsilon} \ket{n+\epsilon' }.
$$
Then the equation $Q J^3= J^3Q$ implies
$$
q_{n+\epsilon}(n+\epsilon) =(n+\epsilon')q_{n+\epsilon}
$$
hence $\epsilon =\epsilon'$.
The equations $Q J^\pm = J^\pm Q$ take the form
\begin{eqnarray*}
q_{n+\epsilon+1} (-j+n+\epsilon) &=& (-j'+n+\epsilon)q_{n+\epsilon},
\\
q_{n+\epsilon}(j+n+\epsilon+1)&=&  (j'+n+\epsilon+1)q_{n+\epsilon+1}.
\end{eqnarray*}
For non-vanishing  $q_{n+\epsilon}$ they can be satisfied only if
$$
j'=j,\;\;\;{\rm or}\;\;\;\;j'=-j-1.
$$
In the second case one gets nontrivial $Q_j$:
\begin{eqnarray*}
Q_jJ^a_j &=& J^a_{-1-j}Q_j,
\\
q^j_{n+\epsilon+1}
&=&
 {j+1+n+\epsilon\over -j+n+\epsilon}q^j_{n+\epsilon}.
\end{eqnarray*}
This determines $Q_{j,\epsilon}$ up to a $(j,\epsilon)$-dependent constant.
We chose this constant by assuming that $Q_{j,\epsilon}$ is given by
\begin{equation}
\label{reflectionI}
q^j_{n+\epsilon}=
{\Gamma(1+j+n+\epsilon)\over \Gamma(-j+n+\epsilon)}.
\end{equation}
Let us observe that
$$
q^j_{n+\epsilon}={1\over q^{-1-j}_{n+\epsilon}}.
$$

\subsection{invariant bilinear forms}

We say that a bilinear form $D$ is invariant on   ${\cal D}_{j',\epsilon'}\times{\cal D}_{j,\epsilon}$
if:
$$
D(J^a f_1,f_2)+D(f_1,J^af_2)=0.
$$
Let $\{F_n=\ket{n + \epsilon}\},\{F'_n=\ket{n + \epsilon'}\}$ be the bases in ${\cal D}_{j,\epsilon}, {\cal D}_{j',\epsilon'}$, respectively.
The condition for $J^3$ reads
$$
D(J^3 F'_n,F_m)+D(F'_n, J^3F_m) =(n+m+ \epsilon+\epsilon')(F'_n,F_m) =0.
$$
Since $\epsilon,\epsilon'\in\{0,{1\over 2}\}$ and $n+m$ is an integer one gets $\epsilon=\epsilon'$ and
$$
D(F'_n,F_m) = 0 \;\;{\rm for}\;\;n\neq - m- 2\epsilon.
$$
The other two conditions
\begin{eqnarray*}
D( F'_m,J^+ F_n) +D(J^+ F'_m, F_n)&=& 0,
\\
D( F'_n,J^- F_m) + D(J^- F'_n, F_m)&=& 0,
\end{eqnarray*}
yield
\begin{eqnarray}
\label{1eq1}
(n+\epsilon - j) D(F'_{-n-1-2\epsilon},F_{n+1}) + (-n-1-2\epsilon +\epsilon -j') D(F'_{-n-2\epsilon},F_n)&=& 0,
\\
\nonumber
(n+2\epsilon-\epsilon -j') D(F'_{-n-1-2\epsilon},F_{n+1})  +   ( -n-1-\epsilon -j) D(F'_{-n-2\epsilon},F_n)&=&0,
\end{eqnarray}
which implies
$
j(1+j)=   j'(1+\bar j') $
and
\begin{equation}
\label{1eq2}
j'= j\;\;\;\;{\rm or}\;\;\;\;j'=- j-1.
\end{equation}
For non-integral $(j,\epsilon)$, in both cases
the bilinear form $D$ is  determined by (\ref{1eq1})
up to an overall normalization:
$$
\begin{array}{rclllll}
 D_{\rm I}(F'_{-n-1-2\epsilon},F_{n+1})&=&D_{\rm I}(F'_{-n-2\epsilon},F_n) &&{\rm for}& j'= -1-j&,
 \\[10pt]
 D_{\rm II}(F'_{-n-1-2\epsilon},F_{n+1})&=&\displaystyle {n+1+\epsilon +j\over n+\epsilon -j}
 D_{\rm II}(F'_{-n-2\epsilon},F_n) &&{\rm for}& j'=j&.
\end{array}
$$
One easily checks that the  forms are related by the reflection map
$
Q_{ j}:{\cal D}_{ j,\epsilon}\to{\cal D}_{-1- j,\epsilon}$:
$$
D_{\rm II}(\;.\;,\;.\;) \propto D_{\rm I}(\;Q_{ j}\;.\;,\;.\;).
$$

\subsection{invariant bilinear forms in the izospin variables}

Suppose $D: {\cal D}_{-1-j',\epsilon} \times {\cal D}_{-1-j,\epsilon} \to \mathbb{C}$ is an invariant bilinear form:
$$
D(J^aF'_n,F_m) +D(F'_n,J^aF_m)=0.
$$
One can consider the formal double series
\begin{equation}
\label{xD}
D(x,y) \equiv D(\ket{x}_{j', \epsilon},\ket{y}_{j,\epsilon} ) = \sum\limits_{n,m \in \mathbb{Z}} x^{j'-m-\epsilon}y^{j-n-\epsilon} D(F'_m,F_n)
.
\end{equation}
The formula for the coefficients
\begin{equation}
\label{xDco}
 D(F'_m,F_n)  = \oint
 {dx\over 2\pi i}{dy\over 2\pi i} x^{-1-j'+m+\epsilon}y^{-1-j+n+\epsilon}D(x,y)
\end{equation}
requires the integrand
to be a well defined function on the product $S^1\times S^1$
of unit circles  in the  complex $x,y $ variables.

It follows from (\ref{xrep}) that $D(x,y)$ should satisfy the equations
\begin{eqnarray*}
 \partial_x D(x,y) +    \partial_yD(x,y) &=& 0,
\\
\Big( -x\partial_x +j' \Big)D(x,y) +    \Big( -y\partial_y + j \Big)D(x,y) &=& 0,
\\
\Big( -x^2\partial_x +2 j' x\Big)D(x,y) +    \Big( -y^2\partial_y +2 j y\Big)D(x,y) &=& 0.
\end{eqnarray*}
The Dirac delta function $\delta(x-y)$ solves these equations if $j'=-1-j$.
In this case formula (\ref{xDco}) yields
\begin{eqnarray*}
 D(F'_m,F_n)  &=& \oint
 {dx\over 2\pi i}{dy\over 2\pi i} x^{-j'-1+m+\epsilon}y^{-j-1+n+\epsilon}\delta(x-y)
 \\
 &=&
 \oint
 {dx\over 2\pi i} x^{-1 + m+n+2 \epsilon}\;=\; \delta_{m+n+2\epsilon,0}.
\end{eqnarray*}
Thus
$$
D_{\rm I}(x,y) \propto \delta(x-y).
$$
For $j'=j$ one has another solution
$$
D(x,y)= (x-y)^{2j}.
$$
Due to the complex values of the exponent this is a multi-valued function. The solution is determined up to a constant possibly depending on $j$.
One has for instance  another solution
$$
 (y-x )^{2j} = (-1)^{2j} (x-y)^{2j}.
$$
In the $x$-formulation one needs a function (or a distribution) such that
$$
x^{-j-1+m+\epsilon}y^{-j-1+n+\epsilon}D(x,y)
$$
 is well defined (single valued) on the product $S^1\times S^1$
of unit circles  in the  complex $x,y $ variables.
In order to analyze the problem we shall introduce a convenient parametrization of $S^1\times S^1$:
$$
M=(-\pi,\pi]\times (-\pi,\pi] \ni (\varphi_1,\varphi_2) \to (x,y)= ({\rm e}^{i\varphi_1},{\rm e}^{i\varphi_2}) \in S^1\times S^1.
$$
In this parametrization
$$
 (x-y )^{2j} = (2i)^{2j} {\rm e}^{i{(\varphi_1+\varphi_2)}j} \left(\sin{\textstyle {\varphi_1-\varphi_2\over 2}}\right)^{2j}
$$
and
$$
x^{-j-1+m+\epsilon}y^{-j-1+n+\epsilon}(x-y )^{2j} = (2i)^{2j}{\rm e}^{i(m+\epsilon-1)\varphi_1}{\rm e}^{i(n+\epsilon-1)\varphi_2}
\left(\sin{\textstyle {\varphi_1-\varphi_2\over 2}}\right)^{2j}.
$$
This expression is singular along the line $\varphi_1=\varphi_2$ which in the space $M$ of parameters separates  two regions:
$$
M_> =\{(\varphi_1,\varphi_2)\in M: \varphi_1>\varphi_2\},\;\;\;M_< =\{(\varphi_1,\varphi_2)\in M: \varphi_1<\varphi_2\}.
$$
Let $N_>, N_<$ be the images of $M_>,M_<$ in $S^1\times S^1$. For the function
$$
S_{j,\epsilon}(x,y) =
\left\{
\begin{array}{rllll}
(x-y )^{2j} &{\rm on}& N_>
\\
(-1)^{2\epsilon}(y-x )^{2j} &{\rm on}& N_<
\end{array}
\right.
$$
one gets
$$
x^{-j-1+m+\epsilon}y^{-j-1+n+\epsilon}S_{j,\epsilon}(x,y) =
(2i)^{2j}\,{\rm sign}^{2\epsilon}(\varphi_1-\varphi_2) {\rm e}^{i(m+\epsilon-1)\varphi_1}{\rm e}^{i(n+\epsilon-1)\varphi_2}
\left|\sin{\textstyle {\varphi_1-\varphi_2\over 2}}\right|^{2j} \, ,
$$
which has periodic boundary conditions on the square M.
Using the change of variables $\eta=\varphi_1+\varphi_2, \theta=\varphi_1-\varphi_2$ and the symmetry properties of $S_{j,\epsilon}(x,y)$
one obtains
\begin{eqnarray*}
D(F_m,F_n) &=&{1\over (2\pi)^2}
\int\limits_{-\pi}^\pi d\varphi_1\int\limits_{-\pi}^\pi d\varphi_2\,(2i)^{2j}{\rm sign}(\varphi_1-\varphi_2)^{2\epsilon}
{\rm e}^{i(m+\epsilon)\varphi_1}{\rm e}^{i(n+\epsilon)\varphi_2}
\left|\sin{\textstyle {\varphi_1-\varphi_2\over 2}}\right|^{2j}
\\
&=&
{1\over (2\pi)^2}\int\limits_{0}^{2\pi} d\theta\int\limits_{-2\pi}^{2\pi} d\eta\,(2i)^{2j}
{\rm e}^{i{m+n+2\epsilon\over 2}\eta}{\rm e}^{i{m-n\over 2}\theta}
\left|\sin{\textstyle {\theta\over 2}}\right|^{2j}
\\
&=&\delta_{m+n+2\epsilon,0} {(2i)^{2j}\over \pi}\int\limits_{0}^{2\pi} d\theta\,
{\rm e}^{i(m+\epsilon)\theta}
\left(\sin{\textstyle {\theta\over 2}}\right)^{2j}
\\
&=&
\delta_{m+n+2\epsilon,0}  \,C_{j,\epsilon}\, {\Gamma(-j+m+\epsilon)\over \Gamma(1+j+m+\epsilon)},
\\
C_{j,\epsilon}
&=&
{i^{2(j+\epsilon)}\over \cos ((j+\epsilon)\pi) \,\Gamma(-2j)},
\end{eqnarray*}
hence
$$
D_{\rm II}(x,y) \propto S_{j,\epsilon}(x,y).
$$
Let us observe that $S_{j,\epsilon}(x,y)$ can be seen as an integral kernel of the reflection map
\begin{equation}
\label{reflectionkernel}
Q_{-1-j}\ket{x}_{j,\epsilon}  = {1\over C_{j,\epsilon}}
\oint {dy\over 2\pi i} S_{j,\epsilon}(x,y)\ket{y}_{-1-j, \epsilon}.
\end{equation}
Indeed
\begin{eqnarray*}
Q_{-1-j}\ket{m+\epsilon}_{-1-j} &=& \oint {dx\over 2\pi i} x^{-j+m+\epsilon-1}Q_{-1-j}\ket{x}_{j,\epsilon}
\\
&=&
{1\over C_{j,\epsilon}}\sum\limits_n \oint {dx\over 2\pi i}{dy\over 2\pi i} x^{-j+m+\epsilon -1} y^{-1-j-n-\epsilon} S_{j,\epsilon}(x,y) \ket{n+\epsilon}_j
\\
&=&
{1\over (2\pi)^2C_{j,\epsilon}}\sum\limits_n\int\limits_{0}^{2\pi} d\theta\int\limits_{-2\pi}^{2\pi} d\eta\,(2i)^{2j}
{\rm e}^{i{m-n\over 2}\eta}{\rm e}^{i{m+n+2\epsilon\over 2}\theta}
\left|\sin{\textstyle {\theta\over 2}}\right|^{2j}\ket{n+\epsilon}_j
\\
&=&
{(2i)^{2j}\over \pi C_{j,\epsilon}}\int\limits_{0}^{2\pi} d\theta\,
{\rm e}^{i(m+\epsilon)\theta}
\left(\sin{\textstyle {\theta\over 2}}\right)^{2j}\ket{m+\epsilon}_j
\\
&=&
{\Gamma(-j+m+\epsilon)\over \Gamma(1+j+m+\epsilon)}\ket{m+\epsilon}_j
\end{eqnarray*}
in line with (\ref{reflectionI}).

\subsection{invariant three-linear forms in the izospin variables}

In the izospin variables the three-linear invariants
satisfy the equations:
\begin{eqnarray*}
 \Big(\partial_{x_1} +\partial_{x_2}+\partial_{x_3}\Big)
 D\Big[\!
\begin{array}{ccc}
\\[-18pt]
\scriptstyle j_3 & \scriptstyle j_2  & \scriptstyle j_1
\\[-8pt]
\scriptstyle \epsilon_3  &  \scriptstyle \epsilon_2  & \scriptstyle \epsilon_1
\\[-8pt]
\scriptstyle x_3  &  \scriptstyle x_2  &  \scriptstyle x_1
\end{array}
\!\Big]
 &=& 0,
\\
 \Big(-x_1\partial_{x_1}+j_1 -x_2\partial_{x_2}+j_2-x_3\partial_{x_3}+j_3\Big)
D\Big[\!
\begin{array}{ccc}
\\[-18pt]
\scriptstyle j_3 & \scriptstyle j_2  & \scriptstyle j_1
\\[-8pt]
\scriptstyle \epsilon_3  &  \scriptstyle \epsilon_2  & \scriptstyle \epsilon_1
\\[-8pt]
\scriptstyle x_3  &  \scriptstyle x_2  &  \scriptstyle x_1
\end{array}
\!\Big]
&=& 0,
\\
 \Big(-x^2_1\partial_{x_1}+2j_1x_1 -x^2_2\partial_{x_2}+2j_2x_2-x^2_3\partial_{x_3}+2j_3x_3\Big)
D\Big[\!
\begin{array}{ccc}
\\[-18pt]
\scriptstyle j_3 & \scriptstyle j_2  & \scriptstyle j_1
\\[-8pt]
\scriptstyle \epsilon_3  &  \scriptstyle \epsilon_2  & \scriptstyle \epsilon_1
\\[-8pt]
\scriptstyle x_3  &  \scriptstyle x_2  &  \scriptstyle x_1
\end{array}
\!\Big]
 &=& 0.
\end{eqnarray*}
Up to a multiplicative,   $(j,\epsilon) $'s dependent constant the solution reads
$$
D\Big[\!
\begin{array}{ccc}
\\[-18pt]
\scriptstyle j_3 & \scriptstyle j_2  & \scriptstyle j_1
\\[-8pt]
\scriptstyle \epsilon_3  &  \scriptstyle \epsilon_2  & \scriptstyle \epsilon_1
\\[-8pt]
\scriptstyle x_3  &  \scriptstyle x_2  &  \scriptstyle x_1
\end{array}
\!\Big]
  = (x_1-x_2)^{j^3_{12}} (x_2-x_3)^{j^1_{23}}(x_3-x_1)^{j^2_{13}},
$$
where $j^k_{mn}=j_m+j_n-j_k$.
For the izospin representation one needs a function such that
$$
x_1^{-j_1-1+m_1+\epsilon_1}x_2^{-j_2-1+m_2+\epsilon_2}x_3^{-j_3-1+m_3+\epsilon_3}
D\Big[\!
\begin{array}{ccc}
\\[-18pt]
\scriptstyle j_3 & \scriptstyle j_2  & \scriptstyle j_1
\\[-8pt]
\scriptstyle \epsilon_3  &  \scriptstyle \epsilon_2  & \scriptstyle \epsilon_1
\\[-8pt]
\scriptstyle x_3  &  \scriptstyle x_2  &  \scriptstyle x_1
\end{array}
\!\Big]
$$
 is well defined  on the product $S^1\times S^1\times S^1$
of unit circles.
In the parametrization:
$$
M=(-\pi,\pi]\times (-\pi,\pi]\times (-\pi,\pi]
\ni (\varphi_1,\varphi_2,\varphi_3) \to (x_1,x_2,x_3)
= ({\rm e}^{i\varphi_1},{\rm e}^{i\varphi_2},{\rm e}^{i\varphi_3}) \in S^1\times S^1\times S^1
$$
one has
\begin{eqnarray*}
D\Big[\!
\begin{array}{ccc}
\\[-18pt]
\scriptstyle j_3 & \scriptstyle j_2  & \scriptstyle j_1
\\[-8pt]
\scriptstyle \epsilon_3  &  \scriptstyle \epsilon_2  & \scriptstyle \epsilon_1
\\[-8pt]
\scriptstyle x_3  &  \scriptstyle x_2  &  \scriptstyle x_1
\end{array}
\!\Big]
  &=& (x_1-x_2)^{j^3_{12}} (x_2-x_3)^{j^1_{23}}(x_3-x_1)^{j^2_{13}}
\\
&=&
 (2i)^{j_{123}} {\rm e}^{i{(\varphi_1j_1+\varphi_2j_2+\varphi_3j_3)}}
 \\
&\times &
\left(\sin{\textstyle {\varphi_1-\varphi_2\over 2}}\right)^{j^3_{12}}
 \left(\sin{\textstyle {\varphi_2-\varphi_3\over 2}}\right)^{j^1_{23}}
 \left(\sin{\textstyle {\varphi_3-\varphi_1\over 2}}\right)^{j^2_{13}},
\end{eqnarray*}
where $j_{123}=j_1+j_2+j_3$ and
\begin{eqnarray*}
&&\hspace{-50pt}
x_1^{-j_1-1+m_1+\epsilon_1}x_2^{-j_2-1+m_2+\epsilon_2}x_3^{-j_3-1+m_3+\epsilon_3}
D\Big[\!
\begin{array}{ccc}
\\[-18pt]
\scriptstyle j_3 & \scriptstyle j_2  & \scriptstyle j_1
\\[-8pt]
\scriptstyle \epsilon_3  &  \scriptstyle \epsilon_2  & \scriptstyle \epsilon_1
\\[-8pt]
\scriptstyle x_3  &  \scriptstyle x_2  &  \scriptstyle x_1
\end{array}
\!\Big]
\\
&=& (2i)^{j_{123}}{\rm e}^{i(m_1+\epsilon_1-1)\varphi_1}{\rm e}^{i(m_2+\epsilon_2-1)\varphi_2}{\rm e}^{i(m_3+\epsilon_3-1)\varphi_3}
\\
&\times &
 \left(\sin{\textstyle {\varphi_1-\varphi_2\over 2}}\right)^{j^3_{12}}
 \left(\sin{\textstyle {\varphi_2-\varphi_3\over 2}}\right)^{j^1_{23}}
 \left(\sin{\textstyle {\varphi_3-\varphi_1\over 2}}\right)^{j^2_{13}}.
\end{eqnarray*}
This expression is singular along the planes $\varphi_1=\varphi_2$, $\varphi_2=\varphi_3$, $\varphi_1=\varphi_3$ intersecting
the parameter cube $M$ into six regions:
$$
\begin{array}{rclcrcl}
M_{321}&=&\{\varphi_3<\varphi_2<\varphi_1\}&, \;\;& M_{312}&=&\{\varphi_3<\varphi_1<\varphi_2\},
\\
M_{132}&=&\{\varphi_1<\varphi_3<\varphi_2\}&,\;\;& M_{123}&=&\{\varphi_1<\varphi_2<\varphi_3\},
\\
M_{213}&=&\{\varphi_2<\varphi_1<\varphi_3\}&,\;\;& M_{231}&=&\{\varphi_2<\varphi_3<\varphi_1\}.
\end{array}
$$
Let $N_{abc}$ denote the images of these regions  in the 3-dimensional torus $N\subset \mathbb{C}^3$.
The singularity surface divides $N$ into two disjoint parts:
$$
N_{\rm A} = N_{321}\cup N_{132} \cup N_{213},\;\;\;\;N_{\rm B} = N_{312}\cup N_{123} \cup N_{231}.
$$
This implies that the space of 3-linear invariant forms is 2-dimensional.
For $\epsilon_i\in \{0,{1\over 2}\}$ satisfying
$$
\epsilon_1+\epsilon_2+\epsilon_3\;\dot{=}\;0,
$$
we define
\bea
S_{\rm A}\Big[\!
\begin{array}{ccc}
\\[-18pt]
\scriptstyle j_3 & \scriptstyle j_2  & \scriptstyle j_1
\\[-8pt]
\scriptstyle \epsilon_3  &  \scriptstyle \epsilon_2  & \scriptstyle \epsilon_1
\\[-8pt]
\scriptstyle x_3  &  \scriptstyle x_2  &  \scriptstyle x_1
\end{array}
\!\Big]
 &=&
\left\{
\begin{array}{rllll}
(-1)^{2(\epsilon_1+\epsilon_3)}(x_1-x_2)^{j^3_{12}} (x_2-x_3)^{j^1_{23}}(x_1-x_3)^{j^2_{13}}
&{\rm on}& N_{321}
\\
(-1)^{2(\epsilon_1+\epsilon_2)}(x_2-x_1)^{j^3_{12}} (x_2-x_3)^{j^1_{23}}(x_3-x_1)^{j^2_{13}}
&{\rm on}& N_{132}
\\
(-1)^{2(\epsilon_2+\epsilon_3)}(x_1-x_2)^{j^3_{12}} (x_3-x_2)^{j^1_{23}}(x_3-x_1)^{j^2_{13}}
&{\rm on}& N_{213}\\
0&{\rm on}& N_{\rm B}
\end{array}
\right.,
\\
S_{\rm B}
\Big[\!
\begin{array}{ccc}
\\[-18pt]
\scriptstyle j_3 & \scriptstyle j_2  & \scriptstyle j_1
\\[-8pt]
\scriptstyle \epsilon_3  &  \scriptstyle \epsilon_2  & \scriptstyle \epsilon_1
\\[-8pt]
\scriptstyle x_3  &  \scriptstyle x_2  &  \scriptstyle x_1
\end{array}
\!\Big]
&=&
\left\{
\begin{array}{rllll}
(-1)^{2(\epsilon_2+\epsilon_3)}(x_2-x_1)^{j^3_{12}} (x_2-x_3)^{j^1_{23}}(x_1-x_3)^{j^2_{13}}
&{\rm on}& N_{312}
\\
(-1)^{2(\epsilon_1+\epsilon_3)}(x_2-x_1)^{j^3_{12}} (x_3-x_2)^{j^1_{23}}(x_3-x_1)^{j^2_{13}}
&{\rm on}& N_{123}
\\
(-1)^{2(\epsilon_1+\epsilon_2)}(x_1-x_2)^{j^3_{12}} (x_3-x_2)^{j^1_{23}}(x_1-x_3)^{j^2_{13}}
&{\rm on}& N_{231}\\
0&{\rm on}& N_{\rm A}
\end{array}
\right. ,
\eea
and the linear combinations
\begin{equation}
\label{Sinvariant}
S_\epsilon = (-1)^{2\epsilon}S_{\rm A} +S_{\rm B},\;\;\;\epsilon =0,{\textstyle {1\over 2}}.
\end{equation}
The function
\begin{eqnarray*}
&&\hspace{-100pt}
x_1^{-j_1-1+m_1+\epsilon_1}x_2^{-j_2-1+m_2+\epsilon_2}x_3^{-j_3-1+m_3+\epsilon_3}S_\epsilon
\Big[\!
\begin{array}{ccc}
\\[-18pt]
\scriptstyle j_3 & \scriptstyle j_2  & \scriptstyle j_1
\\[-8pt]
\scriptstyle \epsilon_3  &  \scriptstyle \epsilon_2  & \scriptstyle \epsilon_1
\\[-8pt]
\scriptstyle x_3  &  \scriptstyle x_2  &  \scriptstyle x_1
\end{array}
\!\Big]
\\
&=& (2i)^{j_1+j_2+j_3}{\rm e}^{i(m_1+\epsilon_1-1)\varphi_1}{\rm e}^{i(m_2+\epsilon_2-1)\varphi_2}{\rm e}^{i(m_3+\epsilon_3-1)\varphi_3}
\\
&\times &{\rm sign}(\varphi_1-\varphi_2)^{2(\epsilon+\epsilon_1+\epsilon_2)}\left|\sin{\textstyle {\varphi_1-\varphi_2\over 2}}\right|^{j^3_{12}}
\\
&\times &
{\rm sign}(\varphi_2-\varphi_3)^{2(\epsilon+\epsilon_2+\epsilon_3)}
 \left|\sin{\textstyle {\varphi_2-\varphi_3\over 2}}\right|^{j^1_{23}}
 \\
&\times &
{\rm sign}(\varphi_3-\varphi_1)^{2(\epsilon+\epsilon_1+\epsilon_3)}
 \left|\sin{\textstyle {\varphi_3-\varphi_1\over 2}}\right|^{j^2_{13}}
\end{eqnarray*}
is periodic in all variables $\varphi_i$ for all $\epsilon, \epsilon_i$ satisfying
$
\epsilon_1+\epsilon_2+\epsilon_3\in \mathbb{Z}.
$
Under this condition the following identity holds:
\begin{eqnarray*}
&&\hspace{-20pt}{\rm sign}(\varphi_1-\varphi_2)^{2(\epsilon+\epsilon_1+\epsilon_2)}
{\rm sign}(\varphi_2-\varphi_3)^{2(\epsilon+\epsilon_2+\epsilon_3)}
{\rm sign}(\varphi_3-\varphi_1)^{2(\epsilon+\epsilon_1+\epsilon_3)}
\\
&=&
(-1)^{\epsilon
+ (\epsilon +\epsilon_1+\epsilon_2){\rm sign}(\varphi_1-\varphi_2)
+ (\epsilon +\epsilon_2+\epsilon_3){\rm sign}(\varphi_2-\varphi_3)
+ (\epsilon +\epsilon_3+\epsilon_1){\rm sign}(\varphi_3-\varphi_1)
}.
\end{eqnarray*}
From the definition of $S_\epsilon$ one gets the identity
\begin{eqnarray}
\label{Sidentity}
(x_3-x_1)^{n^2_{13}}
(x_2-x_3)^{n^1_{23}}
(x_1-x_2)^{n^3_{12}}
 S_{\epsilon}\Big[\!
\begin{array}{ccc}
\\[-18pt]
\scriptstyle j_3 & \scriptstyle j_2  & \scriptstyle j_1
\\[-8pt]
\scriptstyle \epsilon_3  &  \scriptstyle \epsilon_2  & \scriptstyle \epsilon_1
\\[-8pt]
\scriptstyle x_3  &  \scriptstyle x_2  &  \scriptstyle x_1
\end{array}
\!\Big]
&=&
S_{\epsilon\+\frac12 n_{123}}\Big[\!
\begin{array}{ccc}
\\[-18pt]
  \scriptstyle j_3 +n_3
& \scriptstyle j_2   +n_2
& \scriptstyle j_1+n_1
\\[-4pt]
 \scriptstyle \epsilon_3 \+n_3 &
 \scriptstyle \epsilon_2\+ n_2&
 \scriptstyle \epsilon_1\+ n_1
\\[-4pt]
\scriptstyle x_3 &  \scriptstyle x_2  &  \scriptstyle x_1
\end{array}
\!\Big].
\end{eqnarray}

\subsection{invariant three-linear forms in the spin basis }

Let $\{F^i_{n_i}=\ket{n_i+\epsilon_i}\}$ be the base in ${\cal D}_{j_i,\epsilon_i}$ ($ i=1,2,3$).
The 3-linear forms in these bases are given by
\begin{eqnarray*}
&&\hspace{-20pt}
S_{\epsilon}\Big[\!
\begin{array}{ccc}
\\[-18pt]
\scriptstyle -1-j_3 & \scriptstyle -1-j_2  & \scriptstyle -1-j_1
\\[-8pt]
\scriptstyle \epsilon_3  &  \scriptstyle \epsilon_2  & \scriptstyle \epsilon_1
\\[-8pt]
\scriptstyle m_3  &  \scriptstyle m_2  &  \scriptstyle m_1
\end{array}
\!\Big]
\\
&=&
\oint {dx_1\over 2\pi i}\oint {dx_2\over 2\pi i}\oint {dx_3\over 2\pi i}
x_1^{-1-j_1+m_1+\epsilon_1}x_2^{-1-j_2+m_2+\epsilon_2}x_3^{-1-j_3+m_3+\epsilon_3}
S_{\epsilon}\Big[\!
\begin{array}{ccc}
\\[-18pt]
\scriptstyle j_3 & \scriptstyle j_2  & \scriptstyle j_1
\\[-8pt]
\scriptstyle \epsilon_3  &  \scriptstyle \epsilon_2  & \scriptstyle \epsilon_1
\\[-8pt]
\scriptstyle x_3  &  \scriptstyle x_2  &  \scriptstyle x_1
\end{array}
\!\Big]
\\
&=& (2i)^{j_{123}}
\int\limits_{-\pi}^\pi {d\varphi_1\over 2\pi }\int\limits_{-\pi}^\pi {d\varphi_2\over 2\pi }\int\limits_{-\pi}^\pi {d\varphi_3\over 2\pi }
{\rm e}^{i(m_1+\epsilon_1)\varphi_1}{\rm e}^{i(m_2+\epsilon_2)\varphi_2}{\rm e}^{i(m_3+\epsilon_3)\varphi_3}
\\
&\times&
(-1)^{\epsilon
+ (\epsilon +\epsilon_3+\epsilon_2){\rm sign}(\varphi_3-\varphi_2)
+ (\epsilon +\epsilon_1+\epsilon_2){\rm sign}(\varphi_2-\varphi_1)
+ (\epsilon +\epsilon_3+\epsilon_1){\rm sign}(\varphi_1-\varphi_3)
}
\\
&\times &
\left|\sin{\textstyle {\varphi_1-\varphi_2\over 2}}\right|^{j^3_{12}}
\left|\sin{\textstyle {\varphi_2-\varphi_3\over 2}}\right|^{j^1_{23}}
\left|\sin{\textstyle {\varphi_3-\varphi_1\over 2}}\right|^{j^2_{13}},
\end{eqnarray*}
where $j_{123}=j_1+j_2+j_3,\,j^a_{bc}=j_b+j_c-j_a$.
This is the well known integral representation of the Clebsch-Gordan coefficients (CGC) for the principal continuous representations
of ${\rm SL}(2,\mathbb{R})$.
It can be explicitly calculated \cite{Vilenkin,Kerimov,Ribault:2009ui}:
\bea
S_{\epsilon}\Big[\!
\begin{array}{ccc}
\\[-18pt]
\scriptstyle -1-j_3 & \scriptstyle -1-j_2  & \scriptstyle -1-j_1
\\[-8pt]
\scriptstyle \epsilon_3  &  \scriptstyle \epsilon_2  & \scriptstyle \epsilon_1
\\[-8pt]
\scriptstyle m_3  &  \scriptstyle m_2  &  \scriptstyle m_1
\end{array}
\!\Big]
&=&
-\frac{1}{\pi^2} \delta_{\sum_i (m_i + \epsilon_i),0}  \, i^{j_{123}} \,
 \Gamma(1+ j^3_{12})  \Gamma(1+ j^1_{32})  \Gamma(1+ j^2_{13}) \,
 g^{\epsilon}(j_i, m_i+\epsilon_i)
\end{eqnarray*}
 and
\begin{eqnarray}
\nonumber
g^{\epsilon} (j_i, m_i+\epsilon_i)
&=&
   s({\textstyle\frac12} j_{13}^2 - \epsilon +\epsilon_2)  s(j_1-\epsilon_1) s(j_3+\epsilon_3)
    \left( g^{31} (j_i, m_i+\epsilon_i)
+ (-1)^{2\epsilon} g^{13} (j_i, m_i+\epsilon_i)   \right)\, ,
\\[6pt]
\nonumber
 g^{31} (j_i, m_i)
&=&
G\left[
{}^{-j_3-m_3\;\;-j^2_{13}\;\; -j_1+m_1}_{1+j_2-j_3+m_1\;\;1+j_2-j_1-m_3}\right]\, ,
\\[6pt]
\nonumber
 g^{13} (j_i, m_i)
&=&
G\left[
{}^{-j_3+m_3\;\;-j^2_{13}\;\; -j_1-m_1}_{1+j_2-j_3-m_1\;\;1+j_2-j_1+m_3}\right] \, ,
\\[6pt]
\label{Gfunction}
G\left[\,{}^{a\,b\,c}_{\;e\,f}\,\right]
&=&
{\Gamma[a]\,\Gamma[b]\,\Gamma[c]\over \Gamma[e]\,\Gamma[f]} \,{}_3F_2 \Big[\,{}^{a\,b\,c}_{\;e\,f}\,|\,1\,\Big],\;\;\;\;s(x)\equiv \sin(\pi x).
\end{eqnarray}
The  CGC coefficients satisfy
the reflection relation\footnote{We derive this relation in Appendix A.}
\begin{eqnarray}
\nonumber
&& \hspace{-30pt}
\frac{\Gamma( - j_3+ m_3+ \epsilon_3)}{\Gamma( 1+ j_3+ m_3+\epsilon_3)}
S_{\epsilon}\Big[\!
\begin{array}{ccc}
\\[-18pt]
\scriptstyle j_3 & \scriptstyle -1-j_2  & \scriptstyle -1-j_1
\\[-8pt]
\scriptstyle \epsilon_3  &  \scriptstyle \epsilon_2  & \scriptstyle \epsilon_1
\\[-8pt]
\scriptstyle m_3  &  \scriptstyle m_2  &  \scriptstyle m_1
\end{array}
\!\Big]
 \\
 \label{refrel}
   &=& (-1)^{2\epsilon_3} \frac{ \Gamma(- j_{23}^1)  }{ \Gamma(1+ j_{13}^2) }
 \frac{s(-\frac12 -\frac12 j_{23}^1 + \epsilon + \epsilon_{2}) }{  s(\frac12 j_{13}^2 - (\epsilon\+\epsilon_3) + \epsilon_{2}) }
 S_{\epsilon\+ \epsilon_3}\Big[\!
\begin{array}{ccc}
\\[-18pt]
\scriptstyle -1-j_3 & \scriptstyle -1-j_2  & \scriptstyle -1-j_1
\\[-8pt]
\scriptstyle \epsilon_3  &  \scriptstyle \epsilon_2  & \scriptstyle \epsilon_1
\\[-8pt]
\scriptstyle m_3  &  \scriptstyle m_2  &  \scriptstyle m_1
\end{array}
\!\Big]
\\
\nonumber
&=&
\frac{ \Gamma(- j_{13}^2)  }{ \Gamma(1+ j_{23}^1) }
 \frac{s(-\frac12 -\frac12 j_{13}^2 + \epsilon + \epsilon_{1}) }{  s(\frac12 j_{23}^1 - (\epsilon\+\epsilon_3) + \epsilon_{1}) }
S_{\epsilon\+ \epsilon_3}\Big[\!
\begin{array}{ccc}
\\[-18pt]
\scriptstyle -1-j_3 & \scriptstyle -1-j_2  & \scriptstyle -1-j_1
\\[-8pt]
\scriptstyle \epsilon_3  &  \scriptstyle \epsilon_2  & \scriptstyle \epsilon_1
\\[-8pt]
\scriptstyle m_3  &  \scriptstyle m_2  &  \scriptstyle m_1
\end{array}
\!\Big].
\end{eqnarray}
In the izospin variables
it takes the form
\begin{eqnarray}
\nonumber
&&\hspace{-30pt}
{1\over C_{j_3,\epsilon_3}}
\oint {dy\over 2\pi i} S_{j_3,\epsilon_3}(x_3,y)
S_{\epsilon}\Big[\!
\begin{array}{ccc}
\\[-18pt]
\scriptstyle -1-j_3 & \scriptstyle j_2  & \scriptstyle j_1
\\[-8pt]
\scriptstyle \epsilon_3  &  \scriptstyle \epsilon_2  & \scriptstyle \epsilon_1
\\[-8pt]
\scriptstyle y  &  \scriptstyle x_2  &  \scriptstyle x_1
\end{array}
\!\Big]
\\
 \label{refrelizo}
   &=& (-1)^{2\epsilon_3} \frac{ \Gamma(- j_{23}^1)  }{ \Gamma(1+ j_{13}^2) }
 \frac{s(-\frac12 -\frac12 j_{23}^1 + \epsilon + \epsilon_{2}) }{  s(\frac12 j_{13}^2 - (\epsilon\+\epsilon_3) + \epsilon_{2}) }
 S_{\epsilon\+ \epsilon_3}\Big[\!
\begin{array}{ccc}
\\[-18pt]
\scriptstyle j_3 & \scriptstyle j_2  & \scriptstyle j_1
\\[-8pt]
\scriptstyle \epsilon_3  &  \scriptstyle \epsilon_2  & \scriptstyle \epsilon_1
\\[-8pt]
\scriptstyle x_3  &  \scriptstyle x_2  &  \scriptstyle x_1
\end{array}
\!\Big]
\\
\nonumber
&=&
\frac{ \Gamma(- j_{13}^2)  }{ \Gamma(1+ j_{23}^1) }
 \frac{s(-\frac12 -\frac12 j_{13}^2 + \epsilon + \epsilon_{1}) }{  s(\frac12 j_{23}^1 - (\epsilon\+\epsilon_3) + \epsilon_{1}) }
S_{\epsilon\+ \epsilon_3}\Big[\!
\begin{array}{ccc}
\\[-18pt]
\scriptstyle j_3 & \scriptstyle j_2  & \scriptstyle j_1
\\[-8pt]
\scriptstyle \epsilon_3  &  \scriptstyle \epsilon_2  & \scriptstyle \epsilon_1
\\[-8pt]
\scriptstyle x_3  &  \scriptstyle x_2  &  \scriptstyle x_1
\end{array}
\!\Big].
\end{eqnarray}

\subsection{hermitian adjoint representations}

Now we shall turn to the question of a hermitian pairing between representations.
The method of analysis is the same as in the case of ${\rm sl}(2,\mathbb{R})$
but the conditions for the adjoint generators are different.
We say that the ${\rm su}(2)$ representations ${\cal D}_{j,\epsilon}, {\cal D}_{j',\epsilon'}$
are hermitian adjoint
if
there exists a hermitian form such that:
$$
(J^3 f,f')=(f,J^3f'),\;\;\;\;\;\;(J^\pm f,f')=(f,J^\mp f').
$$
This conditions can be satisfied if and only if
\begin{equation}
\label{eq2}
j'=\bar j\;\;\;\;{\rm or}\;\;\;\;j'=-\bar j-1.
\end{equation}
In both cases
the hermitian pairing is  determined
up to an overall normalization:
\begin{equation}
\label{cond}
\begin{array}{rclllll}
(\bar j-\epsilon-n) (F_{n+1},F'_{n+1})_{\rm II}&=&\displaystyle ({\bar j +\epsilon +n+1 } )(F_n,F'_n)_{\rm II} &&{\rm for}& j'=\bar j&,
 \\[10pt]
 (F_{n+1},F'_{n+1})_{\rm I}&=&-(F_n,F'_n)_{\rm I} &&{\rm for}& j'=-1-\bar j&,
\end{array}
\end{equation}
where $\{F_n=\ket{n+\epsilon}\},\{F'_n=\ket{n+\epsilon}\}$ are the bases in ${\cal D}_{j,\epsilon}, {\cal D}_{j',\epsilon}$, respectively.

One easily checks that both hermitian forms are related by the reflection map\\
$Q_{\bar j}:{\cal D}_{\bar j,\epsilon}\to{\cal D}_{-1-\bar j,\epsilon}$:
$$
(\;.\;,\;.\;)_{\rm II} = (\;.\;,\;Q_{\bar j}\;.\;)_{\rm I}.
$$
For non-integral $(j,\epsilon)$
and real $j$  the form $(\;.\;,\;.\;)_{\rm II}$ defines an indefinite scalar product on ${\cal D}_{j,\epsilon}$.

For non-integral $(j,\epsilon)$
and $j$ of the form
$$
j=-{\textstyle{1\over 2}} + is,\;\;\;\;\;s\in \mathbb{R}
$$
$(\;.\;,\;.\;)_{\rm I}$ defines an indefinite scalar product on ${\cal D}_{j,\epsilon}$. We denote by $(\;.\;,\;.\;)_{j,\epsilon}$
this scalar product normalized by the condition
$$
(\,F_0\,,\,F_0\,)_{j,\epsilon}=1.
$$
This product is invariant with respect to the reflection map (\ref{reflectionI})
$$
(\,Q_{j,\epsilon} \chi\,,\, Q_{j,\epsilon} \chi'  \,)_{-1-j,\epsilon}
=
(\,\chi\,,\,\chi'\,)_{j,\epsilon}.
$$

For integral $(l,\epsilon)$ the scalar product $(\;.\;,\;.\;)_{\rm II}$ is positively defined.
For instance in the cases of $l={1\over 2},-{3\over 2}$  condition (\ref{cond}) reads
$$
(F_{0},F_{0})_{\rm II}=(F_{-1},F_{-1})_{\rm II}.
$$
For generic $j\in \mathbb{C}$
one  has  the hermitian conjugate pairs of representations
$$
(j=-\textstyle{1\over 2} +is +\nu, \epsilon),  \;\;\;\;(\tilde j=-\textstyle{1\over 2} +is -\nu,\epsilon), \;\;\;s,\nu\in \mathbb{R}.
$$
Except the case $\nu=0$ the eigenvalues of the  Casimir operator
for this representations are complex conjugate numbers
\begin{eqnarray*}
j(j+1) &=& (-\textstyle{1\over 4}   -s^2 +2i s\nu +\nu^2)
,\;\;\;\;\;\; \tilde j(\tilde j+1)\;=\; (-\textstyle{1\over 4}   -s^2 -2i s\nu +\nu^2).
\end{eqnarray*}
In order to construct a hermitian representations with $\nu\neq 0$ it is necessary to extend the space of representation to the direct sum
$$
{\cal G}_{j,\epsilon}={\cal D}_{j,\epsilon} \oplus {\cal D}_{-1-\bar j,\epsilon}.
$$
On the extended space the generators are defined by
$$
J^a=\left(
\begin{array}{cc}
J^a&0
\\
0&J^a
\end{array}
\right).
$$
This is a hermitian representation of ${\rm su}(2)$ with the indefinite scalar product defined by
$$
(F_n,F_m)=(\tilde F_n,\tilde F_m)=0,\;\;\;\;\;(F_n,\tilde F_m)=(\tilde F_m, F_n) = (F_n,\tilde F_m)_{\rm I} =(-1)^n\delta_{m,n},
$$
where $F_n\in {\cal D}_{j,\epsilon},\,\tilde F_n\in  {\cal D}_{-1-\bar j,\epsilon}$.

\subsection{tensoring representations}

Let us consider the tensor product of the ${\rm su}(2)$ representations:
$
{\cal D}_{j,\epsilon} \otimes {\cal S}_{1\over 2}.
$
The eigenvalues of ${\cal J}^3=J^3+K^3$ are double degenerate on ${\cal D}_{j,\epsilon} \otimes {\cal S}_{1\over 2}$:
\begin{eqnarray*}
{\cal J}^3\ket{n+\epsilon}_j\otimes \ket{{\textstyle {1\over 2}}}_{1\over 2}
&=&
(n+\epsilon+{\textstyle {1\over 2}})\ket{n+\epsilon}_j\otimes \ket{{\textstyle {1\over 2}}}_{1\over 2},
\\
{\cal J}^3\ket{n+1+\epsilon}_j\otimes \ket{-{\textstyle {1\over 2}}}_{1\over 2}
&=&
(n+\epsilon+{\textstyle {1\over 2}})\ket{n+1+\epsilon}_j\otimes \ket{-{\textstyle {1\over 2}}}_{1\over 2}.
\end{eqnarray*}
Diagonalizing  the Casimir operator on these 2-dim eigenspaces one gets
the eigenvalues
$$
(j+{\textstyle{1\over 2}})(j+{\textstyle{3\over 2}}),
\;\;\;\;\;\;
(j-{\textstyle{1\over 2}})(j+{\textstyle{1\over 2}})
$$
and the corresponding eigenvectors
\begin{eqnarray*}
\ket{n+{\textstyle {1\over 2}}+\epsilon}_{j+{1\over 2}} &=&
(1+j+n+\epsilon)\ket{n}_j\otimes\ket{{\textstyle {1\over 2}}}_{1\over 2}
+ (j-n-\epsilon)\ket{n+1}_j\otimes\ket{-{\textstyle {1\over 2}}}_{1\over 2},
\\
\ket{n+{\textstyle {1\over 2}}+\epsilon}_{j-{1\over 2}}
&=&
\ket{n}_j\otimes\ket{{\textstyle {1\over 2}}}_{1\over 2}
-\ket{n+1}_j\otimes\ket{-{\textstyle {1\over 2}}}_{1\over 2}\, ,
\end{eqnarray*}
It follows that
$$
{\cal D}_{j,\epsilon} \otimes {\cal S}_{1\over 2} \simeq {\cal D}_{j+{1\over 2},\bar \epsilon } \oplus {\cal D}_{j-{1\over 2},\bar \epsilon }.
$$
where $\bar\epsilon= \epsilon \dot{+} {1\over 2}$.
The same analysis for
$
{\cal D}_{j,\epsilon} \otimes {\cal S}_{-{3\over 2}}
$
yields the same eigenvalues. The corresponding vectors are slightly different:
\begin{eqnarray}
\nonumber
\ket{n+{\textstyle {1\over 2}}+\epsilon}_{j+{1\over 2}} &=&
(1+j+n+\epsilon)\ket{n}_j\otimes\ket{{\textstyle {1\over 2}}}_{-{3\over 2}}
+ (-j+n+\epsilon)\ket{n+1}_j\otimes\ket{-{\textstyle {1\over 2}}}_{-{3\over 2}},
\\
\label{tenpro}
\ket{n+{\textstyle {1\over 2}}+\epsilon}_{j-{1\over 2}}
&=&
\ket{n}_j\otimes\ket{{\textstyle {1\over 2}}}_{-{3\over 2}}
+\ket{n+1}_j\otimes\ket{-{\textstyle {1\over 2}}}_{-{3\over 2}},
\end{eqnarray}
but the tensor product decomposition takes the same form
$$
{\cal D}_{j,\epsilon} \otimes {\cal S}_{-{3\over 2}}
 \simeq {\cal D}_{j+{1\over 2},\bar \epsilon } \oplus {\cal D}_{j-{1\over 2},\bar\epsilon }.
$$
For the hermitian representations one thus gets
\begin{eqnarray*}
{\cal G}_{j,\epsilon} \otimes {\cal S}_{1\over 2}
&\simeq&
{\cal G}_{j,\epsilon} \otimes {\cal S}_{-{3\over 2}}
\;\simeq\;
{\cal G}_{j+{1\over 2},\bar\epsilon }
\oplus
{\cal G}_{j-{1\over 2},\bar\epsilon }.
\end{eqnarray*}

\section{Nonrational $\hat{\rm su}(2)_\kappa$   WZNW model}

\subsection{relaxed $\hat{\rm su}(2)_\kappa$ modules}

The $\hat{\rm su}(2)_\kappa$ affine algebra at the level $\kappa$ is defined by:
\begin{eqnarray}
\nonumber
\left[ J^3_m, J^3_n\right]
&=&
{\kappa\over 2} \, m  \, \delta_{m+n,0},
\\
\label{affine}
\left[ J^3_m, J^\pm_n\right]
&=&
\pm J^\pm_{m+n},
\\
\nonumber
\left[ J^+_m, J^-_n\right]
&=&
2J^3_{m+n} + \kappa \, m \, \delta_{m+n,0},\;\;\;\;\;\;m,n\in\mathbb{Z}.
\end{eqnarray}
The Sugawara construction
$$
L_m
= {1\over 2(\kappa+2)}\sum_n \left(2:J^3_nJ^3_{m-n}:
+:J^+_nJ^-_{m-n}:
+:J^-_nJ^+_{m-n}: \right)
$$
yields  the associate Virasoro algebra
\begin{eqnarray*}
\left[L_m,L_n\right]&=& (m-n)L_{m+n} +{c\over 12} (m^3-m)\delta_{m+n,0},
\\
\left[L_m,J^a_n \right]&=& -nJ^a_{m+n},
\end{eqnarray*}
with the central charge
$$
c={3\kappa\over 2+\kappa}.
$$

Let $\left\{\ket{m+\epsilon}\right\}_{m\in \mathbb{Z}}$ be the canonical basis in the representation ${\cal D}_{j,\epsilon}$
with the ${\rm su}(2)$ generators denoted by $J^3_0,J^\pm_0$.
 The relaxed $\hat{\rm su}(2)_\kappa$ module $\hat {\cal D}_{j,\epsilon}^{\kappa}$ \cite{Feigin:1997ha} is generated
from the states
$\ket{m+\epsilon}$ satisfying the annihilation conditions:
$$
J^3_n\ket{m+\epsilon} = J^+_n\ket{m+\epsilon} =  J^-_n\ket{m+\epsilon} = 0, \;\;{\rm for}\;n>0,
$$
by the action of the generators:
$
J^3_n,\;J^\pm_n,\; n<0.
$
It has a natural
$\mathbb{Z}$ grading
\begin{eqnarray*}
\hat{\cal D}^\kappa_{j,\epsilon}
&=&
 \bigoplus\limits_{n=0}^\infty \hat{\cal D}^{\kappa,n}_{j,\epsilon},\;\;\;\;
 \hat{\cal D}^{\kappa,0}_{j,\epsilon}={\cal D}_{j,\epsilon},
\end{eqnarray*}
where $\hat{\cal D}^{\kappa,n}_{j,\epsilon}$ are eigenspaces
 of the operator
 $
 L_0-{{j(j+1)}\over \kappa+2}
 $.
For generic $j$ there is no hermitian, non-degenerate bilinear form on $\hat{\cal D}^\kappa_{j,\epsilon}$.
But for the modules
$\hat{\cal D}^\kappa_{j,\epsilon}$ and $\hat{\cal D}^\kappa_{-1-\bar j,\epsilon}$
there is a hermitian pairing defined on the zero level subspaces ${\cal D}_{j,\epsilon}$, ${\cal D}_{-1-\bar j,\epsilon}$
by the form $(\,.\,,\,.\,)_{\rm I}$ and extended to the whole modules by the
the hermitian conjugation rules:
\begin{equation}
\label{hcr}
\left( J^3_n\right)^\dagger =J^3_{-n},\;\;\;\;\;\;\left( J^\pm_n\right)^\dagger =J^\mp_{-n}.
\end{equation}

\subsection{associated highest weight module}

We shall now describe a construction of the highest weight module ${\cal H}_j^\kappa$ associated to the
module $\hat {\cal D}^\kappa_{-1-j,\epsilon}$. To this end let us
define a new set of generators
\begin{eqnarray} \label{Jx}
\nonumber
J^+_n(x)&=& J^+_n -2xJ^3_n -x^2J^-_n ,\\
J^3_n(x)&=& J^3_n+ xJ^-_n, \\ \nonumber
J^-_n(x)&=& J^-_n,
\end{eqnarray}
where $x$ is a complex parameter.
For any $x$ they satisfy the commutation relations of the $\hat {\rm su}(2)$ affine algebra at level $\kappa$:
\begin{eqnarray*}
\left[ J^3_m(x), J^3_n(x)\right]
&=&
{\kappa\over 2}m \delta_{m+n,0},
\\
\left[ J^3_m(x), J^\pm_n(x)\right]
&=&
\pm J^\pm_{m+n}(x),
\\
\left[ J^+_m(x), J^-_n(x)\right]
&=&
2J^3_{m+n}(x) + \kappa m \delta_{m+n,0},\;\;\;\;\;\;m,n\in\mathbb{Z}.
\end{eqnarray*}
The state $\ket{x}_{j,\epsilon} $ (\ref{hws}) is the highest weight state with respect to the algebra $J^a_n(x)$.
Indeed it is annihilated by $J^a_n(x), n>0$ and
\begin{eqnarray*}
J^+_0(x)\ket{x}_{j,\epsilon}  &=&0,
\\
J^3_0(x)\ket{x}_{j,\epsilon}  &=&
j\ket{x}_{j,\epsilon} .
\end{eqnarray*}
The construction above was first introduced in the case of finite dimensional representations ${\cal S}_l$ \cite{Zamolodchikov:1986bd}
and is known as the $x$-representation in the WZNW models (with $x$ called the isospin variable). One has for instance
\begin{eqnarray*}
 \ket{x}_{1\over 2} \equiv
 \ket{x}_{\frac12, \frac12} &=& \ket{\textstyle {1\over 2}}_{-{3\over 2}} +x\ket{\textstyle -{1\over 2}}_{-{3\over 2}},
\end{eqnarray*}
and:
\begin{eqnarray*}
K^+_0(x)\ket{x}_{1\over 2} &=& (K^+_0-2xK^3_0 -x^2K^-_0)(\ket{\textstyle {1\over 2}}_{-{3\over 2}} +x\ket{\textstyle -{1\over 2}}_{-{3\over 2}})
=0,
\\
K^3_0(x)\ket{x}_{1\over 2} &=& (K^3_0 +x K^-_0)(\ket{\textstyle {1\over 2}}_{-{3\over 2}} +x\ket{\textstyle -{1\over 2}}_{-{3\over 2}})
= {\textstyle{1\over 2}}\ket{x}_{1\over 2},
\\
K^-_0(x)\ket{x}_{1\over 2} &=& K^-_0\ket{x}_{1\over 2} = \ket{\textstyle - {1\over 2}}_{-{3\over 2}}.
\end{eqnarray*}

\subsection{spectrum and operator - state correspondence}

We assume that  the $\hat{\rm su}(2)_\kappa$ WZNW model at non-rational level $\kappa$ is based on the representations
$$
\hat{\cal D}^\kappa_{j,\epsilon},\;\;\;\;j=-{\textstyle{1\over 2}}+ is,\;\;\;
s\in \mathbb{R},\;\;\epsilon =0,{\textstyle \frac12}.
$$
The spectrum is diagonal and the space of states is the direct integral/sum of the tensor products of the left and the right relaxed modules
$$
\hat{\cal D}^{\kappa}_{j,\epsilon}\otimes \hat{\cal D}^{\kappa}_{j,\epsilon}
$$
with the indefinite  scalar product $\langle\,.\,,\,.\,\rangle_{j,\epsilon }$ defined by hermitian form  (\ref{cond}) extended by  hermitian conjugation relations
(\ref{hcr}).

The primary fields in the isospin variables can be introduced by the operator state correspondence:
\begin{eqnarray*}
\lim\limits_{z\to 0} \Phi_{j,\epsilon}(x,\bar x;z,\bar z)\ket{0} = \ket{x}_{j,\epsilon}\otimes\ket{\bar x}_{j,\epsilon},
\end{eqnarray*}
where $\ket{x}_{j,\epsilon},\ket{\bar x}_{j,\epsilon}$ are the highest weight states defined in  (\ref{hws}):
$$
\ket{x}_{j,\epsilon}= \sum\limits_{m=-\infty}^\infty x^{j-m-\epsilon} \ket{m+\epsilon}_{-1-j}.
$$

One can decompose $\Phi_{j,\epsilon}(x,\bar x;w,\bar w)$ into the primary fields in the spin bases
$$
\Phi_{j,\epsilon}(x,\bar x;z,\bar z)=
\sum\limits_{m, \bar m=-\infty}^\infty
 V^{-1-j,\epsilon}_{m+\epsilon,\bar m+\epsilon}(z,\bar z) x^{j-m-\epsilon} \, \bar x^{j-\bar m-\epsilon}
$$
for which the  operator-state correspondence takes the form
\begin{eqnarray*}
\lim\limits_{z\to 0} V^{-1-j,\epsilon}_{m+\epsilon,\bar m+\bar\epsilon}(z,\bar z )\ket{0} =
\ket{m+\epsilon}_{-1-j}\otimes \ket{\bar m+\bar\epsilon}_{-1-j}.
\end{eqnarray*}
The properties of the primary fields listed above suggest that they can be seen as tensor products
of their chiral parts
\bea
\Phi_{j,\epsilon} (x,\bar x;z,\bar z)
&=&
\Phi_{j,\epsilon} (x;z) \otimes \Phi_{j,\epsilon} (\bar x;\bar z),
\\[8pt]
 V^{-1-j,\epsilon}_{m+\epsilon,\bar m+\bar\epsilon}(z,\bar z)
 &=&
  V^{-1-j,\epsilon}_{m+\epsilon}(z) \otimes  V^{-1-j,\epsilon}_{\bar m+\epsilon}(\bar z) .
\eea
It should be emphasized that although $\bar z$ is  complex conjugate to the complex variable $z$,
$x$ and $\bar x$ are independent complex variables.

Since the representations ${\cal D}_{j,\epsilon}, {\cal D}_{-1-j,\epsilon}$ are equivalent we declare an identification
of the corresponding primary fields. It means that
in any correlation function the following relation holds
$$
(Q_{-1-j,\epsilon} \otimes Q_{-1-j,\epsilon}) \Phi_{j,\epsilon} (x,\bar x;z,\bar z)
=R_{j,\epsilon}^2\Phi_{1,\epsilon} (x,\bar x;z,\bar z).
$$
In terms of the chiral primary fields one has
$$
\begin{array}{rcl}
Q_{-1-j,\epsilon}  \Phi_{j,\epsilon} (x,z)
&=&R_{j,\epsilon}
\Phi_{j,\epsilon} (x;z),
\\[8pt]
Q_{-1-j,\epsilon}  \Phi_{j,\epsilon} (\bar x,\bar z)
&=&
R_{j,\epsilon}\Phi_{j,\epsilon} (\bar x;\bar z),
\end{array}
$$
where $Q_{-1-j}$ is reflection map  (\ref{reflectionI})
and the coefficient $R_{j,\epsilon}$ is the chiral reflection amplitude. Since $Q_jQ_{-1-j}={\rm id}$,  it satisfies
$
R_{-1-j,\epsilon}R_{j,\epsilon}=1.
$
Using (\ref{reflectionkernel})
one can write the identification in the form
\begin{equation}
\label{identification}
\begin{array}{rcl}
\displaystyle {1\over C_{j,\epsilon}}
\oint {dy\over 2\pi i} S_{j,\epsilon}(x,y)\Phi_{-1-j,\epsilon} (y,z)
&=&R_{j,\epsilon}
\Phi_{j,\epsilon} (x;z),
\\[10pt]
\displaystyle {1\over C_{j,\epsilon}}
\oint {dy\over 2\pi i} S_{j,\epsilon}(\bar x,y)\Phi_{-1-j,\epsilon} (y,\bar z)
&=&
R_{j,\epsilon}\Phi_{j,\epsilon} (\bar x;\bar z).
\end{array}
\end{equation}

\subsection{Ward identities}

Relations (\ref{xrep}) imply the following OPE of the ${\rm su}(2)$ currents with the primary fields
\begin{eqnarray}
\nonumber
J^+(z)\Phi_{j,\epsilon} (x,\bar x;w,\bar w)
&\sim &
{ (-x^2\partial_x +2j x )\Phi_{j,\epsilon}(x,\bar x;w,\bar w)\over z-w},
\\
\label{OPE}
J^3(z)\Phi_{j,\epsilon} (x,\bar x;w,\bar w)
&\sim &
{(- x\partial_x +j )\Phi_{j,\epsilon}(x,\bar x;w,\bar w)\over z-w},
\\
\nonumber
J^-(z)\Phi_{j,\epsilon} (x,\bar x;w,\bar w)
&\sim &
{ \partial_x  \Phi_{j,\epsilon}(x,\bar x;w,\bar w)\over z-w},
\end{eqnarray}
and the corresponding relations for the right currents $
J^+(\bar z),
J^3(\bar z),
J^-(\bar z)$.
It is convenient to introduce the ${\rm su}(2)$ currents with the isospin variables \cite{Zamolodchikov:1986bd}:
\begin{eqnarray}
\nonumber
J^+(x,z) &=&J^+(z) -2x  J^3(z)   - x^2 J^-(z),
 \\
J^3(x,z) &=& J^3(z)   + x J^-(z) ,
 \label{J(x,z)}
 \\
\nonumber
J^-(x,z) &=& J^-(z).
 \end{eqnarray}
The excited states are iteratively defined in the standard manner
$$
J^a_{n}(y) J_M \Phi_{j,\epsilon} (x, z)
=
\oint_z {dw\over 2\pi i} (w-z)^n J^a(y,w)  J_M \Phi_{j,\epsilon} (x, z),
$$
where $J_M$ denotes an arbitrary array of the operators $J^a_{m}(y')$.
The general Ward identities with the ${\rm su}(2)$ currents take the following  form,
\begin{eqnarray*}
&& \hspace{-40pt}
\left\langle J_K \Phi_{j_3,\epsilon_3} (x_3, z_3) J_L \Phi_{j_2,\epsilon_2} (x_2, z_2)  J^a_{-n} J_M \Phi_{j_1,\epsilon_1} (x_1, z_1) \right\rangle
\\
&=&
- \sum_{p=0}^{\infty} (-1)^p \left(^{n+p -1}_{\quad p} \right) \frac{\left\langle J^a_{p}  \, J_K \Phi_{j_3,\epsilon_3} (x_3, z_3)    J_L \Phi_{j_2,\epsilon_2} (x_2, z_2)  J_M \Phi_{j_1,\epsilon_1} (x_1, z_1)  \right\rangle}{(z_3-z_1)^{n+p}}
\\
&&
- \sum_{p=0}^{\infty} (-1)^p \left(^{n+p -1}_{\quad p} \right)
\frac{\left\langle J_K \Phi_{j_3,\epsilon_3} (x_3, z_3) \,   J^a_{p}    J_L \Phi_{j_2,\epsilon_2} (x_2, z_2)  J_M \Phi_{j_1,\epsilon_1} (x_1, z_1)  \right\rangle}{(z_2-z_1)^{n+p}}.
\end{eqnarray*}
In the special case of two primary fields and the $x$-dependent currents we have
\begin{eqnarray*}
&& \hspace{-80pt}
\left\langle  \Phi_{j_3,\epsilon_3} (x_3, z_3) \Phi_{j_2,\epsilon_2} (x_2, z_2)  J^a_{-n}(x_1) J_M(x_1)  \Phi_{j_1,\epsilon_1}(x_1, z_1) \right\rangle
\\
&=&
-  \frac{\left\langle  J^a_{0}(x_1)  \Phi_{j_3,\epsilon_3} (x_3, z_3)  \Phi_{j_2,\epsilon_2} (x_2, z_2)  J_M(x_1) \Phi_{j_1,\epsilon_1}(x_1, z_1) \right\rangle
}{(z_3-z_1)^{n}}
\\&&
-
\frac{\left\langle  \Phi_{j_3,\epsilon_3} (x_3, z_3)  J^a_{0}(x_1)  \Phi_{j_2,\epsilon_2} (x_2, z_2)  J_M(x_1) \Phi_{j_1,\epsilon_1}(x_1, z_1)  \right\rangle
}{(z_2-z_1)^{n}},
\end{eqnarray*}
where the action of the zero modes
can be derived from (\ref{OPE}):
\begin{eqnarray}
\nonumber
J_0^+(x)  \Phi_{j,\epsilon} (y, w )
&=&
- \left( (x-y)^2 \partial_y + 2j (x-y) \right) \Phi_{j,\epsilon} (y, w ),
\\
\label{JxOPE}
J_0^3(x)  \Phi_{j,\epsilon} (y, w )
&=&
\left( (x-y) \partial_y + j \right)  \phi_j(y,w),
\\ \nonumber
J_0^-(x) \Phi_{j,\epsilon} (y, w )
&=&
\partial_y  \Phi_{j,\epsilon} (y, w ).
\end{eqnarray}
In the limit $z_3\to \infty,z_2\to z,z_1\to 0$ one gets further simplification
\begin{eqnarray}
\label{simpleWI}
&& \hspace{-80pt}
\left\langle  \Phi_{j_3,\epsilon_3} (x_3, \infty) \Phi_{j_2,\epsilon_2} (x_2, z)  J^a_{-n}(x_1) J_M(x_1)  \Phi_{j_1,\epsilon_1}(x_1, 0) \right\rangle
\\
\nonumber
&=&
-
\frac{\left\langle  \Phi_{j_3,\epsilon_3} (x_3, \infty)  J^a_{0}(x_1)  \Phi_{j_2,\epsilon_2} (x_2, z)  J_M(x_1) \Phi_{j_1,\epsilon_1}(x_1, 0)  \right\rangle
}{z^{n}}.
\end{eqnarray}

\subsection{2-point functions}

The global Ward identities imply the following general form of the 2-point function
\bea
\langle
\Phi_{j_1,\epsilon_1}(x_1,\bar x_1;z_1,\bar z_1)
\Phi_{j_2,\epsilon_2}(x_2,\bar x_2;z_2,\bar z_2)
\rangle
&=&
\langle
\Phi_{j_1,\epsilon_1}(x_1;z_1)
\Phi_{j_2,\epsilon_2}(x_2;z_2)
\rangle
\langle
\Phi_{j_1,\epsilon_1}(\bar x_1;\bar z_1)
\Phi_{j_2,\epsilon_2}(\bar x_2;\bar z_2)
\rangle,
\\
\langle
\Phi_{j_1,\epsilon_1}(x_1;z_1)
\Phi_{j_2,\epsilon_2}(x_2;z_2)
\rangle
&=&
 (z_1 - z_2)^{-2 \Delta_1} \delta_{\epsilon_1,\epsilon_2}
 \\
 & \times&
\left[ A_{j_2,\epsilon_2} \delta_{-1-j_1, j_2}
\delta(x_1 - x_2)   \right.
+\left.B_{j_2,\epsilon_2}  \delta_{j_1, j_2}  S_{j_2,\epsilon_2}(x_1,x_2)  \right].
\eea
The consistency conditions  with identification (\ref{identification}) read
\bea
R_{j_1,\epsilon_1}\langle
\Phi_{j_1,\epsilon_1}(x_1;z_1)
\Phi_{j_2,\epsilon_2}(x_2;z_2)
\rangle
&=&
{1\over C_{j_1,\epsilon_1}}
\oint {dy\over 2\pi i} S_{j_1,\epsilon_1}(x_1,y)
\langle
\Phi_{-1-j_1,\epsilon_1}(y\,;z_1)
\Phi_{j_2,\epsilon_2}(x_2;z_2)
\rangle ,
\\
R_{j_2,\epsilon_2}\langle
\Phi_{j_1,\epsilon_1}(x_1;z_1)
\Phi_{j_2,\epsilon_2}(x_2;z_2)
\rangle
&=&
{1\over C_{j_2,\epsilon_2}}
\oint {dy\over 2\pi i} S_{j_2,\epsilon_2}(x_2,y)
\langle
\Phi_{j_1,\epsilon_1}(x_1;z_1)
\Phi_{-1-j_2,\epsilon_2}(y\,;z_2)
\rangle .
\eea
They are satisfied if
$$
A_{j,\epsilon} = R_{j,\epsilon}C_{j,\epsilon} B_{j,\epsilon} \, .
$$
We chose the solution
$$
A_{j,\epsilon} = c, \;\;\;\;B_{j,\epsilon}\;=\;{c\over R_{j,\epsilon}C_{j,\epsilon}}, 
$$
which for a given $R_{j,\epsilon}$ fixes the normalization of fields up to an overall $(j,\epsilon)$-independent constant $c$.
With this normalization the chiral 2-point function takes the form
\begin{equation}
\label{2point}
\begin{array}{rcl}
\langle
\Phi_{j_1,\epsilon_1}(x_1;z_1)
\Phi_{j_2,\epsilon_2}(x_2;z_2,)
\rangle
&=&
 c\,(z_1 - z_2)^{-2 \Delta_1} \delta_{\epsilon_1,\epsilon_2}
 \\[6pt]
 & &\hspace{-100pt}\times\;
\left[ \delta_{-1-j_1, j_2}
\delta(x_1 - x_2)
+\; {\delta_{j_1, j_2}\over R_{j_2,\epsilon_2}C_{j_2,\epsilon_2}}  S_{j_2,\epsilon_2}(x_1,x_2)  \right].
\end{array}
\end{equation}

\subsection{3-point functions}

As in the case of the 2-point function the global Ward identities imply factorization into chiral parts up to  $j,\epsilon$-dependent constants
\begin{eqnarray}
\nonumber
&&\hspace{-60pt}\langle
\Phi_{j_3,\epsilon_3}(x_3,\bar x_2;z_3,\bar z_3)
\Phi_{j_2,\epsilon_2}(x_2,\bar x_2;z_2,\bar z_2)
\Phi_{j_1,\epsilon_1}(x_1,\bar x_1;z_1,\bar z_1)
\rangle
\\
\nonumber
&=&
c \left[ \Delta(j_i); z_i \right]c \left[ \Delta(j_i); \bar z_i \right]
\sum\limits_{\epsilon,\epsilon'=0,\frac12}
S_\epsilon\Big[\!
\begin{array}{ccc}
\\[-18pt]
\scriptstyle j_3 & \scriptstyle j_2  & \scriptstyle j_1
\\[-8pt]
\scriptstyle \epsilon_3  &  \scriptstyle \epsilon_2  & \scriptstyle \epsilon_1
\\[-8pt]
\scriptstyle x_3  &  \scriptstyle x_2  &  \scriptstyle x_1
\end{array}
\!\Big]
S_{\epsilon'}\Big[\!
\begin{array}{ccc}
\\[-18pt]
\scriptstyle j_3 & \scriptstyle j_2  & \scriptstyle j_1
\\[-8pt]
\scriptstyle \epsilon_3  &  \scriptstyle \epsilon_2  & \scriptstyle \epsilon_1
\\[-8pt]
\scriptstyle \bar x_3  &  \scriptstyle \bar x_2  &  \scriptstyle \bar x_1
\end{array}
\!\Big]
\,C^{\epsilon,\epsilon'}[j_i;\epsilon_i]\,,
\end{eqnarray}
where
\bea
 c \left[ \Delta_i; z_i \right]
&=&
(z_2 - z_1)^{\Delta_3-\Delta_2-\Delta_1}
 (z_3 - z_1)^{\Delta_2-\Delta_3-\Delta_1}
 (z_3 - z_2)^{\Delta_1-\Delta_3-\Delta_2}
\eea
and $S_\epsilon$ is 3-linear ${\rm su}(2)$ invariant (\ref{Sinvariant}).
We assume the following diagonal form of the constants $C^{\epsilon,\epsilon'}[j_i;\epsilon_i]$:
\begin{equation}
\label{special form}
C^{\epsilon,\epsilon'}[j_i;\epsilon_i]= \delta_{\epsilon \epsilon'} s_{\epsilon}[j_i;\epsilon_i]C[j_3,j_2,j_1],
\end{equation}
where $s_{\epsilon}[j_i;\epsilon_i]$ is a classical part independent of $\kappa$ and $C[j_3,j_2,j_1]$
is a quantum part independent of $\epsilon$'s. With this assumption the 3-point function takes the form
\begin{eqnarray}
\label{3pointf}
&&\hspace{-60pt}\langle
\Phi_{j_3,\epsilon_3}(x_3,\bar x_2;z_3,\bar z_3)
\Phi_{j_2,\epsilon_2}(x_2,\bar x_2;z_2,\bar z_2)
\Phi_{j_1,\epsilon_1}(x_1,\bar x_1;z_1,\bar z_1)
\rangle
\\
\nonumber
&=&
c \left[ \Delta(j_i); z_i \right]c \left[ \Delta(j_i); \bar z_i \right]
\left(\sum\limits_{\epsilon=0,\frac12}
S_\epsilon\Big[\!
\begin{array}{ccc}
\\[-18pt]
\scriptstyle j_3 & \scriptstyle j_2  & \scriptstyle j_1
\\[-8pt]
\scriptstyle \epsilon_3  &  \scriptstyle \epsilon_2  & \scriptstyle \epsilon_1
\\[-8pt]
\scriptstyle x_3  &  \scriptstyle x_2  &  \scriptstyle x_1
\end{array}
\!\Big]
S_{\epsilon}\Big[\!
\begin{array}{ccc}
\\[-18pt]
\scriptstyle j_3 & \scriptstyle j_2  & \scriptstyle j_1
\\[-8pt]
\scriptstyle \epsilon_3  &  \scriptstyle \epsilon_2  & \scriptstyle \epsilon_1
\\[-8pt]
\scriptstyle \bar x_3  &  \scriptstyle \bar x_2  &  \scriptstyle \bar x_1
\end{array}
\!\Big]s_{\epsilon}[j_i;\epsilon_i]\right)
\,C[j_3,j_2,j_1]\,.
\end{eqnarray}
As it was mentioned in the introduction
the
 $\widehat{\rm su}(2)$ WZNW model at level $\kappa<-2$ should be equivalent to
the $H^+_3= SL(2,\mathbb{C})/ SU(2)$ coset model at level $\kappa'=-\kappa>2$.
This suggests that the quantum part $C[j_3,j_2,j_1]$ should be given by the $H^+_3$ structure constants.
With the normalization chosen in \cite{Dabholkar:2007ey} in the case of $\kappa>-2$ (being in line with the normalization used in the
${\rm su}(2)$ minimal models) one has
\begin{itemize}
\item
for $\kappa < -2$,
$
-(\kappa+2)= b^{-2} \, > 0
$,
\begin{equation}
\label{real}
C^{\s}_b[j_3, j_2, j_1] =
\frac{  M^{\s}_b \,  \sqrt{ \prod_{a=1}^3  \Upsilon_b(-b\, 2 j_a) \Upsilon_b(-b(2j_a +1))  } }
{\Upsilon_b(-b( j_{123} +1)) \Upsilon_b(-b j^3_{12} )  \Upsilon_b(-b j^2_{13} ) \, \Upsilon_b(-b j^1_{23} )  },
\end{equation}
\item
for $\kappa > -2$,
$
-(\kappa+2)= - \hat{b}^{-2} \, <0
$.
\begin{equation}
\label{imaginary}
C^{\h}_{\hat b}[j_3, j_2, j_1] =
\frac{  M^\h_{\hat b} \,
\Upsilon_{\hat b}({\hat b}( j_{123} +2))
\Upsilon_{\hat b}({\hat b}( j^3_{12} +1))
\Upsilon_{\hat b}({\hat b}( j^2_{13} +1)) \,
\Upsilon_{\hat b}({\hat b}( j^1_{23}  +1)) }
{
\sqrt{ \prod_{a=1}^3
\Upsilon_{\hat b}(\hat b (2j_a+1)  ) \, \Upsilon_{\hat b}(\hat b (2j_a +2) )} }.
\end{equation}
\end{itemize}
Using the shift relations
\begin{equation}
\label{shift}
\Upsilon_b(x+b)=\gamma(bx)b^{1-2bx}\Upsilon_b(x),\;\;\Upsilon_b(x+b^{-1})=\gamma(b^{-1}x)b^{-1+2b^{-1}x}\Upsilon_b(x),\;\;
\Upsilon_b(Q-x)=\Upsilon_b(x),
\end{equation}
one gets the reflection properties
\bea
C^{\s}_b(-1-j_3, j_2, j_1)
&=&
\sqrt{\gamma(-2j_3-1)\gamma(-2j_3) }
\gamma(j^1_{23} +1)\gamma(j^2_{13} +1)
C^{\s}_b(j_3, j_2, j_1),
\\
C^{\h}_{\hat b}(-1-j_3, j_2, j_1)
&=&
\sqrt{\gamma(-2j_3-1)\gamma(-2j_3) }
\gamma(j^1_{23} +1)\gamma(j^2_{13} +1)
C^{\h}_{\hat b}(j_3, j_2, j_1).
\eea
If we chose (\ref{special form}) with
$$
s_\epsilon[j_i;\epsilon_i] =
{1\over
s({1\over 2}j^3_{12}+\epsilon+\epsilon_3)
s({1\over 2}j^2_{13}+\epsilon+\epsilon_2)
s({1\over 2}j^1_{23}+\epsilon+\epsilon_1)
s({1\over 2}+{1\over 2}j_{123}+\epsilon)},
$$
then
 (\ref{refrelizo}) implies that
3-point functions (\ref{3pointf})
satisfy the simple reflection rule
\bea
&&\hspace{-20pt}
{1\over C^2_{j_3,\epsilon_3}}
\oint {dy\over 2\pi i} S_{j_3,\epsilon_3}(x_3,y)
\oint {d\bar y\over 2\pi i} S_{j_3,\epsilon_3}(\bar x_3,\bar y)
\\
&&\hspace{40pt}\times\;
\langle
\Phi_{-1-j_3,\epsilon_3}(x_3,\bar x_2;z_3,\bar z_3)
\Phi_{j_2,\epsilon_2}(x_2,\bar x_2;z_2,\bar z_2)
\Phi_{j_1,\epsilon_1}(x_1,\bar x_1;z_1,\bar z_1)
\rangle
\\[6pt]
&=&
\sqrt{\gamma(-2j_3-1)\gamma(-2j_3) }
\langle
\Phi_{j_3,\epsilon_3}(x_3,\bar x_2;z_3,\bar z_3)
\Phi_{j_2,\epsilon_2}(x_2,\bar x_2;z_2,\bar z_2)
\Phi_{j_1,\epsilon_1}(x_1,\bar x_1;z_1,\bar z_1)
\rangle \, .
\eea
For the purposes of the off-diagonal extension we shall introduce the chiral 3-point functions
\bea
&&\hspace{-60pt}\langle
\Phi_{j_3,\epsilon_3}(x_1;z_1)
\Phi_{j_2,\epsilon_2}(x_2;z_2)
\Phi_{j_1,\epsilon_1}(x_3;z_3)\rangle_\epsilon^{\scriptscriptstyle \rm A}
\\
&=&
 c \left[ \Delta(j_i); z_i \right]
\widetilde S_\epsilon\Big[\!
\begin{array}{ccc}
\\[-18pt]
\scriptstyle j_3 & \scriptstyle j_2  & \scriptstyle j_1
\\[-8pt]
\scriptstyle \epsilon_3  &  \scriptstyle \epsilon_2  & \scriptstyle \epsilon_1
\\[-8pt]
\scriptstyle x_3  &  \scriptstyle x_2  &  \scriptstyle x_1
\end{array}
\!\Big]
 {\sf C}^{\scriptscriptstyle \rm A}[ j_3,j_2,j_1],
\\
&&\hspace{-60pt}\langle
\Phi_{j_3,\epsilon_3}(\bar x_1;\bar z_1)
\Phi_{j_2,\epsilon_2}(\bar x_2;\bar z_2)
\Phi_{j_1,\epsilon_1}(\bar x_3;\bar z_3)\rangle_\epsilon^{\scriptscriptstyle \rm A}
\\
&=&
 c \left[ \Delta(j_i); \bar z_i \right]
\widetilde S_\epsilon\Big[\!
\begin{array}{ccc}
\\[-18pt]
\scriptstyle j_3 & \scriptstyle j_2  & \scriptstyle j_1
\\[-8pt]
\scriptstyle \epsilon_3  &  \scriptstyle \epsilon_2  & \scriptstyle \epsilon_1
\\[-8pt]
\scriptstyle \bar  x_3  &  \scriptstyle \bar x_2  &  \scriptstyle \bar  x_1
\end{array}
\!\Big]
 \bar {\sf C}^{\scriptscriptstyle \rm A}[ j_3,j_2,j_1], \hspace{20pt}{\scriptscriptstyle \rm A}=\s,\,\h,
\eea
where
\begin{equation}
\label{Stilde}
\widetilde S_\epsilon\Big[\!
\begin{array}{ccc}
\\[-18pt]
\scriptstyle j_3 & \scriptstyle j_2  & \scriptstyle j_1
\\[-8pt]
\scriptstyle \epsilon_3  &  \scriptstyle \epsilon_2  & \scriptstyle \epsilon_1
\\[-8pt]
\scriptstyle x_3  &  \scriptstyle x_2  &  \scriptstyle x_1
\end{array}
\!\Big]
=
S_\epsilon\Big[\!
\begin{array}{ccc}
\\[-18pt]
\scriptstyle j_3 & \scriptstyle j_2  & \scriptstyle j_1
\\[-8pt]
\scriptstyle \epsilon_3  &  \scriptstyle \epsilon_2  & \scriptstyle \epsilon_1
\\[-8pt]
\scriptstyle x_3  &  \scriptstyle x_2  &  \scriptstyle x_1
\end{array}
\!\Big] \sqrt{s_\epsilon[j_i;\epsilon_i] },
\end{equation}
and
branches of the square root are chosen in such a way that formula (\ref{3pointf}) for the full 3-point function  is reproduced.
The chiral splitting of the $j$-dependent parts of the structure constants is motivated by the similar splitting in the Liouville theory \cite{SL-LL}:
\begin{eqnarray}\label{C_hj}
\nonumber
{\sf C}^\s_b(j_3,j_2,j_1)
& = &
\frac{ \sqrt{M^\s_b}
\Gamma_{b}\left(-b (j_{123}+1)\right)\,
\Gamma_{b}\left( -b j^3_{12} \right) \, \Gamma_{b}\left( -b j^2_{13}  \right) \, \Gamma_{b}\left( -b j^1_{23}\right)  }
{
\sqrt{ \prod_{i=1}^3
\Gamma_{b}\left( -2b j_i\right) \Gamma_{b}\left( - b (2j_i +1) \right)  }
}
\, ,
\\[-4pt]
\\[-4pt]
\nonumber
\bar {\sf C}^\s_b(j_3,j_2,j_1)
 &=&
\frac{ \sqrt{M^\s_b}
\Gamma_{b}\left(Q+ b (j_{123}+1)\right)
\Gamma_{b}\left(Q+ b j^3_{12} \right)
\Gamma_{b}\left(Q+ b j^2_{13} \right)
\Gamma_{b}\left(Q+ b j^1_{23} \right)}
{
\sqrt{ \prod_{i=1}^3
 \Gamma_{b}\left(Q +   2b j_i\right) \Gamma_{b}\left(Q+ b(2 j_i+1)\right) }
}\, ,
\end{eqnarray}
\begin{eqnarray*}
{\sf C}^\h_b(j_3,j_2,j_1)
& = &
\frac
{ \sqrt{M^\h_{\hat b}}
\sqrt{ \prod_{i=1}^3
\Gamma_{b}\left( \frac{1}{b}- 2  b j_i\right) \Gamma_{b}\left(\frac{1}{b}-b(2   j_i+1)\right) }
}
{
\Gamma_{b}\left(\frac{1}{b} -b (j_{123}+1) \right)
\Gamma_{b}\left( \frac{1}{b} - b  j^3_{12} \right)
\Gamma_{b}\left( \frac{1}{b} - b  j^2_{13} \right)
\Gamma_{b}\left( \frac{1}{b} - b  j^1_{23} \right) }
\ ,
\\[4pt]
\nonumber
\bar {\sf C}^\h_b(j_3,j_2,j_1)
&=&
\frac
{ \sqrt{M^\h_{\hat b}}
\sqrt{ \prod_{i=1}^3
\Gamma_{b}\left( b(2  j_i+1)\right) \Gamma_{b}\left( b (2j_i+2)\right)  }
}
{
\Gamma_{b}\left(  b +b(j_{123}+1)\right)\,
\Gamma_{b}\left( b + bj^3_{12}  \right) \Gamma_{b}\left( b +j^2_{13} \right)  \Gamma_{b}\left( b +b j^1_{23} \right)  }
\ .
\end{eqnarray*}
As in the case of the SL-LL correspondence there is no canonical splitting into chiral parts.
Our choice is motivated by the relations
\begin{eqnarray} \label{Cbar_hs}
\nonumber
 \bar {\sf C}^\s_b(j_3,j_2,j_1) &=& r(j_i) \,  {\sf C}^\s_b(-j_3-1,-j_2-1,-j_1-1) \ ,
 \\[-4pt]
 \\[-4pt] \nonumber
 \bar {\sf C}^\h_b(j_3,j_2,j_1) &= & r(j_i)  \, {\sf C}^\h_b(-j_3-1,-j_2-1,-j_1-1) \ ,
\end{eqnarray}
where
\begin{equation} \label{Rji}
 r(j_i)=\frac{
\sqrt{ 2 \pi \,  \prod_{i=1}^3
 \Gamma\left(2  j_i+1\right) \Gamma\left(  2 j_i +2\right)  } }{
\Gamma\left( 2+  j_{123}\right)
\Gamma\left( 1+  j_{12} -j_3 \right)  \Gamma\left( 1+  j_{13} -j_2 \right)   \Gamma\left( 1+  j_{23} -j_1 \right)  } \ ,
\end{equation}
which reduce the calculations to one sector.
Relations (\ref{Cbar_hs}) can be easily derived using the shift formulae
\begin{equation}\label{Gamma_shift}
\Gamma_b(x+ b) = \sqrt{2 \pi}  \,  b^{b x - \frac12} \,\frac{  \Gamma_b(x)}{\Gamma(b x)}  ,
\qquad
\Gamma_b(x+ b^{-1}) =  \sqrt{2 \pi} \,  b^{- b^{-1}x + \frac12} \, \frac{\Gamma_b(x)}{\Gamma( b^{-1}x)} \ .
\end{equation}

\subsection{off-diagonal extension}

As we shall see the coset construction requires an off-diagonal extension of the  $\hat{\rm su}(2)_\kappa$   WZW model
to the following class of representations
$$
\hat{\cal D}^\kappa_{j,\epsilon},\;\;\;\;j=-{\textstyle{1\over 2}}+n+ is,\;\;\;
s\in \mathbb{R},\;\;\epsilon =0,{\textstyle \frac12},\;\;\;n\in {\textstyle \frac12}\mathbb{Z}.
$$
The space of states is the direct integral/sum of the tensor products of the left and the right relaxed modules from this class
$$
\hat{\cal D}^{\kappa}_{\jL,\epsilon}\otimes \hat{\cal D}^{\kappa}_{\jR,\epsilon},\;\;\;\jL-\jR=\nL-\nR \in \mathbb{Z}.
$$
Let us note that the spectrum is off-diagonal only with respect to the discrete variable $n$.

For $(\nL,\nR)\neq (0,0)$ there is no hermitian invariant form on
$\hat{\cal D}^{\kappa}_{\jL,\epsilon}\otimes \hat{\cal D}^{\kappa}_{\jR,\epsilon}$.
One can however construct such form on the direct sum
$$
\left(\hat{\cal D}^{\kappa}_{\jL,\epsilon}\otimes \hat{\cal D}^{\kappa}_{\jR,\epsilon}\right)
\oplus
\left(\hat{\cal D}^{\kappa}_{-1-\bar \jL,\epsilon}\otimes \hat{\cal D}^{\kappa}_{-1-\bar \jR,\epsilon}\right)
$$
by analogy with the construction of the hermitian ${\rm su}(2)$ representations given in subsection 2.9.

In the case of the diagonal spectrum 2- and 3-point functions were introduced as products
of their left and right chiral components.
The extension of the chiral 3-point function in the $z$- and the $j$-dependent parts is straightforward.
If we assume the condition
\begin{equation}
\label{ncondition}
n_1+n_2 +n_3\;\dot{=}\;0
\end{equation}
the arguments of all the Barnes gamma functions involved are shifted by integer multiplicities of $b$ or $ \hat b$
and can be explicitly calculated by shift relations (\ref{Gamma_shift}).
In the following we assume condition (\ref{ncondition}).
As we shall see it is always satisfied in relations (\ref{rel}) if we restrict ourselves to the diagonal spectra on their left hand sides.

The extension of the $x$-dependent part is more subtle. One has to preserve its role of the generating function
for the structure constants in the spin basis. This requires definite periodicity conditions on the torus $S^1\times S^1\times S^1$.
Using (\ref{Sidentity}) and the identities for the sine function
\begin{eqnarray*}
s\left( x +{\textstyle \frac12}n_{123}-n_i+(\epsilon\dot{+}{\textstyle \frac12}n_{123}) +(\epsilon_i\dot{+}n_i)\right)
&=&
(-1)^{{ \frac12}n_{123} +(1 +4\epsilon) \delta +n_i -(1 +4\epsilon_i)\delta_i}
s\left( x +\epsilon+\epsilon_i\right),
\\
s\left(x +{\textstyle \frac12}n_{123} +(\epsilon\dot{+}{\textstyle \frac12}n_{123})\right)
&=&
(-1)^{{ \frac12}n_{123} +(1 +4\epsilon) \delta }
s\left( x +\epsilon\right),
\end{eqnarray*}
where
$\delta=\frac12 n_{123}\dot{+}0,\delta_i=n_i\dot{+}0, $
 one gets the identity for
the invariant $S_\epsilon\sqrt{s_\epsilon}$:
\begin{eqnarray}
\label{Fidentity}
&&
\hspace{-20pt}
(x_1 - x_2)^{n_{12}^3} (x_3 - x_1)^{n_{13}^2} (x_2 - x_3)^{n_{23}^1}  \,
  S_\epsilon\Big[\!
\begin{array}{ccc}
\\[-18pt]
\scriptstyle j_3 & \scriptstyle j_2  & \scriptstyle j_1
\\[-8pt]
\scriptstyle \epsilon_3  &  \scriptstyle \epsilon_2  & \scriptstyle \epsilon_1
\\[-8pt]
\scriptstyle   x_3  &  \scriptstyle  x_2  &  \scriptstyle   x_1
\end{array}
\!\Big] \sqrt{s_\epsilon[j_i;\epsilon_i] }
\\
\nonumber
&=&
\sqrt{(-1)^{n_{123} -\sum_i (4\epsilon_i +1)\delta_i}}
 \,
  S_{\epsilon\+\frac12 n_{123}}\Big[\!
\begin{array}{ccc}
\\[-18pt]
\scriptstyle j_3 +n_3& \scriptstyle j_2+n_2  & \scriptstyle j_1+n_1
\\[-8pt]
\scriptstyle \epsilon_3\+n_3  &  \scriptstyle \epsilon_2 \+n_2 & \scriptstyle \epsilon_1\+n_1
\\[-8pt]
\scriptstyle  x_3  &  \scriptstyle  x_2  &  \scriptstyle   x_1
\end{array}
\!\Big] \sqrt{s_{\epsilon\+\frac12 n_{123}}[j_i+n_i;\epsilon_i\dot{+}n_i] }.
\end{eqnarray}
We  use this formula as a motivation for the definition of the off-diagonal extension
of  $x$-dependent part (\ref{Stilde})
\begin{equation}
\label{F}
\widetilde S_{\epsilon\+\frac12 n_{123}}\Big[\!
\begin{array}{ccc}
\\[-18pt]
\scriptstyle j_3 +n_3& \scriptstyle j_2+n_2  & \scriptstyle j_1+n_1
\\[-8pt]
\scriptstyle \epsilon_3\+n_3  &  \scriptstyle \epsilon_2 \+n_2 & \scriptstyle \epsilon_1\+n_1
\\[-8pt]
\scriptstyle  x_3  &  \scriptstyle  x_2  &  \scriptstyle   x_1
\end{array}
\!\Big]
=
(-1)^{\eta(n_3,n_2,n_1)}
(x_1 - x_2)^{n_{12}^3} (x_3 - x_1)^{n_{13}^2} (x_2 - x_3)^{n_{23}^1}  \,
\widetilde S_\epsilon\Big[\!
\begin{array}{ccc}
\\[-18pt]
\scriptstyle j_3 & \scriptstyle j_2  & \scriptstyle j_1
\\[-8pt]
\scriptstyle \epsilon_3  &  \scriptstyle \epsilon_2  & \scriptstyle \epsilon_1
\\[-8pt]
\scriptstyle   x_3  &  \scriptstyle  x_2  &  \scriptstyle   x_1
\end{array}
\!\Big] ,
\end{equation}
where condition (\ref{ncondition}) and the prescription given below (\ref{Stilde}) are assumed.
We also assume that all square root ambiguities are hidden in the sign factor
$(-1)^{\eta(n_3,n_2,n_1)}$. The general form of the function $\eta(n_3,n_2,n_1)$ is not known.
In principle it could be derived from expected equivalence (\ref{3-pkt_gen2}) and the general form
of highest weight states (\ref{ep}). The results of explicit calculations in a number of cases of
low laying states are presented in Appendix 3.

\subsection{Liouville and imaginary Liouville structure constants}

In the symmetric normalization $\Phi_\alpha =\Phi_{Q-\alpha}$
the DOZZ structure constants \cite{Dorn:1994xn,Zamolodchikov:1995aa} for primary fields $\Phi_{\alpha}$ of conformal dimension $\Delta\sL_\alpha= \alpha(Q-\alpha)$
take the form
\begin{eqnarray}
\label{DOZZ:threepoint}
C^{\rm\scriptscriptstyle DOZZ }_b[\alpha_3,\alpha_2,\alpha_1]
& = &
M_b\sL
\frac{
\prod\limits_{i}\sqrt{
\Upsilon_{b}\left(Q- 2 \alpha_i\right)
\Upsilon_{b}\left(-Q+ 2 \alpha_i\right)}
}
{
\Upsilon_{b}\left( \alpha_{123}- Q\right)\,
\Upsilon_{b}\left( \alpha^1_{23}\right)
\Upsilon_{b}\left( \alpha^2_{13}\right)
\Upsilon_{b}\left( \alpha^3_{12}\right) }
\ ,
\end{eqnarray}
where $\alpha_{123}=\alpha_1+\alpha_2+\alpha_3,\ \alpha^3_{12}=\alpha_1+\alpha_2-\alpha_3,$ etc.
and the real parameters $b, Q=b+b^{-1}$ are related to the central charge $c$ of the Liouville theory by
$$
c^{\scriptscriptstyle {\rm L}}  = 1 + 6 Q^2 \, \;\;\;\;\;Q=b + b^{-1}.
$$
For the purpose of
the present paper it is convenient to parameterize  the Liouville primary fields in terms of
$$
j=-{\alpha\over Q},\;\;\;\;\Delta\sL_j = -Q^2 \,j(1+j),
$$
rather then $\alpha$. Using the freedom in decomposing $C^{\rm\scriptscriptstyle DOZZ }_b$ into chiral structure constants \cite{SL-LL}
we define
\begin{eqnarray*}
C^{\rm\scriptscriptstyle L }_b[j_3,j_2,j_1]
&=&
{\sf C}^{\rm \scriptscriptstyle L}_b(j_3,j_2,j_1)
\bar {\sf C}^{\rm \scriptscriptstyle L}_b(j_3,j_2,j_1),
\\[8pt]
\nonumber
{\sf C}^{\rm \scriptscriptstyle L}_b(j_3,j_2,j_1)
& = &
\frac{\sqrt{M_b\sL}\,
\Gamma_{b}\left( -Q (j_{123}+1)\right)\,
\Gamma_{b}\left( -Q j^3_{12}\right) \,
\Gamma_{b}\left( -Q j^2_{13}\right) \,
\Gamma_{b}\left( -Q j^1_{23}\right) }
{
\sqrt{ \prod_{i=1}^3
\Gamma_{b}\left( -2Q j_i\right) \Gamma_{b}\left( -Q(2j_i+1)\right) }}
\ ,
\\
\nonumber
\bar {\sf C}^{\rm \scriptscriptstyle L}_b(j_3,j_2,j_1)
&=&
\frac{ \sqrt{M_b\sL}\,
\Gamma_{b}\left( Q+ Q (j_{123}+1)\right)\,
\Gamma_{b}\left( Q+ Q j^3_{12}\right) \,
\Gamma_{b}\left( Q+ Q j^2_{13}\right)\,
\Gamma_{b}\left( Q+ Q j^1_{23}\right) }
{
\sqrt{ \prod_{i=1}^3
\Gamma_{b}\left(Q+ 2 Qj_i\right) \Gamma_{b}\left(2Q+ 2 Q j_i\right)}}.
\end{eqnarray*}
Let us note  that $ {\sf C}^{\rm \scriptscriptstyle L}_b(-j_3-1,-j_2-1,-j_1-1) =  \bar {\sf C}^{\rm \scriptscriptstyle L}_b(j_3,j_2,j_1)$.

In  the imaginary Liouville theory \cite{Zamolodchikov:2005fy,Zamolodchikov:2005sj} with purely imaginary parameter $b = - i \hat b$ and the central charge
$$
c^{\scriptscriptstyle {\rm IL}}  = 1 - 6 \hat Q^2 \, \;\;\;\;\;\;\hat Q= \hat b^{-1}  - \hat b
$$
it is convenient to use the  parametrization
$$
j=-{\alpha\over \hat Q},\;\;\;\;\Delta\sIL_j = \hat Q^2 \,j(1+j).
$$
The corresponding structure constants are given by
\bea
{C}\sIL_{\hat b}[j_3, j_2, j_1]
&=&
{\sf C}\sIL_{\hat b}[j_3, j_2, j_1]\bar {\sf C}\sIL_{\hat b}[j_3, j_2, j_1],
\\[8pt]
{\sf C}\sIL_{\hat b}[j_3, j_2, j_1] &=&
\frac{ \sqrt{M\sIL_{\hat b}} \,
\sqrt{ \prod_{a=1}^3
\Gamma_{\hat b}(\hat b - 2\hat Q j_a  ) \,
\Gamma_{\hat b}(\hat b - \hat Q(2 j_a +1) )}  }{
\Gamma_{\hat b}(\hat b-\hat Q( j_{123} +1))
\Gamma_{\hat b}(\hat b-\hat Q j^3_{12})
\Gamma_{\hat b}(\hat b-\hat Q j^2_{13}) \,
\Gamma_{\hat b}(\hat b-\hat Q j^1_{23})  },
\\
\bar {\sf C}\sIL_{\hat b}[j_3, j_2, j_1] &=&
\frac{ \sqrt{M\sIL_{\hat b}} \,
\sqrt{ \prod_{a=1}^3
\Gamma_{\hat b}(\hat b^{-1} + 2\hat Q j_a  ) \,
\Gamma_{\hat b}(\hat b^{-1} + \hat Q(2 j_a +1) )}  }
{
\Gamma_{\hat b}(\hat b^{-1} +\hat Q( j_{123} +1))
\Gamma_{\hat b}(\hat b^{-1} +\hat Q j^3_{12})
\Gamma_{\hat b}(\hat b^{-1} +\hat Q j^2_{13}) \,
\Gamma_{\hat b}(\hat b^{-1} +\hat Q j^1_{23})  }.
\eea

\section{Nonrational $\hat{\rm su}(2)_\kappa$   cosets}

\subsection{GKO construction}

The GKO construction \cite{GKO} is based on the observation that any
representation of the algebra $\hat{\rm su}(2)_\kappa \oplus\hat{\rm su}(2)_1 $
is also a representation of the two mutually commuting  algebras: $\hat{\rm su}(2)_{\kappa +1}, {\rm Vir}_c$.
The corresponding generators are given by
\begin{eqnarray}
\label{calJ}
{\cal J}^a_n
&=&
J^a_n +K^a_n,\;\;\;\;\;\;a=\pm,3,
\\
\label{L}
L^{\scriptscriptstyle {\rm V}}_n&=& \frac{1}{\kappa+3} L^{\kappa}_m + \frac{\kappa}{\kappa+3} L^1_m - \frac{1}{\kappa+3} A_m,
\end{eqnarray}
where
\begin{eqnarray*}
A_m &=&  \sum_{r=-\infty}^{\infty}  \left( J^+_{m-r} K^-_r  +  J^-_{m-r} K^+_r  + 2 J^3_{m-r} K^3_r \right),
\end{eqnarray*}
and $L^{\kappa}_m$, $L^1_m$ are the Virasoro generators related by the Sugawara construction to the algebras $J^a_n$ and $K^a_n$:
\begin{eqnarray*}
 L^{\kappa}_m &=& \frac{1}{\kappa+2} \sum_{r=-\infty}^{\infty} \left( : J^+_{m-r} J^-_r : + : J^-_{m-r} J^+_r : + :2 J^3_{m-r} J^3_r : \right),
 \\
 L^1_m &=& {1\over 6}\sum_{n=-\infty}^{\infty}  \left( : K^+_{m-r} K^-_r : + : K^-_{m-r} K^+_r : + :2 K^3_{m-r} K^3_r : \right).
\end{eqnarray*}
The Virasoro generators related to the algebra ${\cal J}^a_n$ can be expressed as
\bea
L_m^{\kappa+1}
 &=&
 \frac{\kappa+2}{\kappa+3} L^{\kappa}_m + \frac{3}{\kappa+3} L^1_m + \frac{1}{\kappa+3} A_m,
\eea
and they satisfy the following  relations with the other Virasoro generators
\begin{eqnarray}\label{modes_combL+}
L_m^{\kappa+1} + L_m^{\scriptscriptstyle {\rm V}}
 &=&
  L^{\kappa}_m + L^1_m,
\\ \nonumber
L_m^{\kappa+1} - (\kappa+2)  L_m^{\scriptscriptstyle {\rm V}}
&=&   \frac{3- \kappa^2 - 2\kappa}{\kappa+3}  L^1_m + A_m.
\end{eqnarray}
The first equation is equivalent to the  condition for the central charges of the corresponding algebras
\begin{eqnarray}\label{central charges}
c^{\kappa+1}  + c^{\scriptscriptstyle {\rm V}} = c^\kappa + 1,
\end{eqnarray}
where $c^\kappa = {3\kappa \over \kappa + 2} $. In this paper we will consider the case where $c^{\scriptscriptstyle {\rm V}} $ is the central charge of the Liouville theory
$$
c^{\scriptscriptstyle {\rm L}}  = 1 + 6 \left(b + b^{-1} \right)^2 \, ,
$$
 or the imaginary Liouville theory with purely imaginary parameter $b = - i \hat b$
$$
c^{\scriptscriptstyle {\rm IL}}  = 1 - 6 \left(\hat b - \hat b^{-1} \right)^2 \, .
$$
Matching condition (\ref{central charges})
implies the relation between the level $\kappa$ and the  Liouville parameter $b$,
$$
\kappa_1 = - \frac{ 3 b + 2 b^{-1}}{ b+ b^{-1} }, \qquad
\kappa_2 = - \frac{ 3 b^{-1} + 2 b}{ b+ b^{-1} }.
$$
We choose $\kappa = \kappa_2$ and assume $b<1$.
 Then the levels of the $\hat{\rm su}(2)_\kappa$ and $\hat{\rm su}(2)_{\kappa+1} $  are  on the opposite sides of
 the $\kappa=-2$ barrier:
$$
\kappa < -2 <\kappa +1.
$$
We shall parameterize the theories by
 $b\sa$ and $\hat b\sB$, respectively
\begin{eqnarray}\label{parameters_b}
\nonumber
 (b\sa)^2 = - \frac{1}{\kappa +2} = 1+ b^{2},  &\qquad &
 ( \hat b\sB)^2 =  \frac{1}{\kappa +3} = 1+ b^{-2} ,
\\
  b\sa = \sqrt{bQ} , & \qquad &
\hat b\sB =    \sqrt{b^{-1}Q}    ,\hspace{40pt}Q =  b+ b^{-1}.
\end{eqnarray}
In the case of the imaginary Liouville theory, for $\hat b <1$ ,
$$
\kappa =  - \frac{ 3 \hat b^{-1} - 2 \hat  b}{ \hat  b^{-1} - \hat  b }  < -3 \,  .
$$
The corresponding ${\rm su}(2)$ theories are parameterized by
 \begin{eqnarray}
 \label{parameters_bIm}
 (b\sa)^2 = - \frac{1}{\kappa +2} = 1- \bh^2, &\qquad  &
 (b\sB)^2 = - \frac{1}{\kappa +3} = \bh^{-2} - 1,
\\ \nonumber
 b\sa = \sqrt{\bh \hat Q}, & \qquad &
b\sB =  \sqrt{\bh^{-1} \hat Q} ,\hspace{40pt}\hat Q = \bh^{-1} - \bh.
\end{eqnarray}

\subsection{decomposing representations}

In this paper we restrict ourselves to the real spectrum of the $\hat{\rm su}(2)_\kappa$ model
on the l.h.s of the expected relations
\bea
\widehat{\rm su}(2)_{\kappa} \ \otimes\ \widehat{\rm su}(2)_{1}
&\sim &
\hbox{Liouville}\ \otimes_P\ \widehat{\rm su}(2)_{\kappa+1},
\\
\widehat{\rm su}(2)_{\kappa} \ \otimes\ \widehat{\rm su}(2)_{1}
&\sim &
\hbox{imaginary Liouville}\ \otimes_P\ \widehat{\rm su}(2)_{\kappa+1}.
\eea
The aim is therefore to decompose  the representations
$$
\hat {\cal D}^\kappa_{j,\epsilon} \otimes \hat{\cal S}^1_{-1},\;\;\;\;\;\;\;
\hat {\cal D}^\kappa_{j,\epsilon} \otimes \hat{\cal S}^1_{-{3\over 2}},\;\;\;\;\;j=-{1\over 2}+is,\;s\in\mathbb{R},
$$
of the algebra $\widehat{\rm su}(2)_{\kappa}\oplus \widehat{\rm su}(2)_{1}$
into irreducible representations of the algebra $\hat{\rm su}(2)_{\kappa +1}\oplus {\rm Vir}_c$.

The general form of these decompositions can be derived by analyzing the
characters of the representations involved.
For the representations $\hat{\cal D}^\kappa_{j,\epsilon}$, $\hat{\cal S}^1_{-1}, \hat{\cal S}^1_{-{3\over 2}}$
and for the Virasoro Verma module  with the weight $\Delta$ and the central charge $c$ they are given by
\begin{eqnarray*}
\chi^{\kappa}_{j, \epsilon} (q,y)
 &=&
  \left(\eta(q)\right)^{-3}\, q^{{1\over 4}+j(1+j) \over \kappa+2} \sum_{m \in \mathbb{Z}}  y^{m+\epsilon},
\\
\chi^1_{-1} (q,y) &=& \frac{1}{\eta(q)} \sum_{n \in \mathbb{Z} }q^{n^2}  y^n,
\\
\chi^1_{-\frac32} (q,y) &=& \frac{1}{\eta(q)} \sum_{n \in \mathbb{Z} }q^{(n-\frac12)^2}  y^{n-\frac12},
\\
\chi^{\rm Vir}_{\Delta, c }(q)  &=&  \frac{1}{\eta(q)} \,   q^{ \Delta- \frac{c-1}{24}},
\end{eqnarray*}
respectively.
One has the following decompositions
\begin{equation}
\label{decomposition}
\begin{array}{rcl}
\chi^\kappa_{j,\epsilon} (q,y) \chi^1_{-1} (q,y)
&=& \displaystyle
\eta(q)^{-4}\, q^{{{1\over 4}+j(j+1) \over \kappa+2}} \sum_{n \in \mathbb{Z}} q^{n^2} \, \sum_{p \in \mathbb{Z}}  y^{p+ \epsilon}
\\[8pt]
&=&\displaystyle
\sum_{n \in \mathbb{Z}} \chi^{\rm Vir}_{\Delta_n,c} (q)  \chi^{\kappa+1}_{j_n,\epsilon} (q,y),
\\[8pt]
\chi^\kappa_{j,\epsilon} (q,y) \chi^1_{-\frac32} (q,y)
 &=& \displaystyle
\eta(q)^{-4}\, q^{{{1\over 4}+j(j+1) \over k+2}} \sum_{n \in \mathbb{Z}- \frac12}   q^{n^2} \, \sum_{p \in \mathbb{Z}}  y^{p+ \epsilon-\frac12}
\\[8pt]
&=& \displaystyle
\sum_{n \in \mathbb{Z}- \frac12}
\chi^{\rm Vir}_{\Delta_n,c} (q)  \chi^{\kappa+1}_{j_n,\bar \epsilon} (q,y),
\end{array}
\end{equation}
where the parameters $\Delta_n, j_n$
are subject to the condition
\begin{equation}
\label{character_cond}
 \Delta_n + {j_n(1+j_n) \over \kappa +3}
 =
 {j(1+j) \over \kappa+2}+n^2,\;\;\;\;\; n\in {\textstyle {1\over 2}}\mathbb{Z}.
\end{equation}
It follows from (\ref{modes_combL+}) that $n^2$ and $n^2-{1\over 4}$
are the levels with respect to the operator $L^\kappa_0 +L_0^1$ in
$\hat {\cal D}^\kappa_{j,\epsilon} \otimes \hat{\cal S}^1_{-1}$ and $\hat {\cal D}^\kappa_{j,\epsilon} \otimes \hat{\cal S}^1_{-{3\over 2}}$, respectively.
We conjecture that the relevant solution to (\ref{character_cond}) is given by
\footnote{In the present paper we use this conjecture for $n=0,\pm{1\over 2},\pm 1$.
It is checked in these cases by
 explicit calculations.
The proof of the general case seems to require more advanced techniques. We hope to come back to this point in a separate publication.}
\begin{equation}
\label{conjecture}
j_n = j+ n \, ,
\end{equation}
which implies
$$
\Delta_{n} =
\Delta^{\scriptscriptstyle {\rm L}}_{j + \frac{n}{bQ}} =
-Q^2 \left( j + \frac{n}{bQ}\right)  \left( 1 + j + \frac{n}{bQ}\right) \, ,
$$
for the Liouville weights, and
$$
\Delta_{n} =
\Delta^{\scriptscriptstyle {\rm IL}}_{j + \frac{n}{\hat b \hat Q}} =
\hat Q^2 \left( j + \frac{n}{\hat b \hat Q}\right)  \left( 1 + j + \frac{n}{\hat b \hat Q}\right) \, ,
$$
for the imaginary Liouville weights.

In the case of Liouville theory decomposition of characters (\ref{decomposition}) and conjecture (\ref{conjecture}) lead to  the following decomposition of representations
\bea
\hat {\cal D}^\kappa_{-1-j,\epsilon} \otimes \hat{\cal S}^1_{-1}
&=&
\bigoplus\limits_{n\in \mathbb{Z}} \hat {\cal D}^{\kappa+1}_{-1-j-n,\epsilon} \otimes {\cal V}_{\Delta^{\scriptscriptstyle {\rm L}}_{j + \frac{n}{bQ}} ,c}
,
\;\;\;\;
\hat {\cal D}^{\kappa}_{-1-j,\epsilon} \otimes \hat{\cal S}^1_{-{3\over 2}}
\;=\;
\bigoplus\limits_{n\in \mathbb{Z}+{1\over 2}} \hat {\cal D}^{\kappa+1}_{-1-j-n,\bar \epsilon} \otimes {\cal V}_{\Delta^{\scriptscriptstyle {\rm L}}_{j + \frac{n}{\hat b \hat Q}},c},
\eea
where we used the relation $\Delta^{\scriptscriptstyle {\rm L}}_{-1-j - \frac{n}{\hat b \hat Q}}=\Delta^{\scriptscriptstyle {\rm L}}_{j + \frac{n}{\hat b \hat Q}}$ and
${\cal V}_{\Delta_n,c}$ denotes the Virasoro Verma module with the weight $\Delta_n$ and the central charge $c$.
The subspaces $\hat {\cal D}^{\kappa+1}_{-1-j-n,\epsilon}\otimes {\cal V}_{\Delta^{\scriptscriptstyle {\rm L}}_{j + \frac{n}{\hat b \hat Q}},c}$
are generated by the algebras ${\cal J}^a_k$, $L^{\scriptscriptstyle {\rm V}}_k$ from the subspaces ${\cal D}_{-1-j-n, \epsilon}\otimes \big|\,\Delta^{\scriptscriptstyle {\rm L}}_{j + \frac{n}{\hat b \hat Q}}\big\rangle$
which can be described in terms of  families $\{\ket{m+\epsilon}^\star_{-1-j-n}\}$ of ${\cal J}^3_0$ eigenstates such that
\begin{equation}
\label{eq}
\begin{array}{rcl}
{\cal J}^a_k\ket{m+\epsilon}^\star_{-1-j-n}
&=&
L^{\scriptscriptstyle {\rm L}}_k\ket{m+\epsilon}^\star_{-1-j-n} = 0,\;\;k>0,
\\[6pt]
L^{\scriptscriptstyle {\rm L}}_0\ket{m+\epsilon}^\star_{-1-j-n}
&=&
\Delta^{\scriptscriptstyle {\rm L}}_{-n}\ket{m+\epsilon}^\star_{-1-j-n},
\\
{\cal J}_0^+\ket{m+\epsilon}^\star_{-1-j-n}&=& (m+\epsilon  +j+1-n)\ket{m+\epsilon+1}^\star_{-1-j-n},
\\
{\cal J}_0^-\ket{m+\epsilon}^\star_{-1-j-n}&=&( -m-\epsilon +j+1-n)\ket{m+\epsilon-1}^\star_{-1-j-n}.
\end{array}
\end{equation}
A family with these properties can be obtained by looking for the states
$$
\ket{x}^\star_{j+n,\epsilon}=\sum\limits_{m\in \mathbb{Z}} x^{j+n-m- \epsilon}\ket{m+\epsilon}^\star_{-1-j-n}\;\;
,
$$
 satisfying
\begin{equation}
\label{ep}
\begin{array}{rcllrllll}
{\cal J}^+_0(x)\ket{x}^\star_{j+n,\epsilon} &=& {\cal J}^a_k(x)\ket{x}^\star_{j+n,\epsilon} \;=\;
L^{\scriptscriptstyle {\rm L}}_k\ket{x}^\star_{j+n,\epsilon}\;=\;0,\;\;\;\;k>0,
\\[8pt]
{\cal J}^3_0(x)\ket{x}^\star_{j+n,\epsilon}
&=&
({j+n})
\ket{x}^\star_{j+n,\epsilon},
\;\;\;\;\;
L^{\scriptscriptstyle {\rm L}}_0\ket{x}^\star_{j+n,\epsilon}\;=\;\Delta^{\scriptscriptstyle {\rm L}}_{n}\ket{x}^\star_{j+n,\epsilon}.
\end{array}
\end{equation}
For $n=0$ the state in question is simply given by
$$
\ket{x}^\star_{j,\epsilon}= \ket{x}_{j,\epsilon}\otimes \ket{x}_{0}.
$$
For $n=\pm{1\over 2},\pm 1$ these states can be
obtained directly from the definition. For higher $n$ the calculations become prohibitively lengthy
and we are not aware of any general construction of such states.
Let us first consider the state
\begin{eqnarray*}
\ket{x}^\star_{j+{1\over 2}, \bar\epsilon} &=&\ket{x}_{ j ,  \epsilon} \otimes \ket{x}_{1\over 2}
 \\
 &=&
\sum x^{j+{1\over 2}-m-  \bar  \epsilon} \left(
\ket{m-  \epsilon}_{-1-j}\otimes\ket{\textstyle {1\over 2}}_{-{3\over 2}}+
\ket{m+1- \epsilon}_{-1-j}\otimes\ket{\textstyle -{1\over 2}}_{-{3\over 2}}
\right).
\end{eqnarray*}
This is a highest weight state with respect to the algebra ${\cal J}^a_k(x)=J^a_k(x) +K^a_k(x)$
and the Virasoro algebra $L^{\scriptscriptstyle {\rm L}}_k$ such that
\[
\begin{array}{rcllrllll}
{\cal J}^3_0(x)\ket{x}^\star_{j+{1\over 2}, \epsilon}
&=&
(j+{\textstyle {1\over 2}})
\ket{x}^\star_{j+{1\over 2}, \epsilon}
,
& \quad
L^{\scriptscriptstyle {\rm L}}_0\ket{x}^\star_{j+{1\over 2}, \epsilon}&=&\Delta^{\scriptscriptstyle {\rm L}}_{1\over 2}\ket{x}^\star_{j+{1\over 2}, \epsilon}.
\end{array}
\]
One checks that the family
$$
\ket{m+ \bar\epsilon}_{-1-j-{1\over 2}}^\star =
\ket{m- \epsilon}_{-1-j}\otimes\ket{\textstyle {1\over 2}}_{-{3\over 2}}+
\ket{m+1-  \epsilon}_{-1-j}\otimes\ket{\textstyle -{1\over 2}}_{-{3\over 2}}
$$
satisfies conditions (\ref{eq}) in agreement with our previous calculations (\ref{tenpro}).

One can also easily verify that the state
\begin{eqnarray*}
\ket{x}^\star_{j-{1\over 2} , \epsilon}
&= &
\left[J^-_0(x) -2j K^-_0(x) \right]\ket{x}^\star_{j+{1\over 2}, \epsilon}
\\
&=&
J^-_0\ket{x}_{j, \bar \epsilon} \otimes \ket{x}_{1\over 2} -2j\ket{x}_{j, \bar \epsilon} \otimes K^-_0\ket{x}_{1\over 2}
\end{eqnarray*}
is a highest weight state satisfying
\[
\begin{array}{rcllrllll}
{\cal J}^3_0(x)\ket{x}^\star_{j-{1\over 2} , \epsilon}
&=&
(j-{\textstyle {1\over 2}})
\ket{x}^\star_{j-{1\over 2} , \epsilon}
,
& \quad
L^{\scriptscriptstyle {\rm L}}_0\ket{x}^\star_{j-{1\over 2} , \epsilon}&=&\Delta^{\scriptscriptstyle {\rm L}}_{-{1\over 2}}\ket{x}^\star_{j-{1\over 2} , \epsilon}
\end{array} .
\]
In the case of $n=\pm 1$
the solutions to conditions (\ref{ep}) take the form
\begin{eqnarray*}
\ket{x}^\star_{j+1 , \epsilon} &=& \left(J^+_{-1}(x)\ket{x}_{j , \epsilon}   -(\kappa-2j)K^+_{-1}(x)\right) \ket{x}_{j , \epsilon}  \otimes \ket{x}_0
,
\\
\ket{x}^\star_{j-1, \epsilon}
&=&\Big( J^+_{-1}(x)  (J^-_0)^2 + 2 (2 j - 1) J^{3}_{-1}(x)  J^-_0 - 2j(2j-1) J^-_{-1} \Big)\ket{x}_{j , \epsilon}  \otimes \ket{x}_0
\\
&-& (\kappa+2j+2)\Big( K^+_{-1}(x)  (J^-_0)^2 + 2 (2 j - 1) K^{3}_{-1}(x)  J^-_0 - 2j(2j-1)K^-_{-1} \Big)\ket{x}_{j , \epsilon}  \otimes \ket{x}_0
.
\end{eqnarray*}
The chiral fields corresponding to states  (\ref{ep})
have a general form of descendants in the product theory of $\hat{\rm su}(2)_\kappa$ and $\hat{\rm su}(2)_1$
\bea
 \Phi^\star_{j+n, \epsilon\+ n} (x, z) =  M(j, \epsilon,n) \,  \mathcal{O}_{n}(J^a(x),K^a(x)) \,  \left( \Phi\sa_{j, \epsilon}
 \otimes  \Phi^{\scriptscriptstyle 1}_{ \epsilon'} \right) (x,z),
\eea
where $\epsilon'\dot{=} n$ and $M(j, \epsilon,n) $ is a normalization.
In particular, for $n_i =0, \pm \frac12, \pm 1$ they read
\begin{eqnarray}
\Phi_{j, \epsilon }^\star (x,z)  &=& \left(\Phi\sa_{j, \epsilon} \otimes \Phi^{\scriptscriptstyle 1}_{ 0}\right) (x,z) \ ,
\nonumber
\\
\label{Phi*first}
\Phi_{j+\frac12,\bar \epsilon }^\star (x,z)  &=& M(j,\epsilon, {\textstyle \frac12})
 \left( \Phi\sa_{j, \epsilon} \otimes \Phi^{\scriptscriptstyle 1}_{ \frac12} \right) (x,z) \ ,
\\ \nonumber
\Phi_{j-\frac12, \bar \epsilon }^\star (x,z)  &=& M(j,\epsilon,-{\textstyle \frac12}) \left(J^-_0(x) -2j K^-_0(x) \right) \left( \Phi\sa_{j,   \epsilon } \otimes \Phi^{\scriptscriptstyle 1}_{ \frac12} \right) (x,z) \ ,
\\ \nonumber
\Phi_{j+1,\epsilon}^\star (x,z)
&=& M(j,\epsilon,1) \left(J^+_{-1}(x)  -(\kappa-2j)K^+_{-1}(x)\right) \left( \Phi\sa_{j,\epsilon}\otimes \Phi^{\scriptscriptstyle 1}_0 \right) (x,z) \ ,
\\ \nonumber
\Phi_{j-1,\epsilon}^\star (x,z)
&=& M(j,\epsilon,-1) \Big(- J^+_{-1}(x)  (J^-_0)^2 - 2 (2 j - 1) J^{3}_{-1}(x)  J^-_0 + 2j(2j-1) J^-_{-1}
\\ \nonumber
&& \hspace{-60pt} + (\kappa+2j+2)\left( K^+_{-1}(x)  (J^-_0)^2 + 2 (2 j - 1) K^{3}_{-1}(x)  J^-_0 - 2j(2j-1)K^-_{-1} \right) \Big)
 \left( \Phi\sa_{j,\epsilon}\otimes \Phi^{\scriptscriptstyle 1}_0 \right) (x,z) \ .
\end{eqnarray}
The  normalization in the $\widehat{\rm su}(2)_1$ theory is chosen such that
\begin{equation}
\label{structure1}
\begin{array}{rcl}
\langle \Phi^1_{0} (x_3, z_3) \, \Phi^1_{0} ( x_2,z_2) \, \Phi^1_{0} ( x_1, z_1) \rangle
&=& 1,
\\[6pt]
\langle \Phi^1_{0} (x_3, z_3)  \Phi^1_{\frac12} ( x_2,z_2)  \Phi^1_{\frac12} ( x_1, z_1) \rangle
&=&
(x_1 - x_2) \,  (z_2 - z_1)^{-\frac12},
\\[6pt]
\langle \Phi^1_{\frac12} (x_3, z_3)  \Phi^1_{0} ( x_2,z_2)  \Phi^1_{\frac12} ( x_1, z_1) \rangle
&=&
(x_3 - x_1) \,  (z_3 - z_1)^{-\frac12},
\\[6pt]
\langle \Phi^1_{\frac12}(x_3, z_3)  \Phi^1_{\frac12} ( x_2,z_2)  \Phi^1_{0} ( x_1, z_1) \rangle
&=&
(x_2 - x_3) \,  (z_3 - z_2)^{-\frac12}.
\end{array}
\end{equation}
\subsection{ checks for $n=0$}

We conjecture the relation between chiral correlators in the product theory of $\hat{\rm su}(2)_\kappa$ and $\hat{\rm su}(2)_1$ on one side, and chiral correlators of $\hat{\rm su}(2)_{\kappa+1}$ and Liouville theory on the other side, (\ref{3-pkt_gen2}):
\begin{eqnarray} \label{conjecture_L}
\nonumber
&& \hspace{-20pt}
\langle \Phi^\star_{j_3+n_3, \epsilon_3 \+n_3} (x_3, z_3) \Phi^\star_{j_2+n_2,\epsilon_2 \+n_2 } ( x_2,z_2)
\Phi^\star_{j_1+n_1,\epsilon_1 \+n_1 } ( x_1, z_1) \rangle_\epsilon
 =
\langle \Phi_{j_3+\frac{n_3 }{bQ}} (z_3) \Phi_{j_2+\frac{n_2}{bQ} } ( z_2) \Phi_{j_1+\frac{n_1}{bQ}} (z_1)  \rangle\sL
 \\
&&
\times
 \langle  \Phi\sB_{j_3+n_3,  \epsilon_3 \+n_3} (x_3, z_3) \Phi\sB_{j_2+n_2,  \epsilon_2 \+n_2 } ( x_2,z_2)  \Phi\sB_{j_1+n_1, \epsilon_1 \+n_1 } ( x_1, z_1)  \rangle^\h_{\epsilon\+\frac12 n_{123}} \, .
\end{eqnarray}
In the case  of all $n_i=0$ the conjecture
takes its simplest form
\begin{eqnarray}
\label{n=0}
&& \hspace{-80pt}
\langle \Phi^\star_{j_3, \epsilon_3} (x_3, z_3) \Phi^\star_{j_2,\epsilon_2 } ( x_2,z_2) \Phi^\star_{j_1,\epsilon_1 } ( x_1, z_1) \rangle_\epsilon
\\
\nonumber
&&
 =
\langle \Phi_{j_3 } (z_3) \Phi_{j_2 } ( z_2) \Phi_{j_1} (z_1)  \rangle\sL
\,
 \langle  \Phi\sB_{j_3,  \epsilon_3} (x_3, z_3)  \Phi\sB_{j_2,  \epsilon_2 } ( x_2,z_2)  \Phi\sB_{j_1, \epsilon_1 } ( x_1, z_1)  \rangle^\h_\epsilon\, .
\end{eqnarray}
On both sides of this relation:
\bea
\langle \Phi^\star_{j_3, \epsilon_3} (x_3, z_3) \Phi^\star_{j_2,\epsilon_2 } ( x_2,z_2) \Phi^\star_{j_1,\epsilon_1 } ( x_1, z_1) \rangle_\epsilon
&=&
\langle \Phi\sa_{j_3} (x_3, z_3) \Phi\sa_{j_2} ( x_2,z_2) \Phi\sa_{j_1} ( x_1, z_1) \rangle^\s_\epsilon
\\
&=&  c \left[ \Delta\sa(j_i); z_i \right] \, \widetilde S_\epsilon\Big[\!
\begin{array}{ccc}
\\[-18pt]
\scriptstyle j_3 & \scriptstyle j_2  & \scriptstyle j_1
\\[-8pt]
\scriptstyle \epsilon_3  &  \scriptstyle \epsilon_2  & \scriptstyle \epsilon_1
\\[-8pt]
\scriptstyle x_3  &  \scriptstyle x_2  &  \scriptstyle x_1
\end{array}
\!\Big]   \,
 {\sf C}^{\scriptscriptstyle {\rm S}}_{ b\sa}[ j_3,j_2,j_1],
\eea
and
\bea
&& \hspace{-30pt}
\langle \Phi_{j_3 } (z_3) \Phi_{j_2 } ( z_2) \Phi_{j_1} (z_1)  \rangle\sL
\,
 \langle \Phi\sB_{j_3,  \epsilon_3} (x_3, z_3) \Phi\sB_{j_2,  \epsilon_2 } ( x_2,z_2) \Phi\sB_{j_1, \epsilon_1 } ( x_1, z_1)  \rangle^\h_\epsilon
\\
&& =  c \left[\Delta\sL(j_i  ) + \Delta\sB (j_i   ); z_i \right] \,  \widetilde S_\epsilon\Big[\!
\begin{array}{ccc}
\\[-18pt]
\scriptstyle j_3 & \scriptstyle j_2  & \scriptstyle j_1
\\[-8pt]
\scriptstyle \epsilon_3  &  \scriptstyle \epsilon_2  & \scriptstyle \epsilon_1
\\[-8pt]
\scriptstyle x_3  &  \scriptstyle x_2  &  \scriptstyle x_1
\end{array}
\!\Big]   \,
{\sf C}^{\scriptscriptstyle {\rm L}}_{ b}[ j_3,j_2,j_1]  \, {\sf C}^{\scriptscriptstyle {\rm IS}}_{ \hat b\sB}[ j_3,j_2,j_1],
\eea
we have the same 3-linear ${\rm su}(2)$ invariant $ \widetilde S_\epsilon$. The $z_i$-dependent terms agree due to (\ref{parameters_b})
and equation (\ref{n=0}) reduces to the relation between $j$-dependent parts
\begin{equation}
\label{relconst}
 {\sf C}^{\scriptscriptstyle {\rm S}}_{ b\sa}[ j_3,j_2,j_1]  =  {\sf C}^{\scriptscriptstyle {\rm L}}_{ b}[ j_3,j_2,j_1]
 \,  {\sf C}^{\scriptscriptstyle {\rm IS}}_{\hat b\sB}[ j_3,j_2,j_1]  \, .
\end{equation}
Using the identity for Barnes gamma functions
\begin{eqnarray}
\label{Gamma_LH}
{\Gamma_{b\sa}(-b\sa j ) \Gamma_{\hat  b\sB}\left(\frac{1}{\hat  b\sB} - \hat b\sB j \right) \over \Gamma_{b}(-Q j ) }
&=& {  \Gamma_{b\sa}(b\sa)
  \Gamma_{\hat b\sB}\left(\frac{1}{\hat  b\sB} + \hat b\sB  \right) \over  \Gamma_{b}(Q)}
\left( {b^{-1} Q  } \right)^{ {Q^2 j(j+1)\over 4} }
b^{-\frac12 b Q j(j+1)  -\frac12 j-\frac12},
\end{eqnarray}
one can show that (\ref{relconst}) indeed holds if the relative normalization is given by
$$
\frac{M^{\scriptscriptstyle {\rm IS}}_{\hat b\sB } M^{\scriptscriptstyle {\rm L}}_{ b } }
{M^{\scriptscriptstyle {\rm S}}_{ b\sa } }
=
\left(\frac{\Gamma_{b\sa}(b\sa ) \,\Gamma_{\hat  b\sB}\left(\frac{1}{\hat  b\sB} + \hat b\sB \right)}
{\Gamma_{b}\left(\frac{1}{b} +b \right)}\right)^2
\left(b^{-1}Q\right)^{-{Q^2\over 2}} b^{b^2+\frac12}  \ .
$$
Under the same condition equations (\ref{barnesliouville}) and (\ref{Cbar_hs}) imply the relation for the right structure constants
$$
\bar {\sf C}^{\scriptscriptstyle {\rm S}}_{ b\sa}[ j_3,j_2,j_1]  = \bar {\sf C}^{\scriptscriptstyle {\rm L}}_{ b}[ j_3,j_2,j_1]
 \, \bar {\sf C}^{\scriptscriptstyle {\rm IS}}_{\hat b\sB}[ j_3,j_2,j_1]  \, .
$$

In  the case of imaginary Liouville theory
the counterpart of relation (\ref{n=0}) reads
\begin{eqnarray}
\label{in=0}
&& \hspace{-80pt}
\langle \Phi^\star_{j_3, \epsilon_3} (x_3, z_3) \Phi^\star_{j_2,\epsilon_2 } ( x_2,z_2) \Phi^\star_{j_1,\epsilon_1 } ( x_1, z_1) \rangle_\epsilon
\\
\nonumber
&&
 =
\langle  \Phi_{j_3 } (z_3)  \Phi_{j_2 } ( z_2)  \Phi_{j_1} (z_1)  \rangle\sIL
\,
 \langle  \Phi\sB_{j_3,  \epsilon_3} (x_3, z_3)  \Phi\sB_{j_2,  \epsilon_2 } ( x_2,z_2)  \Phi\sB_{j_1, \epsilon_1 } ( x_1, z_1)  \rangle^\s_\epsilon\, .
\end{eqnarray}
For  parameters of the theories  as in  (\ref{parameters_bIm}),
 the l.h.s  is the same as in (\ref{n=0}), while on the r.h.s one has correlators from the imaginary Liouville theory
  and the $\hat{\rm su}(2)_{\kappa+1}$ WZNW model with $\kappa<-3$. By the same argument as in the Liouville case, relation (\ref{in=0}) reduces to the relation
  between $j$-dependent parts of the structure constants
\begin{equation}
\label{irelconst}
{\sf C}^{\scriptscriptstyle {\rm S}}_{ b\sa}[ j_3,j_2,j_1]  =
{\sf C}\sIL_{\hat b}[ j_3,j_2,j_1]\, {\sf C}^{\scriptscriptstyle {\rm S}}_{ b\sB}[ j_3,j_2,j_1].
\end{equation}
Using another identity
\begin{eqnarray}
\label{Gamma_ImH}
{ \Gamma_{b\sB}(-b\sB j) \over \Gamma_{b\sa}(-b\sa j)  \Gamma_{\bh}( \bh- \hat Q j  )}
&=&
{  \Gamma_{b\sB}(b\sB) \over \Gamma_{b\sa}(b\sa) \Gamma_{\bh}\left({\bh}^{-1}\right)}
\left( \bh^{-1} \hat Q   \right)^{\frac{\hat Q^2 j( j + 1)}{4}}
\bh^{ \frac{\hat Q}{2\hat b}j(j+1) + \frac12 j+\frac12},
\end{eqnarray}
one can show that equality (\ref{irelconst}) and its right counterpart hold if the normalizations of fields are related by
$$
\frac {M^{\scriptscriptstyle {\rm S}}_{ b\sa }}
 {M^{\scriptscriptstyle {\rm S}}_{b\sB}  M^{\scriptscriptstyle {\rm IL}}_{\hat  b } }
=
\left(\frac{ \Gamma_{b\sB}(b\sB )}{\Gamma_{b\sa}(b\sa )  \Gamma_{\bh}( \bh^{-1})}\right)^2
\left( \bh^{-1} \hat Q   \right)^{-\frac{\hat Q^2 }{2}}
\bh^{ -{\hat b}^{-2}+\frac32}  .
$$
We assume in the following that the relative normalizations are fixed by relations (\ref{relconst}) and (\ref{irelconst}).
One can therefore safely drop the normalization constants $M^{\scriptscriptstyle {\rm S}}_{ b },
 M^{\scriptscriptstyle {\rm IS}}_{\hat b}, M^{\scriptscriptstyle {\rm L}}_{  b } ,  M^{\scriptscriptstyle {\rm IL}}_{\hat  b } $ in subsequent formulae.
Identities (\ref{Gamma_LH}), (\ref{Gamma_ImH}) are derived in Appendix B.

\subsection{reformulation of the general case}

In the general case the r.h.s. of   \eqref{conjecture_L} can be calculated explicitly using shift relations (\ref{shift})
\begin{eqnarray*}
\nonumber
&& \hspace{-30pt}
\langle \Phi_{j_3+\frac{n_3 }{bQ}} (z_3) \Phi_{j_2+\frac{n_2}{bQ} } ( z_2) \Phi_{j_1+\frac{n_1}{bQ}} (z_1)  \rangle\sL
\\
& &\times\;
  \langle
\Phi\sB_{j_3+n_3, \epsilon_3\+ n_3} (x_3, z_3)
\Phi\sB_{j_2+n_2, \epsilon_2\+ n_2} (x_2, z_2)
\Phi\sB_{j_1+n_1, \epsilon_1\+ n_3} (x_1, z_1)  \rangle^\h_{\epsilon\+\frac12 n_{123}}
 \\
  \nonumber
& =&
c \left[{\textstyle \Delta\sL(j_i +  \frac{n_i}{bQ} ) + \Delta\sB (j_i +n_i  ); z_i }\right] \,
  \widetilde S_{\epsilon\+\frac12 \epsilon_{123}}\Big[\!
\begin{array}{ccc}
\\[-18pt]
\scriptstyle j_3+ n_3 & \scriptstyle j_2+n_2  & \scriptstyle j_1+n_1
\\[-7pt]
\scriptstyle \epsilon_3\+ n_3 &  \scriptstyle \epsilon_2\+ n_2  & \scriptstyle  \epsilon_1\+ n_1
\\[-8pt]
\scriptstyle x_3  &  \scriptstyle x_2  &  \scriptstyle x_1
\end{array}
\!\Big]
\\
&&\times\;
 {\sf C}^\h_{\hat b\sB}[ j_3+n_3,j_2+n_2,j_1+n_1] \,
 {\sf C}\sL_b\left[{\textstyle j_3+  \frac{n_3 }{bQ},j_2+  \frac{n_2}{bQ}, j_1+  \frac{n_1}{bQ} }\right]
 \\ \nonumber
 &=&
\Big( \prod_{i=1}^3 N(j_i,n_i) \Big)\, l(j_{123}+1, n_{123}) \, l(j_{12}^3, n_{12}^3) \, l(j_{13}^2, n_{13}^2) \, l(j_{23}^1, n_{23}^1)\\ \nonumber
 & & \times\;
 c \left[ \Delta\sa(j_i)+ n_i^2; z_i \right] \,
  \widetilde S_{\epsilon\+\frac12 \epsilon_{123}}\Big[\!
\begin{array}{ccc}
\\[-18pt]
\scriptstyle j_3+ n_3 & \scriptstyle j_2+n_2  & \scriptstyle j_1+n_1
\\[-7pt]
\scriptstyle \epsilon_3\+n_3 &  \scriptstyle \epsilon_2\+n_2  & \scriptstyle  \epsilon_1\+n_1
\\[-8pt]
\scriptstyle x_3  &  \scriptstyle x_2  &  \scriptstyle x_1
\end{array}
\!\Big]
 {\sf C}^{\scriptscriptstyle {\rm IS}}_{\hat b\sB}[ j_3,j_2,j_1] \,  {\sf C}\sL_b[ j_3,j_2,j_1]\, ,
\end{eqnarray*}
where
\begin{eqnarray}
\nonumber
l(x, n) &=&
  \left\lbrace
  \begin{array}{lllll}
  \prod_{p=2}^n \prod_{q=1}^{p-1}
\left( x -  p (\kappa +2) + q (\kappa +3)\right) & , && n>1,
\\[4pt]
1& , && n=0,1,
\\[4pt]
 \prod_{p=0}^{|n|-1}  \prod_{q=0}^{p }
\left( x +  p (\kappa +2) - q (\kappa +3)\right) & , && n<0,
  \end{array}
  \right.
\\[6pt]
\label{Njn}
N(j,n) &=& (-1)^{n(2n-1)} \,  \left( l(2j, 2n) \, l(2j+1, 2n) \right)^{-\frac12}.
\end{eqnarray}
Applying (\ref{Fidentity}) and (\ref{relconst}) one can rewrite the r.h.s. of    \eqref{conjecture_L} in the following form
\begin{eqnarray}
\label{cor_LhatB}
\nonumber
&& \hspace{-30pt}
\langle \Phi_{j_3+\frac{n_3 }{bQ}} (z_3) \Phi_{j_2+\frac{n_2}{bQ} } ( z_2) \Phi_{j_1+\frac{n_1}{bQ}} (z_1)  \rangle\sL
\\
&&\times\;
  \langle
\Phi\sB_{j_3+n_3, \epsilon_3\+ n_3} (x_3, z_3)
\Phi\sB_{j_2+n_2, \epsilon_2\+ n_2} (x_2, z_2)
\Phi\sB_{j_1+n_1, \epsilon_1\+ n_3} (x_1, z_1)  \rangle^\h_{\epsilon\+\frac12 n_{123}}
\\
\nonumber
&=&
(-1)^{\eta(n_3,n_2,n_1)}\Big(\prod_{i=1}^3 N(j_i,n_i) \Big)
\\
\nonumber
&& \times\;l(j_{123}+1, n_{123}) \, l(j_{12}^3, n_{12}^3) \, l(j_{13}^2, n_{13}^2) \, l(j_{23}^1, n_{23}^1)
\\ \nonumber
&& \times\;
(z_2 - z_1)^{n_3^2-n_1^2-n_2^2} (z_3 - z_1)^{n_2^2-n_1^2-n_3^2} (z_3 - z_2)^{n_1^2-n_2^2-n_3^2}
\\ \nonumber
&& \times\;
(x_1 - x_2)^{n_1+n_2-n_3} (x_3 - x_1)^{n_1+n_3-n_2} (x_2 - x_3)^{n_2+n_3-n_1}
\\
\nonumber
&& \times\;
\langle
 \Phi\sa_{j_3, \epsilon_3} (x_3, z_3)
 \Phi\sa_{j_2, \epsilon_2} (x_2, z_2)
 \Phi\sa_{j_1, \epsilon_1} (x_1, z_1)
\rangle^\s_\epsilon .
\end{eqnarray}
The proof of  \eqref{conjecture_L} reduces therefore to showing that one can chose  normalizations of fields such that the following relation holds
\begin{eqnarray}
\label{cosetfactor}
&&\hspace{-30pt}
{\langle \Phi^\star_{j_3+n_3, \epsilon_3\+n_3} (x_3, z_3) \Phi^\star_{j_2+n_2,\epsilon_2\+ n_2 } ( x_2,z_2) \Phi^\star_{j_1+n_1,\epsilon_1\+n_1 } ( x_1, z_1) \rangle_\epsilon
\over
\langle
  \Phi\sa_{j_3, \epsilon_3} (x_3, z_3)
  \Phi\sa_{j_2, \epsilon_2} (x_2, z_2)
  \Phi\sa_{j_1, \epsilon_1} (x_1, z_1)
\rangle^\s_\epsilon}
\\
\nonumber
&=&
(-1)^{\eta(n_3,n_2,n_1)}\Big(\prod_{i=1}^3 N(j_i,n_i) \Big)
\\
\nonumber &&\times\;\,l(j_{123}+1, n_{123}) \, l(j_{12}^3, n_{12}^3) \, l(j_{13}^2, n_{13}^2) \, l(j_{23}^1, n_{23}^1)
\\ \nonumber
&& \times\;
(z_2 - z_1)^{n_3^2-n_1^2-n_2^2} (z_3 - z_1)^{n_2^2-n_1^2-n_3^2} (z_3 - z_2)^{n_1^2-n_2^2-n_3^2}
\\ \nonumber
&& \times\;
(x_1 - x_2)^{n_1+n_2-n_3} (x_3 - x_1)^{n_1+n_3-n_2} (x_2 - x_3)^{n_2+n_3-n_1}.
\end{eqnarray}
As it was mentioned in the Introduction we call the object on the l.h.s of (\ref{cosetfactor}) the coset factor.\footnote{Its counterpart in the SL-LL correspondence
is called the blow up factor for reason coming from the 4-dim side of the AGT relation.}

In the case of the imaginary Liouville theory the counterpart of conjecture  \eqref{conjecture_L} takes the form
\begin{eqnarray}
\label{3-pkt_gen2_IL}
\nonumber
&& \hspace{-40pt}
\langle \Phi^\star_{j_3+n_3, \epsilon_3\+n_3} (x_3, z_3)
 \Phi^\star_{j_2+n_2,\epsilon_2\+ n_2 } ( x_2,z_2) \Phi^\star_{j_1+n_1,\epsilon_1\+n_1 } ( x_1, z_1) \rangle_\epsilon
\\
&& =\;
\langle 
\Phi_{j_3+\frac{n_3}{\hat b \hat Q}} (z_3) 
\Phi_{j_2+\frac{n_2}{\hat b \hat Q}} ( z_2)
\Phi_{j_1+\frac{n_1}{\hat b \hat Q} } (z_1)  \rangle\sIL
  \\
&& \nonumber
\times\;
 \langle  \Phi\sB_{j_3+n_3,  \epsilon_3 \+n_3} (x_3, z_3)  \Phi\sB_{j_2+n_2,  \epsilon_2 \+n_2} ( x_2,z_2)
 \Phi\sB_{j_1+n_1, \epsilon_1 \+n_1 } ( x_1, z_1)  \rangle^\s_{\epsilon\+\frac12 n_{123}} \, .
\end{eqnarray}
One can follow the same steps as in the Liouville case using (\ref{irelconst}) instead of (\ref{relconst}).
The result is exactly the same as in (\ref{cosetfactor}). The only difference is that in the  Liouville case
$\kappa$ is in the range $-3<\kappa<-2$ while for the imaginary Liouville $\kappa<-3$.
In both cases the $\widehat{\rm su}(2)_\kappa$ WZNW model is on the real side of the $\kappa =-2$ barrier
so the analytic form of the coset factor should be the same in agreement with conjectured relation (\ref{cosetfactor}).
Checking (\ref{cosetfactor}) verifies therefore both relations (\ref{rel}).

\subsection{ checks for $n=\pm\frac12$}

We choose locations of fields
$
z_3=\infty,z_2=z,z_1=0
$
for which  the Ward identities take their simple form (\ref{simpleWI}).
Due to the general condition $\epsilon_1+\epsilon_2+\epsilon_3\,\dot{=}\,0$
there are three subcases of the 3-point function containing two fields with  $j \pm \frac12$ and one primary field.
The simplest case reads
\bea
&& \hspace{-20pt}
\langle  \Phi^\star_{j_3, \epsilon_3} ( x_3, \infty)
\Phi^\star_{j_2+\frac12,\epsilon_2\+\frac12} ( x_2,z)
\Phi^\star_{j_1+\frac12,\epsilon_1\+\frac12} ( x_1,0) \rangle_\epsilon
\\
&& =
 M(j_2,\epsilon_2,{\textstyle \frac12})  M(j_1,\epsilon_1,{\textstyle \frac12})
 \langle  \Phi\sa_{j_3,\epsilon_3} (x_3,\infty)  \Phi\sa_{j_2, \epsilon_2} ( x_2,z)  \Phi\sa_{j_1,  \epsilon_1} (x_1,0) \rangle^\s_\epsilon
 \langle  \Phi^{\scriptscriptstyle 1}_{ \frac12}  ( x_2,z)   \Phi^{\scriptscriptstyle 1}_{ \frac12} (x_1,0) \rangle
\\
&&=
 M(j_2,\epsilon_2,{\textstyle \frac12})  M(j_1,\epsilon_1,{\textstyle \frac12})
 (x_1 - x_2)z^{-{1\over 2}}
\langle  \Phi\sa_{j_3,\epsilon_3} (x_3,\infty)  \Phi\sa_{j_2, \epsilon_2} ( x_2,z)  \Phi\sa_{j_1,  \epsilon_1} (x_1,0) \rangle^\s_\epsilon.
\eea
In the calculation of the next one
\bea
&& \hspace{-20pt}
\langle  \Phi^\star_{j_3, \epsilon_3} ( x_3, \infty)    \Phi^\star_{j_2-\frac12,\epsilon_2\m\frac12} ( x_2,z)   \Phi^\star_{j_1+ \frac12,\epsilon_1\+\frac12} ( x_1,0) \rangle_\epsilon
\\
&=&
 M(j_2,\epsilon_2,-{\textstyle \frac12})  M(j_1,\epsilon_1,{\textstyle \frac12})
\Big[ \langle
\Phi\sa_{j_3, \epsilon_3} (x_3,\infty) J^-_0   \Phi\sa_{j_2,\epsilon_2} (x_2,z)
\Phi\sa_{j_1, \epsilon_1} (x_1,0)  \rangle^\s_\epsilon
\langle  \Phi^{\scriptscriptstyle 1}_{ \frac12}  ( x_2,z)   \Phi^{\scriptscriptstyle 1}_{ \frac12} (x_1,0) \rangle
\\ &&
-\; 2j_2
\langle  \Phi\sa_{j_3, \epsilon_3} (x_3,\infty)   \Phi\sa_{j_2, \epsilon_2} ( x_2,z)  \Phi\sa_{j_1, \epsilon_1} (x_1,0) \rangle^\s_\epsilon
 \langle K^-_0  \Phi^{\scriptscriptstyle 1}_{ \frac12}  ( x_2,z)   \Phi^{\scriptscriptstyle 1}_{ \frac12} (x_1,0) \rangle
 \Big]
\\
&=& -
M(j_2,\epsilon_2,-{\textstyle \frac12})  M(j_1,\epsilon_1,{\textstyle \frac12})      \,  (j_2+j_3- j_1) \,
\frac{ (x_3 - x_1)}{ (x_2 - x_3)}  z^{-{1\over 2}}\,
  \langle  \Phi\sa_{j_3, \epsilon_3} (x_3,\infty)   \Phi\sa_{j_2, \epsilon_2} ( x_2,z)  \Phi\sa_{j_1, \epsilon_1} (x_1,0) \rangle^\s_\epsilon
\eea
one uses the action of the zero modes (\ref{JxOPE}) and the identity
$$
\partial_{x_2} \widetilde S_\epsilon\Big[\!
\begin{array}{ccc}
\\[-18pt]
\scriptstyle j_3 & \scriptstyle j_2  & \scriptstyle j_1
\\[-8pt]
\scriptstyle \epsilon_3  &  \scriptstyle \epsilon_2  & \scriptstyle \epsilon_1
\\[-8pt]
\scriptstyle x_3  &  \scriptstyle x_2  &  \scriptstyle x_1
\end{array}
\!\Big]
=
\left(
{j_2+j_3-j_1 \over x_2-x_3} -{j_1+j_2-j_3\over x_1-x_2}
\right)
\widetilde S_\epsilon\Big[\!
\begin{array}{ccc}
\\[-18pt]
\scriptstyle j_3 & \scriptstyle j_2  & \scriptstyle j_1
\\[-8pt]
\scriptstyle \epsilon_3  &  \scriptstyle \epsilon_2  & \scriptstyle \epsilon_1
\\[-8pt]
\scriptstyle x_3  &  \scriptstyle x_2  &  \scriptstyle x_1
\end{array}
\!\Big] .
$$
The third case is slightly more complicated:
\bea
&& \hspace{-40pt}
\langle
\Phi^\star_{j_3, \epsilon_3} ( x_3, \infty)
\Phi^\star_{j_2-\frac12,\epsilon_2\m\frac12} ( x_2,z)
\Phi^\star_{j_1-\frac12,\epsilon_1\m\frac12} ( x_1,0) \rangle_\epsilon
\;=\;
M(j_2,\epsilon_2,-{\textstyle \frac12})  M(j_1,\epsilon_1,-{\textstyle \frac12})
\\
&&\times\,  \Big[
\langle  \Phi\sa_{j_3, \epsilon_3} (x_3,\infty) J^-_0   \Phi\sa_{j_2, \epsilon_2} ( x_2,z) J^-_0  \Phi\sa_{j_1, \epsilon_1} (x_1,0) \rangle^\s_\epsilon
\langle  \Phi^{\scriptscriptstyle 1}_{ \frac12}  ( x_2,z)   \Phi^{\scriptscriptstyle 1}_{ \frac12} (x_1,0) \rangle
\\ &&\;\;\; -
 2j_2
\langle  \Phi\sa_{j_3} (x_3,\infty)  \Phi\sa_{j_2} ( x_2,z) J^-_0 \Phi\sa_{j_1} (x_1,0) \rangle^\s_\epsilon
 \langle K^-_0  \Phi^{\scriptscriptstyle 1}_{ \frac12}  ( x_2,z)   \Phi^{\scriptscriptstyle 1}_{ \frac12} (x_1,0) \rangle
\\ && \;\;\;-
 2j_1
\langle  \Phi\sa_{j_3} (x_3,\infty) J^-_0 \Phi\sa_{j_2} ( x_2,z)  \Phi\sa_{j_1} (x_1,0) \rangle^\s_\epsilon
 \langle   \Phi^{\scriptscriptstyle 1}_{ \frac12}  ( x_2,z)  K^-_0  \Phi^{\scriptscriptstyle 1}_{ \frac12} (x_1,0) \rangle
 \Big]
\\
&=&
 M(j_2,\epsilon_2,-{\textstyle \frac12})  M(j_1,\epsilon_1,-{\textstyle \frac12}) { (j_1+j_2+j_3+1) (j_1+j_2- j_3)\over x_1 - x_2}
 z^{-{1\over 2}}
 \\
&& \;\;\;\times
\langle  \Phi\sa_{j_3} (x_3,\infty)  \Phi\sa_{j_2} ( x_2,z)  \Phi\sa_{j_1} (x_1,0) \rangle^\s_\epsilon\, .
\eea
Choosing the fields normalization equal to the factor \eqref{Njn},
$$
M(j,\epsilon,\pm{\textstyle \frac12}) = N(j,\pm{\textstyle \frac12})
$$
one gets the agreement with the conjecture (\ref{cosetfactor}) in all the cases.

\subsection{checks for $n=\pm 1$}

Let us first consider  correlators containing one field $\Phi^\star_{j \pm 1, \epsilon}$.   For $n=1$, using Ward identities \eqref{simpleWI},  one obtains
\bea
&& \hspace{-35pt}
\langle  \Phi^\star_{j_3,\epsilon_3} (x_3, \infty)  \Phi^\star_{j_2,\epsilon_2} ( x_2,z)
  \Phi^\star_{j_1+1,\epsilon_1}( x_1,0) \rangle^\s_\epsilon
\\
&=&  M(j_1,\epsilon_1,1)
 \langle  \Phi\sa_{j_3,\epsilon_3} (x_3, \infty)  \Phi\sa_{j_2,\epsilon_2} ( x_2,z)
  J^+_{-1} (x_1)\, \Phi\sa_{j_1,\epsilon_1}( x_1,0) \rangle^\s_\epsilon
  \\
&=&
-   \frac{1}{z} M(j_1,\epsilon_1,1) \langle  \Phi\sa_{j_3,\epsilon_3} (x_3, \infty)   J^+_{0} (x_1)  \Phi\sa_{j_2,\epsilon_2} ( x_2,z)   \Phi\sa_{j_1,\epsilon_1}( x_1,0) \rangle^\s_\epsilon
 \\
 &=&
    - M(j_1,\epsilon_1,1)
   (j_2 + j_3 - j_1) \,
\frac{ (x_1 - x_2) (x_3 - x_1)}{ (x_2 - x_3)} z^{-1}
\langle  \Phi\sa_{j_3,\epsilon_3} (x_3, \infty)  \Phi\sa_{j_2,\epsilon_2} ( x_2,z)
  \Phi\sa_{j_1,\epsilon_1}( x_1,0) \rangle^\s_\epsilon,
\eea
where the last equation is based on the identity
\bea
  &&\hspace{-70pt} \left( (x_1-x_2)^2\partial_{x_2} +  2 j_2 (x_1-x_2)  \right)
\widetilde S_\epsilon\Big[\!
\begin{array}{ccc}
\\[-18pt]
\scriptstyle j_3 & \scriptstyle j_2  & \scriptstyle j_1
\\[-8pt]
\scriptstyle \epsilon_3  &  \scriptstyle \epsilon_2  & \scriptstyle \epsilon_1
\\[-8pt]
\scriptstyle x_3  &  \scriptstyle x_2  &  \scriptstyle x_1
\end{array}
\!\Big]
\\
 &=&  (j_2 + j_3 - j_1) \,
\frac{ (x_1 - x_2) (x_1 - x_3)}{ (x_2 - x_3)}
\widetilde S_\epsilon\Big[\!
\begin{array}{ccc}
\\[-18pt]
\scriptstyle j_3 & \scriptstyle j_2  & \scriptstyle j_1
\\[-8pt]
\scriptstyle \epsilon_3  &  \scriptstyle \epsilon_2  & \scriptstyle \epsilon_1
\\[-8pt]
\scriptstyle x_3  &  \scriptstyle x_2  &  \scriptstyle x_1
\end{array}
\!\Big].
\eea
Calculations in the case of $n=-1$  are more tedious
\bea
&& \hspace{-20pt}
\langle  \Phi^\star_{j_3,\epsilon_3} (x_3, \infty)  \Phi^\star_{j_2,\epsilon_2} ( x_2,z)
  \Phi^\star_{j_1-1,\epsilon_1}( x_1,0) \rangle_\epsilon
  \\
&=&
   M(j,\epsilon,-1) \Big[-
   \langle  \Phi\sa_{j_3,\epsilon_3} (x_3, \infty)  \Phi\sa_{j_2,\epsilon_2} ( x_2,z)   J^+_{-1}(x_1)  (J^-_0)^2 \Phi\sa_{j_1,\epsilon_1} (x_1,0) \rangle^\s_\epsilon
  \\
  &&
  \hspace{60pt}- 2 (2 j_{1} - 1)
\langle  \Phi\sa_{j_3,\epsilon_3} (x_3, \infty)  \Phi\sa_{j_2,\epsilon_2} ( x_2,z)   \left( J^{3}_{-1}(x_1)  J^-_0 - j_{1} J^-_{-1}  \right) \Phi\sa_{j_1,\epsilon_1} (x_1,0) \rangle^\s_\epsilon \Big]
\\
&=&
   M(j,\epsilon,-1)z^{-1} \Big[  \langle \Phi\sa_{j_3,\epsilon_3}( x_3, \infty) J^+_{0}(x_1)\Phi\sa_{j_2,\epsilon_2} ( x_2,z) \partial_{x_1}^2 \Phi\sa_{j_1,\epsilon_1}( x_1,0) \rangle^\s_\epsilon
\\
&&
\hspace{60pt}+  2 (2 j_{1} - 1)
\langle \Phi\sa_{j_3,\epsilon_3}( x_3, \infty) J^3_{0}(x_1)\Phi\sa_{j_2,\epsilon_2} ( x_2,z) \partial_{x_1} \Phi\sa_{j_1,\epsilon_1}( x_1,0) \rangle^\s_\epsilon
\\
&&
\hspace{60pt}- 2j_{1}(2j_{1}-1)
\langle \Phi\sa_{j_3,\epsilon_3}( x_3, \infty) J^-_{0}(x_1)\Phi\sa_{j_2,\epsilon_2} ( x_2,z) \Phi\sa_{j_1,\epsilon_1}( x_1,0) \rangle^\s_\epsilon \Big]
\\
&=&- M(j_1,\epsilon_1,-1) \, (j_1 + j_2 - j_3) (j_1 - j_2 + j_3) (1 + j_1 + j_2 + j_3)
\frac{(x_2 - x_3)}{ (x_1 - x_2)(x_3 - x_1)} z^{-1}
 \\
 &&\hspace{60pt}
 \times \,  \langle
 \Phi\sa_{j_3,\epsilon_3} (x_3, \infty)
 \Phi\sa_{j_2,\epsilon_2} ( x_2,z)
 \Phi\sa_{j_1,\epsilon_1}( x_1,0) \rangle^\s_\epsilon.
\eea
The relation relevant at the last step reads
\bea
&& \hspace{-10pt}
\left(
 \left( (x_1-x_2)^2\partial_{x_2} +  2 j_2 (x_1-x_2)  \right) \partial_{x_1}^2
 -
   2 (2 j_{1} - 1) \left( \left((x_1- x_2) \partial_{x_2}  +  j_2 \right) \,   \partial_{x_1}
   -  j_{1}   \partial_{x_2} \right) \right) \widetilde S_\epsilon\Big[\!
\begin{array}{ccc}
\\[-18pt]
\scriptstyle j_3 & \scriptstyle j_2  & \scriptstyle j_1
\\[-8pt]
\scriptstyle \epsilon_3  &  \scriptstyle \epsilon_2  & \scriptstyle \epsilon_1
\\[-8pt]
\scriptstyle x_3  &  \scriptstyle x_2  &  \scriptstyle x_1
\end{array}
\!\Big]
\\
&&
=
- (j_1 + j_2 - j_3) (j_1 - j_2 + j_3) (1 + j_1 + j_2 + j_3)
\frac{(x_2 - x_3)}{ (x_1 - x_2)(x_1 - x_3)} \, \widetilde S_\epsilon\Big[\!
\begin{array}{ccc}
\\[-18pt]
\scriptstyle j_3 & \scriptstyle j_2  & \scriptstyle j_1
\\[-8pt]
\scriptstyle \epsilon_3  &  \scriptstyle \epsilon_2  & \scriptstyle \epsilon_1
\\[-8pt]
\scriptstyle x_3  &  \scriptstyle x_2  &  \scriptstyle x_1
\end{array}
\!\Big].
\eea
The calculations above are in perfect agreement with conjecture (\ref{cosetfactor}) if we assume:
$$
M(j,\epsilon,\pm 1) = N(j,\pm 1).
$$

Finally, one can consider the correlators of the form
$$
\langle \Phi^\star_{j_3 \pm \frac12, \epsilon_3} ( x_3, \infty)   \Phi^\star_{j_2 \pm 1, \epsilon_2} ( x_2,z)  \Phi^\star_{j_1\pm \frac12,\epsilon_1} ( x_1,0) \rangle \, ,
$$
when the  Ward identities for $\widehat{\rm su}(2)_1$ currents yield a nontrivial contribution and the already established normalizations are assumed.
We checked all examples of this type. The calculations are similar to these already presented, but more lengthy. With the sign factor given by the tables in Appendix C they agree  with conjecture (\ref{cosetfactor}) in each case.

\section{Conclusions}

In this paper we have formulated exact relations between
$\widehat{\rm su}(2)$ WZNW model with non-rational level on the one hand side and the Liouville or the imaginary Liouville field theories on the other.
These relations can be seen as  continuous spectra counterparts of the GKO construction of the minimal models.
We have found strong evidences that they are correct.
There are, however, still many questions left open.

First of all one would like to prove decomposition (\ref{decorep})
and find an explicit construction of the excited states generating representations on the r.h.s. of (\ref{decorep}).
In an analogous problem in the SL-LL correspondence the free field realization of the Virasoro and super Virasoro Verma modules proved to be helpful.
Whether the Wakimoto realization might be relevant in the present context is still not clear to us.
Another approach is to explore the relation between these excited states and the null states in the relaxed (spin basis) and the highest weight (izospin variables) modules and its behavior under the reflection map.
The next step would be to calculate the general form of the coset factor (\ref{cosetfactor}).
This point was the most difficult part of the proof of the SL-LL relation and one may expect similar difficulties in the present case.
The last step - the proof of equivalence of the 3-point functions for descendants of the fields $\Phi^*_{j+n,\epsilon}$ with respect
to the algebra $\widehat{\rm su}(2)_{\kappa+1}\otimes {\rm Vir}$ is then possible by slight generalizations of the arguments used in the SL-LL case.

Extending the equivalence to $n$-point correlation functions ($n>3$) on the sphere requires a better understanding of the $\widehat{\rm su}(2)_\kappa$ model introduced in Section 3. The main conjecture of the present paper seems to be a good motivation for further investigations of this model even though
it is based on non-unitary representations.
There are some obvious steps in this direction: the analysis of the 4-point invariants and the structure of the 4-point conformal blocks,
the derivation of the $j$-dependent part of the structure constants by Teschner's method and the factorization properties of the 4-point functions.
In a slightly more general context an analytic expression for the fusion matrix would be an essential step in  a deeper understanding of
the scheme of fig.2.

Another group of  questions concerns relations of the model of Section 3 to other WZNW models with continuous spectra
like the $H^+_3$ coset model based on different class of representations. It would be also interesting to explore the relation to the
$\widehat{\rm sl}(2,\mathbb{R})$ model based on the principal unitary series of ${\rm sl}(2,\mathbb{R})$ representations
which differs from the present one only by the invariant hermitian form.

There are interesting problems related to the GSO construction itself. Motivated by some aspects of various generalizations of the AGT relation \cite{Nishioka:2011jk,Wyllard:2011mn} and the coset constructions of rational CFT
 one can look for non-rational counterparts of  (\ref{rationalcoset}) with integer $p$ bigger then 1.
This leads to the following conjectures
$$
\begin{array}{rcrl}
\widehat{\rm su}(2)_{\kappa} \ \otimes\ \widehat{\rm su}(2)_{2}
\hskip 8pt &\sim &\hskip 8pt
\hbox{$N=1$ super-Liouville}\ \otimes_P\ \widehat{\rm su}(2)_{\kappa+2}\,,
\\
\widehat{\rm su}(2)_{\kappa} \ \otimes\ \widehat{\rm su}(2)_{p}
\hskip 8pt &\sim & \hskip 8pt
\hbox{para-Liouville}\ \otimes_P\ \widehat{\rm su}( 2)_{\kappa+p}\,,&\;\;\;\;p>2.
\end{array}
$$
As the structure constants of the N=1 super-Liouville \cite{Rashkov:1996jx,Poghosian:1996dw} and the
para-Liouville theories \cite{Bershtein:2010wz} are already known this is
a perfect ground to test the coset construction.

Even more interesting would be a generalization to the symmetry algebras $\widehat{\rm su}(N), N>2$. In this case one should expect the
relations involving  Toda \cite{Fateev:2007ab,Fateev:2008bm} and the para-Toda field theories
$$
\begin{array}{rcrl}
\widehat{\rm su}(N)_{\kappa} \ \otimes\ \widehat{\rm su}(N)_{1}
\hskip 8pt &\sim &\hskip 8pt
\hbox{ Toda }\ \otimes_P\ \widehat{\rm su}(N)_{\kappa+1}\,,
\\
\widehat{\rm su}(N)_{\kappa} \ \otimes\ \widehat{\rm su}(N)_{p}
\hskip 8pt &\sim & \hskip 8pt
\hbox{para-Toda}\ \otimes_P\ \widehat{\rm su}( N)_{\kappa+p}\,,&\;\;\;\;p>1.
\end{array}
$$
A challenging  problem is to analyze whether
the recently proposed Toda field theory structure constants \cite{Mitev:2014isa,Isachenkov:2014eya}  fit the general scheme of Fig.1 and Fig.2.
This however requires the  $\widehat{\rm su}(N)$ WZNW structure constants which to our knowledge  are not yet known.

Let us observe that for $n_i=0$ equation (\ref{3-pkt_gen2}) relates
the $\widehat{\rm su}(2)$ WZNW and the Liouville structure constants. If the conjecture holds for higher point functions as well it would
allow to calculate $n$-point functions of primary fields of the Liouville theory in terms of correlators of the $\widehat{\rm su}(2)$ WZNW model:
\begin{eqnarray*}
\langle \Phi_{j_k} (z_k)\dots \Phi_{j_2} ( z_2) \Phi_{j_1} (z_1)  \rangle\sL
&=&
{\langle
\Phi\sa_{j_k,\epsilon_k} (x_k,z_k) \dots
\Phi\sa_{j_2,\epsilon_2} (x_2,z_2)
\Phi\sa_{j_1,\epsilon_1} (x_1, z_1) \rangle^\s
\over
 \langle
\Phi\sB_{j_k, \epsilon_k} (x_k, z_k) \dots
\Phi\sB_{j_2,  \epsilon_2 } ( x_2,z_2)
\Phi\sB_{j_1, \epsilon_1 } ( x_1, z_1)  \rangle^\h  }\, ,
\end{eqnarray*}
where  all the variables of the right sector are suppressed. In this particular case the symmetries on both sides
allow for independent calculations.
There are however situations
when the symmetry of the coset  is not strong enough to fix the structure constants of the theory
(e.g. the Toda field theories) or it is not known at all (e.g. most of the para-Liouville field theories).
Solving the $\widehat{\rm su}(N)$ WZNW model would  automatically provide solutions to its various cosets.

The coset construction for non-rational CFT models seems to be a powerful tool for analyzing
basic common structures of large classes of models.
This provides a strong motivation to investigate the $\widehat{\rm su}(2)$ WZNW models with non-rational levels
and their generalizations for other groups.

\appendix

\section{reflection relation}

Using the property of the $G$-function defined in (\ref{Gfunction})
\bea
G\left[
{}^{a\;\;b\;\; c}_{e\;\;f}\right]
 &=&  \frac{\Gamma(b ) \Gamma(c ) \Gamma(e+f-a-b-c) }{\Gamma(f-a) \Gamma(e-b ) \Gamma(e-c ) } \, G\left[
{}^{a\;\;e-b\;\;e- c}_{e\;\;e+f-b-c}\right] \, .
\eea
one can find the transformations of functions $g^{31}, g^{13}$ under the reflection $j_3 \to -j_3 -1$
\bea
G\left[
{}^{-j_1-m_1\;\;\;1+ j^1_{23}\;\;\;1+j_3+ m_3}_{1+j_2- j_1+m_3\;\;2+j_2+j_3-m_1}\right]
 &=&  \frac{\Gamma(1+j^1_{23} ) \Gamma(1+j_3+m_3 ) \Gamma(1+j^3_{12} ) }{\Gamma(2+ j_{123}) \Gamma( - j_3+m_3)\Gamma(-j^2_{13})}
G\left[
{}^{-j_1-m_1\;\;\;-j_3+m_3\;\;\; -j^2_{13}}_{1+j_2-j_1+m_3\;\;1-j_3+j_2-m_1}\right],
\\
G\left[
{}^{-j_1+m_1\;\;\;1+ j^1_{23}\;\;\;1+j_3- m_3}_{1+j_2- j_1-m_3\;\;2+j_2+j_3+m_1}\right]
 &=&  \frac{\Gamma(1+j^1_{23} ) \Gamma(1+j_3-m_3 ) \Gamma(1+j^3_{12} ) }{\Gamma(2+ j_{123}) \Gamma( - j_3-m_3)\Gamma(-j^2_{13})}
G\left[
{}^{-j_1+m_1\;\;\;-j_3-m_3\;\;\; -j^2_{13}}_{1+j_2-j_1-m_3\;\;1-j_3+j_2+m_1}\right].
\eea
This leads to the relation
\bea
g^{\epsilon} (-j_3-1, j_2,j_1; m_i) &=& \frac{\Gamma(1+j^1_{23} ) \Gamma(1+j^3_{12} ) }{\Gamma(-j^2_{13} ) \Gamma(2+ j_{123})}
\,
 s\left(\textstyle -\frac12 -\frac12 j_{23}^1 - \epsilon +\epsilon_2\right) \, s(j_1-\epsilon_1) s(-j_2-1+\epsilon_2)
 \\
 &\times &
    \left(   \frac{\Gamma( 1+ j_3-m_3)}{\Gamma( - j_3-m_3)} g^{31}
+ (-1)^{2\epsilon}  \frac{\Gamma( 1+ j_3+ m_3)}{\Gamma( - j_3+ m_3)} \, g^{13}   \right)
\eea
and
\bea
&& \hspace{-30pt}
S_{\epsilon}\Big[\!
\begin{array}{ccc}
\\[-18pt]
\scriptstyle j_3 & \scriptstyle -1-j_2  & \scriptstyle -1-j_1
\\[-8pt]
\scriptstyle \epsilon_3  &  \scriptstyle \epsilon_2  & \scriptstyle \epsilon_1
\\[-8pt]
\scriptstyle m_3  &  \scriptstyle m_2  &  \scriptstyle m_1
\end{array}
\!\Big]
=
 \frac{ \Gamma(- j_{23}^1)  }{ \Gamma(1+ j_{13}^2) }
 \sum_{\epsilon'}\left[
 \frac{s(-\frac12 -\frac12 j_{23}^1 + \epsilon + \epsilon_{2}) }{ 2 s(\frac12 j_{13}^2 - \epsilon' + \epsilon_{2}) }
 \right.
 \\
 &&
\hspace{10pt}\left.\times   \Bigg(   \frac{\Gamma( 1+ j_3-m_3-\epsilon_3)}{\Gamma( - j_3-m_3-\epsilon_3)}
+ (-1)^{2(\epsilon+ \epsilon')}  \frac{\Gamma( 1+ j_3+ m_3+\epsilon_3)}{\Gamma( - j_3+ m_3+ \epsilon_3)}   \Bigg)
S_{\epsilon'}\Big[\!
\begin{array}{ccc}
\\[-18pt]
\scriptstyle -1-j_3 & \scriptstyle -1-j_2  & \scriptstyle -1-j_1
\\[-8pt]
\scriptstyle \epsilon_3  &  \scriptstyle \epsilon_2  & \scriptstyle \epsilon_1
\\[-8pt]
\scriptstyle m_3  &  \scriptstyle m_2  &  \scriptstyle m_1
\end{array}
\!\Big]\right]\, .
\eea
Taking into account
the identity
$$
\frac{\Gamma( - j_3+ m_3+ \epsilon_3)}{\Gamma( 1+ j_3+ m_3+\epsilon_3)}  \frac{\Gamma( 1+ j_3-m_3-\epsilon_3)}{\Gamma( - j_3-m_3-\epsilon_3)}
=
(-1)^{2\epsilon_3}\, ,
$$
one gets the first equation of (\ref{refrel})
\bea
&& \hspace{-30pt}
\frac{\Gamma( - j_3+ m_3+ \epsilon_3)}{\Gamma( 1+ j_3+ m_3+\epsilon_3)}
S_{\epsilon}\Big[\!
\begin{array}{ccc}
\\[-18pt]
\scriptstyle j_3 & \scriptstyle -1-j_2  & \scriptstyle -1-j_1
\\[-8pt]
\scriptstyle \epsilon_3  &  \scriptstyle \epsilon_2  & \scriptstyle \epsilon_1
\\[-8pt]
\scriptstyle m_3  &  \scriptstyle m_2  &  \scriptstyle m_1
\end{array}
\!\Big]
 \\
   &=& (-1)^{2\epsilon_3} \frac{ \Gamma(- j_{23}^1)  }{ \Gamma(1+ j_{13}^2) }
 \frac{s(-\frac12 -\frac12 j_{23}^1 + \epsilon + \epsilon_{2}) }{  s(\frac12 j_{13}^2 - (\epsilon\+\epsilon_3) + \epsilon_{2}) }
 S_{\epsilon\+ \epsilon_3}\Big[\!
\begin{array}{ccc}
\\[-18pt]
\scriptstyle -1-j_3 & \scriptstyle -1-j_2  & \scriptstyle -1-j_1
\\[-8pt]
\scriptstyle \epsilon_3  &  \scriptstyle \epsilon_2  & \scriptstyle \epsilon_1
\\[-8pt]
\scriptstyle m_3  &  \scriptstyle m_2  &  \scriptstyle m_1
\end{array}
\!\Big].
\eea
One checks that
the coefficient has expected properties with respect to the exchange $1\leftrightarrow 2$:
\begin{eqnarray*}
\frac{ \Gamma(- j_{23}^1)  }{ \Gamma(1+ j_{13}^2) }
 \frac{s(-\frac12 -\frac12 j_{23}^1 + \epsilon + \epsilon_{2}) }{  s(\frac12 j_{13}^2 - (\epsilon\+\epsilon_3) + \epsilon_{2}) }
 &=&
 (-1)^{2\epsilon_3}
\frac{ \Gamma(- j_{13}^2)  }{ \Gamma(1+ j_{23}^1) }
 \frac{s(-\frac12 -\frac12 j_{13}^2 + \epsilon + \epsilon_{1}) }{  s(\frac12 j_{23}^1 - (\epsilon\+\epsilon_3) + \epsilon_{1}) },
\end{eqnarray*}
which yields the second  equation of (\ref{refrel}).

 \section{Gamma Barnes identities}

We shall start from the identities for the Barnes gamma function relevant in the Liouville case
\begin{eqnarray}
\nonumber
{\Gamma_{b\sa}(-b\sa j ) \Gamma_{\hat  b\sB}\left(\frac{1}{\hat  b\sB} - \hat b\sB j \right) \over \Gamma_{b}(-Q j ) }
&=& a_1(j),
\\
\label{barnesliouville}
{\Gamma_{b\sa}(\frac{1}{b\sa}+b\sa( j+1) ) \Gamma_{\hat  b\sB}( \hat b\sB (j+1) ) \over \Gamma_{b}(Q( j+1) ) }&=&a_1(j),
\\
\nonumber
 &&\hspace{-140pt}a_1(j)\;\equiv\; {  \Gamma_{b\sa}(b\sa)
  \Gamma_{\hat b\sB}\left(\frac{1}{\hat  b\sB} + \hat b\sB  \right) \over  \Gamma_{b}(Q)}
\left( {b^{-1} Q  } \right)^{ {Q^2 j(j+1)\over 4} }
b^{-\frac12 b Q j(j+1)  -\frac12 j-\frac12}\, ,
\end{eqnarray}
where
\bea
b\sa &=& \sqrt{b Q},\;\;\;\hat b\sB \;=\; \sqrt{b^{-1}Q},\;\;\;Q=b+b^{-1}.
\eea
Using the shift relations
\begin{eqnarray*}
\nonumber
\Gamma_b(x+b)
& = &
\sqrt{2\pi}\,b^{bx-\frac12}\,\Gamma^{-1}\left(bx\right)\,\Gamma_b(x),
\\[-6pt]
\label{Gamma:b:shift}
\\[-6pt]
\nonumber
\Gamma_b\left(x+b^{-1}\right)
& = &
\sqrt{2\pi}\,b^{-x/b+\frac12}\,\Gamma^{-1}\left(x/b\right)\,\Gamma_b(x),
\end{eqnarray*}
one checks that all the left hand sides of equations (\ref{barnesliouville}) have the same properties with respect to the shift
$j\to j - \frac{ b}{ Q} $:
\bea
&&\hspace{-100pt}{\Gamma_{b\sa}(-b\sa j + b\sa - \frac{1}{b\sa} ) \Gamma_{\hat  b\sB}(\frac{2}{\hat  b\sB} - \hat b\sB j  ) \over \Gamma_{b}(-Q j + b ) }
\\
&=&
 b^{  b^2 j + \frac{b}{Q}} \, \left(\sqrt{1+ b^2}\right)^{- \frac12-j bQ }   {\Gamma_{b\sa}(-b\sa j ) \Gamma_{\hat  b\sB}(\frac{1}{\hat  b\sB} - \hat b\sB j ) \over \Gamma_{b}(-Q j ) },
\\
&&\hspace{-100pt}
\frac{\Gamma_{b\sa}\left(\frac{2}{b\sa}+ b\sa j\right)\Gamma_{b\sB}\left(b\sB +b\sB j - \frac{1}{b\sB}\right)}
{\Gamma_b(Q+Qj-b)}
\\
&=&
 b^{  b^2 j + \frac{b}{Q}} \, \left(\sqrt{1+ b^2}\right)^{- \frac12-j bQ }
\frac{\Gamma_{b\sa}\left(\frac{1}{b\sa}+ b\sa+ b\sa j\right)\Gamma_{b\sB}\left(b\sB +b\sB j \right)}
{\Gamma_b(Q+Qj)} ,
\\
 a_1\left(j-\frac{ b}{ Q}\right) &=&   b^{  b^2 j + \frac{b}{Q}} \, \left(\sqrt{1+ b^2}\right)^{- \frac12-j bQ }   a_1(j),
\end{eqnarray*}
and to the shift  $j\to j - \frac{1}{  b Q} $:
\bea
&&\hspace{-100pt}
{\Gamma_{b\sa}(-b\sa j  + \frac{1}{b\sa} ) \Gamma_{\hat  b\sB}(- \hat b\sB j + \hat b\sB ) \over \Gamma_{b}(-Q j + b^{-1} ) }
\\
&=&
b^{\frac12 + j} \,  \left(\sqrt{1+ b^2}\right)^{- \frac12-j\frac{Q}{b}  }
\,
{\Gamma_{b\sa}(-b\sa j ) \Gamma_{\hat  b\sB}(\frac{1}{\hat  b\sB} - \hat b\sB j ) \over \Gamma_{b}(-Q j ) },
\\
&&\hspace{-100pt}
\frac{\Gamma_{b\sa}\left({b\sa}+ b\sa j\right)\Gamma_{b\sB}\left( b\sB j + \frac{1}{b\sB}\right)}
{\Gamma_b\left(Q+Qj-\frac{1}{b}\right)}
\\
&=&
b^{\frac12 + j} \,  \left(\sqrt{1+ b^2}\right)^{- \frac12-j\frac{Q}{b}  }
\,
\frac{\Gamma_{b\sa}\left(\frac{1}{b\sa}+ b\sa+ b\sa j\right)\Gamma_{b\sB}\left(b\sB +b\sB j \right)}
{\Gamma_b(Q+Qj)} ,
\\
 a_1\left(j-\frac{ 1}{b Q}\right) &=&
  b^{\frac12 + j} \,  \left(\sqrt{1+ b^2}\right)^{- \frac12-j\frac{Q}{b}   }
\, a_1(j).
\end{eqnarray*}
For non-rational $b$ it  yields the proof of (\ref{barnesliouville}) up to $j$-independent  factors. They can be found
calculating the sides of (\ref{barnesliouville}) at $j=-1$ and $j=0$, respectively.

Multiplying the first two equations of (\ref{barnesliouville}) side by side one gets the identity for the upsilon functions
\bea
\frac{\Upsilon_{b\sa}(-b\sa j ) \Upsilon_{\hat  b\sB}\left(\frac{1}{\hat  b\sB} - \hat b\sB j \right)}
{\Upsilon_{b}(-Q j )}
&=&
\frac{\Upsilon_{b\sa}(b\sa ) \Upsilon_{\hat  b\sB}\left(\frac{1}{\hat  b\sB} + \hat b\sB \right)}
{\Upsilon_{b}\left(\frac{1}{b} +b \right)}
\left( {b^{-1} Q  } \right)^{{Q^2 j(j+1)\over 2} }
b^{-b Q j(j+1)  - j-1} .
\eea

Using the same method one can prove the identities
\begin{eqnarray*}
{ \Gamma_{b\sB}(-b\sB j) \over \Gamma_{b\sa}(-b\sa j)  \Gamma_{\bh}( - \hat Q j + \bh)}
&=&
{ \Gamma_{b\sB}\left(\frac{1}{b\sB}+ b\sB (j+1) \right) \over \Gamma_{b\sa}(\frac{1}{b\sa}+b\sa( j+1) )  \Gamma_{\bh}(  \hat Q( j+1)  + \bh)}
\;=\; a_2(j),
\\
\nonumber
a_2(j) &=& {  \Gamma_{b\sB}(b\sB) \over \Gamma_{b\sa}(b\sa) \Gamma_{\bh}\left(\frac{1}{\bh}\right)}
\left( \bh^{-1} \hat Q   \right)^{\frac{\hat Q^2 j( j + 1)}{4}}
\bh^{ \frac{\hat Q}{2\hat b}j(j+1) + \frac12 j+\frac12},
\end{eqnarray*}
where
\bea
b\sa &=& \sqrt{\hat b\hat Q},\;\;\;b\sB \;=\;\sqrt{\hat b^{-1}\hat Q},\;\;\;\hat Q\;=\;\hat b^{-1} -\hat b.
\eea
They imply the identity
$$
\frac{\Upsilon_{b\sa}(-b\sa j)  \Upsilon_{\bh}( - \hat Q j + \bh)}{ \Upsilon_{b\sB}(-b\sB j)}
=
\frac{\Upsilon_{b\sa}(b\sa )  \Upsilon_{\bh}( \bh)}{ \Upsilon_{b\sB}(b\sB )}
\left( \bh^{-1} \hat Q   \right)^{\frac{\hat Q^2 j( j + 1)}{2}}
\bh^{ \frac{\hat Q}{\hat b}j(j+1) +  j+1} .
$$

 \section{sign factor $(-1)^{\eta(n_3,n_2,n_1)}$}

\begin{tabular}{|c|c|c|c|}
\hline
$n_1$ &$ n_2$ &$ n_3$ & $\eta(n_3,n_2,n_1)$ \\
\hline
& & & \\[-13pt]
\hline
$\scriptstyle{ 1/2} $ & $\scriptstyle{ 1/2} $ & 0 & 0  \\
\hline
0 & $ \scriptstyle{ 1/2}  $ & $\scriptstyle{ 1/2} $ & 0  \\
\hline
$ \scriptstyle{ 1/2}  $ & 0 & $\scriptstyle{ 1/2} $  & 0  \\
\hline
& & & \\[-14pt]
\hline
$ \scriptstyle{ 1/2} $ &   $-\scriptstyle{ 1/2} $ & 0 & 1  \\
\hline
0 & $ \scriptstyle{ 1/2}  $ &  $-\scriptstyle{ 1/2} $ & 1  \\
\hline
$- \scriptstyle{ 1/2}  $ & 0 & $\scriptstyle{ 1/2} $  & 1  \\
\hline
& & & \\[-14pt]
\hline
$ - \scriptstyle{ 1/2}  $ & $\scriptstyle{ 1/2} $ & 0 & 0  \\
\hline
0 &  $- \scriptstyle{ 1/2}  $ & $\scriptstyle{ 1/2} $ & 0  \\
\hline
$ \scriptstyle{ 1/2}  $ & 0 & $-\scriptstyle{ 1/2} $  & 0  \\
\hline
& & & \\[-14pt]
\hline
$ -\scriptstyle{ 1/2}  $ & $-\scriptstyle{ 1/2} $ & 0 & 0  \\
\hline
0 & $ -\scriptstyle{ 1/2}  $ & $ -\scriptstyle{ 1/2} $ & 0  \\
\hline
$- \scriptstyle{ 1/2}  $ & 0 & $- \scriptstyle{ 1/2} $  & 0  \\
\hline
\end{tabular}
\hspace{60pt}
\begin{tabular}{|c|c|c|c|}
\hline
$n_1$ &$ n_2$ &$ n_3$ & $\eta(n_3,n_2,n_1)$ \\
\hline
& & & \\[-13pt]
\hline
$ \pm 1 $ & 0 & 0 & 1  \\
\hline
0 & $ \pm 1 $  & 0 & 1  \\
\hline
0 & 0 & $ \pm 1 $ &  1  \\
\hline
& & & \\[-13pt]
\hline
$ \scriptstyle{ 1/2}  $ & $\scriptstyle{ 1/2} $ & $\pm 1$  & 1  \\
\hline
$\pm 1$ & $ \scriptstyle{ 1/2}  $ & $\scriptstyle{ 1/2} $ & 1  \\
\hline
$ \scriptstyle{ 1/2}  $ & $\pm 1$ & $\scriptstyle{ 1/2} $  & 1  \\
\hline
& & & \\[-14pt]
\hline
$  \scriptstyle{ 1/2}  $ &   $-\scriptstyle{ 1/2} $ &  $\pm 1$ & 1  \\
\hline
 $\pm 1$ & $ \scriptstyle{ 1/2}  $ &  $ -\scriptstyle{ 1/2} $ & 1  \\
\hline
$- \scriptstyle{ 1/2}  $ &  $\pm 1$ & $\scriptstyle{ 1/2} $  & 1  \\
\hline
& & & \\[-14pt]
\hline
$ - \scriptstyle{ 1/2}  $ & $\scriptstyle{ 1/2} $ &  $\pm 1$ & 0 \\
\hline
 $\pm 1$ &  $- \scriptstyle{ 1/2}  $ & $\scriptstyle{ 1/2} $ & 0   \\
\hline
$ \scriptstyle{ 1/2}  $ &  $\pm 1$ & $ -\scriptstyle{ 1/2} $  & 0  \\
\hline
& & & \\[-14pt]
\hline
$ -\scriptstyle{ 1/2}  $ & $-\scriptstyle{ 1/2} $ & $\pm 1$  & 1   \\
\hline
$\pm 1$  & $ -\scriptstyle{ 1/2}  $ & $ -\scriptstyle{ 1/2} $ & 1   \\
\hline
$- \scriptstyle{ 1/2}  $ &  $\pm 1$ & $- \scriptstyle{ 1/2} $  & 1  \\
\hline
\end{tabular}

\acknowledgments

PS is grateful to  SLAC National Accelerator Laboratory for hospitality.

\end{document}